\documentclass[fleqn,usenatbib]{mnras}

\usepackage{newtxtext,newtxmath}

\usepackage[T1]{fontenc}
\usepackage{ae,aecompl}

\usepackage{graphicx}	
\usepackage{amsmath}	
\usepackage[dvipsnames]{xcolor}
\usepackage{multirow}
\usepackage{delarray}
\usepackage{color}
\usepackage{here}
\usepackage{float}
\usepackage{tensor}
\usepackage{xspace}
\usepackage{orcidlink}
\usepackage{nicefrac}
\usepackage[multiple]{footmisc}

\newcommand{\xvect}{\mathbf{x}}
\newcommand{\dif}{\mathrm{d}}

\newcommand{\be}{\begin{equation}} \newcommand{\ee}{\end{equation}}

\newcommand{\solarmass}{M_{\rm \sun}}
\newcommand{\msun}{\solarmass}

\newcommand{\pc}{\mathrm{pc}}

\newcommand{\vvector}{\mathbf{v}}
\newcommand{\Bvector}{\mathbf{B}}
\newcommand{\grad}{\mathbf{\nabla}}
\newcommand{\appropto}{\mathrel{\v{
  \offinterlineskip\halign{\hfil$##$\cr
    \propto\cr\noalign{\kern2pt}\sim\cr\noalign{\kern-2pt}}}}}

\defcitealias{hopkins2015_gizmo}{H15}
\defcitealias{hopkins_gizmo_mhd}{HR16}
\defcitealias{Springel_2005_gadget}{S05}
\defcitealias{truelove_1997_dens_condition}{T97}
\defcitealias{guszejnov_isothermal_mhd}{Paper 0}
\defcitealias{grudic_starforge_methods}{Paper I}
\defcitealias{starforge_jets_imf}{Paper 2}

\usepackage[normalem]{ulem}

\title[STARFORGE]{STARFORGE: Toward a comprehensive numerical model of star cluster formation and feedback}

\author[Grudi\'{c} et al.]{
Michael Y. Grudi\'{c}\orcidlink{0000-0002-1655-5604}$^{1}$\thanks{mike.grudic@northwestern.edu},
D\'avid Guszejnov\orcidlink{0000-0001-5541-3150}$^{2}$\thanks{guszejnov@utexas.edu},
Philip F. Hopkins\orcidlink{0000-0003-3729-1684}$^{3}$,
\newauthor
Stella S. R. Offner\orcidlink{0000-0003-1252-9916}$^{2}$, and
Claude-Andr{\'e} Faucher-Gigu{\`e}re\orcidlink{0000-0002-4900-6628}$^{1}$
\\
$^{1}${CIERA and Department of Physics and Astronomy, Northwestern University, 1800 Sherman Ave, Evanston, IL 60201, USA}\\
$^{2}$Department of Astronomy, The University of Texas at Austin, TX 78712, USA \\
$^{3}$TAPIR, Mailcode 350-17, California Institute of Technology, Pasadena, CA 91125, USA \\
}

\date{\today \vspace{-0.6cm}}

\pubyear{2020}

\begin{document}
\label{firstpage}
\pagerange{\pageref{firstpage}--\pageref{lastpage}}
\maketitle

\begin{abstract}
We present STARFORGE (STAR FORmation in Gaseous Environments): a new numerical framework for 3D radiation MHD simulations of star formation that simultaneously follow the formation, accretion, evolution, and dynamics of individual stars in massive giant molecular clouds (GMCs) while accounting for stellar feedback, including jets, radiative heating and momentum, stellar winds, and supernovae. We use the {\small GIZMO} code with the MFM mesh-free Lagrangian MHD method, augmented with new algorithms for gravity, timestepping, sink particle formation and accretion, stellar dynamics, and feedback coupling. We survey a wide range of numerical parameters/prescriptions for sink formation and accretion and find very small variations in star formation history and the IMF (except for intentionally-unphysical variations). Modules for mass-injecting feedback (winds, SNe, and jets) inject new gas elements on-the-fly, eliminating the lack of resolution in diffuse feedback cavities otherwise inherent in Lagrangian methods. The treatment of radiation uses {\small GIZMO}'s radiative transfer solver to track 5 frequency bands (IR, optical, NUV, FUV, ionizing), coupling direct stellar emission and dust emission with gas heating and radiation pressure terms. We demonstrate accurate solutions for SNe, winds, and radiation in problems with known similarity solutions, and show that our jet module is robust to resolution and numerical details, and agrees well with previous AMR simulations. STARFORGE can scale up to massive ($>10^5 \msun$) GMCs on current supercomputers while predicting the stellar ($\gtrsim 0.1\msun$) range of the IMF, permitting simulations of both high- and low-mass cluster formation in a wide range of conditions.
\end{abstract}



\section{Introduction}
\label{sec:intro}
 Many physical mechanisms are important in star formation (SF). It is initiated by the formation of radiatively-cooled, gravitationally-unstable cores of gas and dust from magnetized, supersonic, turbulent flows found in giant molecular clouds (GMCs) \citep{larson_law, maclow_star_formation_ism,mckee_sf_theory, girichidis_2020_sf_review}. These cores collapse to protostars, and once formed, protostars and stars  influence the surrounding gas flow in via feedback: the injection of mass, momentum and energy into the ISM in the form of radiation, accretion-powered collimated bipolar outflows (hereafter simply ``jets"), radiatively-driven stellar winds, and supernova (SN) explosions, which may ultimately limit the total stellar mass that can form. The accretion of individual stars is eventually truncated by feedback, gas exhaustion, or dynamical interactions with gas clumps or other stars, setting their final masses \citep{krause2020}. Hence, the problem of SF is an intricate, tightly-coupled interaction of gravity, magnetohydrodynamics (MHD), atomic and molecular physics, radiation, stellar physics, and feedback.

A basic requirement of any star formation theory is to explain the hallmark phenomena of SF, including the stellar initial mass function (IMF), the (in-)efficiency of SF, and the properties of stellar clusters and associations \citep{sf_big_problems}. These phenomena must emerge from the various processes at work in GMCs, so it is important to disentangle the physics' respective roles. This has yet to be accomplished, partly because the wide range of length-scales and multitude of physics involved make SF very challenging to model. 

\subsection{Requirements for a complete dynamical model of star formation and feedback}
While some progress has been made with simpler models that consider only e.g. turbulence and gravity \citep{padoan_nordlund_1997_imf,padoan_nordlund_2002_imf, krumholz05a, hc08, padoan_nordlund_2011_imf,excursion_set_ism,hennebelle_theory}, other physics are likely to be important. In particular, {\it feedback} is important for understanding the end-point of star formation (the disruption of GMCs), and its implications for other questions such as the IMF and stellar multiplicity have only begun to be explored.  Many analytic and semi-analytic calculations of the effects of feedback in GMCs have been performed (for reviews see \citealt{mckee_sf_theory, krumholz_2019_cluster_review}), yielding useful dimensional arguments and analytic insights. But GMCs are complex, turbulent, inherently three-dimensional entities that evolve on their internal crossing timescale \citep{larson_law,maclow_star_formation_ism}. Thus, even under the gross simplification of treating GMCs as isolated entities (i.e. neglecting galactic environment), (semi-)analytic predictions inevitably hinge on many highly-uncertain assumptions. 
With so much parameter freedom it is difficult to say whether a given model is correct for the {\it right reasons}, limiting physical insight and ultimately predictive power. Direct numerical simulations of star formation are a necessary tool to resolve these uncertainties.

In the past two decades, great progress has been made incorporating stellar feedback into direct numerical simulations of star-forming GMCs (for reviews see \citealt{dale_feedback_review,krumholz_2019_cluster_review}). But these studies have shown that further progress on the key questions of star formation requires next-generation simulations that do all of the following:
\begin{enumerate}
    \item \label{req1} {\bf Resolve individual star formation:} 
    Many simulations of star cluster formation do not attempt to resolve the formation and ongoing accretion of individual stars across the entire stellar mass range, instead relying on a sub-grid SF prescription that either enforces a certain underlying IMF directly or is fine-tuned to recover the observed one \citep{colin:2013.rhd.sf.sims, sormani_imf_sampling, howard_2017_imf_sampling,kim_2017_rhd_methods,grudic_2016,su_imf_sampling, lahen_2019_cluster_formation,he_2019_gmc_fb,wall_2020_amuse_torch_fb}. But there is an infinite number of ways to do this, each with different implicit assumptions about how star formation works, and the choice of prescription can have a major effect upon simulation results \citep{elephant}, limiting predictive power. Simulations should ideally attempt to resolve the formation and accretion of individual stars (or sink particles), and to recover a realistic IMF self-consistently from physical (not numerical) processes. This is obviously necessary anyway if one wishes to study the physical origins of the IMF and stellar multiplicity.
    \item \label{req2} {\bf Follow stellar dynamics:} SF simulations that do not integrate stellar orbits explicitly generally discretize the stellar mass formed into a collisionless fluid represented by gravitationally-softened particles \citep[e.g.][]{grudic_2016, Li_Vogelsberger_2019_GMC_disrupt, lahen_2019_cluster_formation}, which can produce qualitatively-correct star cluster density profiles \citep{grudic_2017,lahen_2020_cluster_structure}, but have the severe limitation that the collisionless description (and phase-space density conservation) breaks down on mass scales $M_{\rm \star}\lesssim100M_\odot$, so if cluster formation is a hierarchical assembly from smaller masses \citep[e.g.][]{bonnell:2003.hierarchical}, then individual stellar dynamics is {\it always} important at some stage in the process. A simulation must also treat dynamics on the scale of binary separations to accurately predict stellar multiplicity, let alone phenomena such as common disk accretion \citep[e.g.][]{munoz_2019_binary_accretion, Lee_2019_binary_MHD,duffell_2020_binary_accretion}.

    \item \label{req3} {\bf Follow MHD, chemistry, and cooling:} Obviously, following the dynamics of GMCs, star formation, and accretion requires gas-dynamical simulations, and stars cannot form if gas cannot radiatively cool. Moreover, the ISM is magnetized, and this fact can easily have important implications for star formation. The magnetic field can act as an additional source of support, potentially stabilizing otherwise-unstable cores \citep{chandrasekhar_grav_collapse,Mouschovias_Spitzer_1976_magnetic_collapse}, affecting the IMF \citep{price_bate_2007_mhd_sf,guszejnov_isothermal_mhd}, the rate of star formation \citep{federrath_klessen_2012, federrath_2015_inefficient_sf}, and altering the pressure balance, morphology, growth of instabilities, and transport of energy in feedback bubbles \citep{krumholz_2007_mhd_hii,offner_2018_mhd_feedback,Krumholz_2019_IMF_magnetic_field}.
    
    \item \label{req4}{\bf Scale up to massive GMCs}: Current star formation simulations that do both \ref{req1} and \ref{req2} have focused upon lower-mass systems, simulating gas masses of $100-1000 \msun$ \citep{Jones_Bate_2018_radiation,Wurster_2019_no_magnetic_break_catastrophe,Lee_Hennebelle_2018_IC,federrath_2015_inefficient_sf, li_2018_mhd_jets_sf, Cunningham_2018_feedback, Colman_Teyssier_2019_tidal_screening}, producing $\sim 10-100 M_\odot$ in stars. Low-mass clusters are important to model, as they can be readily compared to well-studied sites of star formation in the Solar neighborhood \citep[e.g.][]{Evans_2014_MW_GMC_SFR}, but the overwhelming majority of SF in our Galaxy occurs in massive complexes with gas mass $>>10^5 M_\odot$ \citep{mckee_williams_ob_assoc, murray_rahman_ob_assoc}. 
    Simulated low-mass clusters are also less likely to host massive ($\gtrsim 10\msun$) stars, and hence cannot be used to study massive SF. \footnote{Some works have simulated massive cluster and star formation with nominally individual star particles, incorporating SN \citep{ Padoan_2017_SN_driving_SFE,padoan_2019_massive_sf} and photoionization \citep{Gavagnin_2017_SF_feedback} feedback, but at fairly modest ($\sim 500-1500\rm AU$) resolution compared to state-of-the-art low-mass SF simulations. At this resolution it is only possible to follow the widest binaries, and the predicted IMF may suffer bias or low-mass incompleteness.} 

    \item \label{req5}{\bf Account for all major feedback channels:} 3D MHD simulations of SF haven't generally incorporated all known dynamically-important feedback mechanisms (jets, winds, full-spectrum radiation, and SNe) acting in concert. A comprehensive treatment of feedback is needed because different feedback channels are effective on different scales, and can interact nonlinearly. For example, direct radiation pressure from a massive star is ineffective if it couples deep within the star's potential well \citep{krumholz_2018_rad_pressure}, and radiation pressure in general may be subdominant to protostellar outflows for regulating the growth of individual massive stars \citep{rosen_2020_jets_radiation}. But by regulating accretion or punching optically-thin holes, outflows could help photons to couple their momentum farther away from the star, eventually allowing them to disrupt the host GMC \citep{fall2010,Murray_2010_GMC_disruption,hopkins_2012_galaxy_structure, raskutti_2016, grudic_2016, kim_2018_gmc_raytrace, hopkins_grudic_2019, FIRE_RT}. Meanwhile, jets can be a powerful feedback mechanism that can regulate star formation on the $\lesssim 1 \rm pc$ scales of individual cores and dense clumps \citep{matzner_mckee_2000_jets, nakamura_jets,wang_2010_jets, Cunningham_2011_outflow_sim, federrath_2015_inefficient_sf, Offner_Chaban_2017_jets_sfe, Cunningham_2018_feedback}, but may have only weak effects upon the gas kinematics at larger (i.e. $\gtrsim 10 \rm pc$) scales within the GMC \citep{murray_2018_jets}. Many other synergies between feedback mechanisms can also be theorized. 
\end{enumerate}

\subsection{Enter STARFORGE}
In this work we introduce the STARFORGE (STAR FORmation in Gaseous Environments) project\footnote{\url{http://www.starforge.space}}, a new initiative to perform next-generation 3D star cluster formation simulations in massive GMCs. The STARFORGE numerical framework that we have implemented in the {\small GIZMO} code \citep{hopkins2015_gizmo} (hereafter \citetalias{hopkins2015_gizmo}) simultaneously follows the formation, accretion, and dynamics of individual stars in massive GMCs, with optional physics modules capable of accounting for all of the most widely-discussed stellar feedback mechanisms (jets, radiative heating and momentum, stellar winds, and supernovae), satisfying the requirements \ref{req1}-\ref{req5} laid out above. In \citet{guszejnov_isothermal_mhd} (hereafter \citetalias{guszejnov_isothermal_mhd}), we used numerical simulations (using an early version of the methods presented here) to show that the simple recipe of isothermal MHD turbulence and gravity fails to yield a realistic IMF and star formation history in Milky Way conditions, motivating the need for additional physics implemented in STARFORGE. In the present paper (Paper 1), we present and test the numerical methods of STARFORGE, permitting simulations that combine all of the important SF physics discussed here into a realistic simulation of GMC evolution and star cluster formation. In \citet{starforge_jets_imf} (hereafter \citetalias{starforge_jets_imf}) we use the algorithms presented here to explore the effects of protostellar jets upon SF across an unprecedented parameter space of GMC properties and jet model parameters.

This paper is structured as follows. In \S\ref{sec:corephysics}, we present the ``core" algorithms that any 3D star cluster formation simulation must have in some form: MHD, gravity, sink particle methods, and an integration scheme that couples gas and stars stably and achieves acceptable accuracy in stellar dynamics. In \S\ref{sec:thermo} we describe the treatment of cooling, chemistry, and thermodynamics (treating the opacity limit and protostellar heating either with self-consistent radiative transfer, or simple inexpensive approximations). In \S\ref{sec:feedback} we describe and test algorithms for the numerical coupling of stellar feedback in the form of jets, winds, SNe, and radiation. In \S\ref{sec:discussion} we explore various potential applications of these methods beyond isolated GMC simulations, and also enumerate the many caveats and uncertain assumptions inherent in simulating SF and feedback in this manner, motivating future work. In \S\ref{sec:conclusion} we summarize our findings and outline the programme of the STARFORGE project.

\section{Core algorithms for star formation}
\label{sec:corephysics}
The STARFORGE framework is implemented in the {\small GIZMO} multi-method, multi-physics N-body and MHD simulation code \citep{hopkins2015_gizmo}\footnote{\url{http://www.tapir.caltech.edu/~phopkins/Site/GIZMO.html}}. {\small GIZMO} was selected for the project for several reasons. It implements second-order, Galilean-invariant, Lagrangian meshless finite-volume MHD methods (\citet{hopkins_gizmo_mhd}, hereafter \citetalias{hopkins_gizmo_mhd}), which have several useful advantages for SF problems (discussed in \S\ref{sec:mhd}). It includes a gravity solver (\S\ref{sec:gravity}) that is spatially adaptive (solved consistently with the MHD discretization) with near-ideal scaling up to $10^6$ cores \citep{hopkins2017_fire2}. In addition to solving the MHD equations, {\small GIZMO}'s meshless discretization and reconstruction schemes provide a flexible framework for solving additional, non-core physics such as diffusion, conduction, and non-ideal MHD terms \citep{hopkins_2017_diffusion}, radiative transfer \citep{hopkins_grudic_2019, FIRE_RT}, and stellar feedback \citep{Hopkins_2018_sne_feedback}. All equations are integrated according to a flexible, hierarchical powers-of-two timestepping scheme (\S\ref{sec:timestepping}) that makes it possible to follow processes over a wide range of timescales, from the $\sim 10\rm Myr $ lifetime of a GMC to a $\lesssim 1 \rm yr$ binary orbit.

Conceptually, our approach follows previous Lagrangian 3D star formation simulations \citep{klessen00a, bate2003}: we discretize the mass of the GMC and the surrounding medium into discrete elements of mass $\Delta m$, and integrate their evolution in time according to the MHD equations. Eventually the self-gravitating MHD equations can no longer be followed self-consistently at the centres of runaway core collapse, so we replace these centres with sink particles \citep{Bate_1995_accretion} nominally representing individual protostars. These sink particles interact with the gas via gravity, accretion, and optionally feedback, with feedback rates determined by a sub-grid model of (proto-)stellar evolution based upon that used in \citet{Offner_2009_radiative_sim}. We target a MHD resolution scale on the order of a few $10 \rm AU$, comparable to state-of-the-art low-mass star cluster formation simulations. We defer physics on smaller scales (e.g. disk formation, accretion, jet launching, and protostellar evolution) to a sub-grid approach, acknowledging the various caveats that this entails (\S \ref{sec:caveats}).

We provide a glossary of the various numerical resolution-related quantities in Table \ref{tab:glossary}.

\begin{table*}
    \centering
    \begin{tabular}{c|c|c|c}
    Symbol & Meaning & Notes or expression & Fiducial value \\
    \hline
     $\Delta m$ & Normal gas cell mass resolution & Numerical parameter & $10^{-3}\msun$\\
     
     $\Delta m_{\rm w}$ & Wind cell mass resolution & Numerical parameter  & $10^{-4}\msun$\\
     
     $h$ & Volume-equivalent spherical cell radius & Eq. \ref{eq:h} & $0.02\mathrm{pc}\,\Delta m_{\rm -3}^{1/3} n_{\rm  3}^{-1/3}$ \\

     $\Delta x$ & Volume-equivalent Cartesian cell length & $\left(\Delta m/\rho\right)^{1/3} $ & $0.03\mathrm{pc}\,\Delta m_{\rm -3}^{1/3} n_{\rm  3}^{-1/3}$ \\
     
     
     $N_\mathrm{NGB}$ & Effective neighbor number & Numerical parameter &  32 \\ 
     
     $H$ & Kernel radius of compact support & $\left(\frac{3 N_{\rm NGB}\Delta m}{4 \uppi \rho}\right)^{1/3} \approx 2 \Delta x$ & $ 0.06 \mathrm{pc}\,\Delta m_{\rm -3}^{1/3} n_{\rm  3}^{-1/3}$ \\ 
     
     
     $f_{\rm J}$ & Number of Jeans lengths per cell length &  Eq. \ref{eq:jeansnumber}  &  0.03 $n_{\rm 3}^{1/6}\,c_{\rm s,0.2}^{-1}\,\Delta m_{\rm-3}^{1/3}$\\ 
     
     $f_{\rm J,max}$ &  $f_{\rm J}$ value for marginal Jeans resolution & Heuristic; problem-dependent; see \S\ref{sec:jeansresolution}  & 1/2 \\ 
     
     $\Delta x_{\rm J}$ & Minimum Jeans-resolved cell length & Eq. \ref{eq:deltax_J} & $30 \rm AU\times\Delta m_{\rm-3}\,c_{\rm s,0.2}^{-2}$\\
     
     $\rho_{\rm J}$ & Maximum Jeans-resolved density & Eq. \ref{eq:rhoJ} & $3 \times 10^{-14} \mathrm{g\,cm^{-3}} \times  \Delta m_{\rm-3}^{-2} c_{\rm s,0.2}^{-6}$ \\
     
      $t_{\rm acc}$ & Accretion smoothing timescale & Eq. \ref{eq:tacc} & $500 \rm yr \times \Delta m_{\rm -3}$ \\
     
     $S_{\rm \star}$ & Sink particle force softening compact support radius & Numerical parameter & $18 \rm AU  \times \Delta m_{\rm -3}$\\
     
     
     $R_\mathrm{sink}$ & Sink particle maximum accretion radius & Eq. \ref{eq:Rsink} & $18 \rm AU  \times \Delta m_{\rm -3}$ \\ 
    \end{tabular}
    \caption{Glossary of numerical resolution-related quantities in STARFORGE simulations. $\Delta m_{\rm -3}$ is the mass resolution $\Delta m$ in units of the fiducial $10^{-3}\msun$ resolution, $n_{\rm 3} = X_{\rm H} \rho/m_{\rm p} \approx 0.7 \rho/m_{\rm p}$ is the local number density of H in units of $10^3 \rm cm^{-3}$, and $c_{\rm s,0.2}$ is the minimum gas isothermal soundspeed in units of $0.2 \rm km\,s^{-1}$.}
    \label{tab:glossary}
\end{table*}

\subsection{Magnetohydrodynamics}
\label{sec:mhd}


The default MHD solver used by STARFORGE simulations is the Meshless Finite Mass (MFM) method presented in \citetalias{hopkins_gizmo_mhd}\footnote{MFM is our method of choice, but all STARFORGE methods are compatible with any quasi-Lagrangian MHD method implemented in {\small GIZMO}, including MFV and SPMHD, enabling easy comparisons.}, which we will briefly summarize. This method discretizes the fluid into a finite numberof gas cells of mass $\Delta m_i$, each representing a domain of volume $V_i = \Delta m_i/\rho_i$ as determined by the kernel\footnote{Following \citetalias{hopkins_gizmo_mhd}, we adopt the M4 cubic spline as the default kernel partition function $W_{g g^{\prime}}=W(|{\bf x}_{g^{\prime}}-{\bf x}_{g}|,\,H_{g})$, with kernel radius of compact support $H_{g}$ defined recursively by $H_{g} = 2\,\Delta x_{g}$ where $\Delta x_{g}$ is the kernel-weighted mean cell separation: $\Delta x_{g} \equiv V_{g}^{1/3} = (\bar{n}^{\rm cells}_{g})^{-1/3} = [\sum_{g^{\prime}} W_{g g^{\prime}}]^{-1/3}$.} volume partition described in \citetalias{hopkins2015_gizmo}. This partition defines the effective face areas ${\bf A}_{g g^{\prime}}$ between each interacting pair of gas cells $g$ and $g^{\prime}$,\footnote{Throughout this work we adopt index notation for gas cells and sinks where $i$ and $j$ denote any element regardless of type, $g$ and $g^{\prime}$ denote gas cells, and $s$ and $s^{\prime}$ denote sink particles.} between which the conservative MHD equations are evolved in standard finite-volume fashion:

\begin{equation}
    \frac{\dif}{\dif t}\left(V \mathbf{U}\right)_{g} = \sum_{g'} \mathbf{A}_{ gg'} \cdot \mathbf{F}_{ gg'},
    \label{eq:conservative}
\end{equation}
where $(V\,{\bf U})_{g}$ gives the usual conserved quantities (mass, momentum, energy, ...) integrated over the volumetric domain of the cell, and ${\bf F}_{g g^{\prime}}$ is the tensor of their fluxes. The fluxes are obtained by solving the appropriate (HLLD) Riemann problem using the fluid states reconstructed at the interface according to a slope-limited, second-order least-squares gradient estimator, evaluated in the frame moving with the interface to ensure Galilean invariance. In MFM, the interfaces are defined and move such that the mass flux vanishes identically, so the method follows the motion of constant-mass, quasi-Lagrangian fluid elements. Cells exchange conserved quantities ensuring machine-precision conservation in this operation. Magnetic field divergence errors are controlled by augmenting Eq. \ref{eq:conservative} with the usual \citet{powell1999} and \citet{dedner2002} source terms and using the \citet{Hopkins_2016_divb_cleaning} constrained gradient method for obtaining the consistent fluid reconstruction operator that minimizes the numerically-unstable terms.

Because the volume partition associated with each cell can have complicated shapes (see Hopkins 2015 for discussion), it is useful to define an effective cell size $\Delta x_g \equiv V_{g}^{1/3} \equiv (\Delta m_{g}/\rho_{g})^{1/3}$ (the equivalent cell side-length for a cubic cell of the same volume and mass):
\begin{align}
\Delta x_{g} \equiv \left( \frac{\Delta m_{g}}{\rho_{g}} \right)^{1/3} \approx 0.03\rm pc \left(\frac{\Delta m}{10^{-3}\msun}\right)^{1/3} \left(\frac{n_{\mathrm{H},\mathit{g}}}{10^3 \rm cm^3}\right)^{-1/3},
\label{eq:h}
\end{align}
and volume-equivalent spherical radius 
\begin{align}
h_{g} \equiv \left( \frac{3\Delta m_{g}}{4 \uppi \rho_{g}} \right)^{1/3} \approx 0.02\rm pc \left(\frac{\Delta m}{10^{-3}\msun}\right)^{1/3} \left(\frac{n_{\mathrm{H},\mathit{g}}}{10^3 \rm cm^3}\right)^{-1/3},
\end{align}
where the latter expressions are given in terms of the typical STARFORGE mass resolution of $10^{-3}\msun$ and the number density of H atoms $n_{\mathrm{H},\mathit{g}} \approx 0.7 \rho_g/m_{\rm p}$.
We emphasize, as discussed in \citetalias{hopkins_gizmo_mhd}, that MFM has little in common with smoothed-particle MHD (SPMHD) -- MFM is formally a member of the class of Arbitrary Lagrangian-Eulerian (ALE) finite-volume Godunov methods, much more closely related to Voronoi moving-mesh methods \citep[e.g.][]{arepo,tess}, and in fact reduces to a Voronoi-mesh method in the limit of sharply-peaked kernel functions with exact volume quadrature.

Meshless, Lagrangian, Galilean-invariant MHD methods have several advantages for simulating SF in GMCs with feedback. In Lagrangian methods, Galilean invariance implies that the timestep required does not depend upon the bulk flow velocity $v$ as $\Delta t \propto \Delta x /v$ (as is required for stable advection in Eulerian fixed-grid methods), so larger timesteps are possible in the highly supersonic flows of GMCs, and the presence of very high ($\gtrsim 100 \rm km\,s^{-1}$) bulk velocities in accretion flows or winds does not incur such a high cost, an issue often encountered by Eulerian simulations combining high velocities with high spatial refinement levels. 

Galilean and rotational invariance also ensure that structures formed in the simulation (e.g. dense cores and clumps) to evolve internally in a manner independent of of their mean bulk velocity and orientation with respect to the coordinate axes (to machine precision). Numerical errors are velocity-independent, and a significant source numerical diffusivity in supersonic flows in Eulerian methods (the grid advection operation, \citealt{robertson_2010_eulerian_galilean, pontzen_2020_AMR_diffusion}) is absent. These advantages can enable more rapid convergence of phenomena involving highly supersonic flows, large density contrasts, angular momentum conservation, and coupling to self-gravity, all of which are highly relevant in SF. We refer the reader to \citetalias{hopkins2015_gizmo}, \citetalias{hopkins_gizmo_mhd}, and \citet{Hopkins_2016_divb_cleaning} for demonstrations of the performance of the MFM MHD method in a wide variety of standard test problems.

\subsubsection{Non-ideal MHD terms}
By default, STARFORGE simulations solve the equations of ideal MHD, but it is also possible to include additional terms in the momentum, energy, and induction equations, including Spitzer anistropic conduction and Braginskii viscosity \citep[e.g.][]{su_2017_mhdcv, hopkins_2020_whatabout}, Ohmic resistivity, ambipolar diffusion, and the Hall effect \citep[e.g.][]{hopkins_2017_diffusion}. These terms are implemented numerically by operator-splitting with the ideal MHD update cycle, as described in \citet{hopkins_2017_diffusion}. The effects of these non-ideal terms in star formation will be the subject of a future study.

\subsubsection{Coupled dust-gas dynamics}
Physically, dust grains are coupled to gas aerodynamically and hence do not necessarily move with the gas \citep{draine_salpter_1979_grains}, and this can have important effects for GMC physics and star formation \citep{hopkins_2014_totallymetal,hopkins_lee_2016_dust}. Our default setup assumes a constant dust-to-metals ratio for cooling, radiative transfer, etc, but compatible with 
STARFORGE modules are fully compatible with {\small GIZMO}'s dust dynamics module \citep{hopkins_lee_2016_dust,lee_2017_charged_grains,moseley_2019_grains}, which follows dust tracer particles in a Monte Carlo sampling of phase space and grain size, with an arbitrary grain size distribution, including Stokes, Epstein, and Coulomb drag, Lorentz forces with collisional, photoelectric, and cosmic ray charging, and gas back-reaction. 
The effect of these physics on star formation will be the subject of future work.

\subsection{Gravity}
\label{sec:gravity}

We compute the gravitational field $\mathbf{g} = -\grad \Phi$, tidal tensor $\mathbf{T}=-\grad \grad \Phi$ (where $\Phi = \grad^{-2} 4 \uppi G \rho$ is the gravitational potential) and the gravitational jerk $\mathbf{j}$ at the location of every gas cell and sink particle in the simulation using a modified version of the massively-parallel, approximate tree-force algorithm introduced in \citet{Springel_2005_gadget} (hereafter \citetalias{Springel_2005_gadget}). This algorithm recursively subdivides the simulation domain into an oct-tree structure, and uses the monopole approximation of the field contribution of the contents of a tree node, unless an the opening criterion is satisfied, in which case the opening criteria are re-evaluated recursively for all sub-nodes and forces evaluated accordingly. We use the acceleration-based opening criterion introduced in \citetalias{Springel_2005_gadget} (which requires that the quadrupole error term $\mathbf{a}_{\rm Q} \sim G M L^2/r^4$ of a node is less than a specified fraction of the total field $\mathbf{g}$), but also always require the original \citet{barneshut} opening criterion: a tree node is always opened if it subtends an angle $\theta \equiv \frac{L}{r} > \Theta$, where $L$ is the side length of the node, $r$ is the distance between the node centre of mass and the target point for field evaluation, and $\Theta=0.5$ is the maximum opening angle. This ensures that a dense sub-system of a hierarchically-structured system that dominates its own field (e.g. a dense clump or a binary) still has some control over the accuracy of the force contribution from surrounding material, which is still needed to predict its centre-of-mass motion \citep{grudic_2020_cluster_formation}. $\mathbf{T}$ and $\mathbf{j}$ are computed in the same pass through the gravity tree as $\mathbf{g}$, summing the respective monopole contributions of tree nodes and particles according to the same opening criterion \citep{vogelsberger:2008.gde, tidaltimestep}. Gravitational forces are updated for gas cells only as frequently as required per the \citet{grudic_adaptive_gravity} adaptive criterion, using $\mathbf{j}$ to construct a predictor of $\mathbf{g}$ between updates. This generally decreases overall cost of calling the gravity solver by a factor of 2 or better.

\subsubsection{Softening}
\label{sec:softening}
We  use a softened form of the gravitational force law for sink particles or gas cells that fall within each other's respective softening radii $S_{\rm i}$ and $S_{\rm j}$, i.e. $r_\mathrm{\rm ij} < \mathrm{max}\left(S_{\rm i}, S_{\rm j}\right)$, ensuring that all interactions are anti-symmetric, conserving total linear and angular momentum. Softening for gravitational interactions between gas cells is fully adaptive, i.e. we set $S_{\rm i}=H_{\rm i}$ so that the gravitational force resolution is always scaled to be consistent with the cell volume partitioning assumed when solving the MHD equations. This prevents various unphysical effects that are seen if the hydro and gravitational resolution scales are mismatched \citep[][]{bate_1997_resolution}. At the second-order consistency of our MHD method, \citetalias{hopkins2015_gizmo} showed that this can be done by using the same spherically-symmetric compact spline softening scheme as SPH, with the same additional terms and symmetrization to ensure conservation of momentum and energy \citep{price_2007_adaptive_softening}. The full form of the pairwise force law between gas cells is given in \citetalias{hopkins2015_gizmo} Eqs. H8-H10.

For interactions between sink particles, it would be ideal to use the full, unsoftened $1/r^2$ force law to be able to follow stellar dynamics on all scales down to stellar radii. Unfortunately this is not presently possible for our code, because binaries with arbitrarily-close separations (down to surface contact) can form and harden dynamically \citep{heggie1975}, potentially requiring very short timesteps. In theory this workload could be accomplished by our hierarchical individual timestepping scheme (\S\ref{sec:timestepping}), but in practice the massively-parallel architecture of {\small GIZMO} is such that global overheads tend to eventually bottleneck timesteps involving a small subset of the total particle number (e.g. two sink particles in a very short period binary). Therefore we adopt a finite, {\it fixed} softening radius $S_{\rm \star}$ for sink particles in a given simulation, allowing us to follow collisional dynamics accurately on spatial scales $\gtrsim S_{\rm \star}$, while limiting the effective hardness of binaries having separation $\lesssim S_{\rm \star}$.

Lastly, softened interactions between fixed-softening sinks and adaptively-softened gas cells must be handled specially, because the respective softenings can differ by orders of magnitude -- e.g. if a star with $S_{\rm \star} \sim 20\rm AU$ is moving through a diffuse part of the GMC where $n_{\rm H} \sim 10 \rm cm^{-3}$, and hence a gas cell would have size $\sim 0.1 \rm pc$ (Eq. \ref{eq:h}). In such a case using the same \citet{price_2007_adaptive_softening} symmetrization as gas-gas interactions (averaging the forces) would result in unphysical noise, because the interaction between the gas and star on the scale of $S_\star$ is totally unresolved, but the back-reaction on the star depends sensitively upon its position with respect to the cell centre. Instead, we take the {\it maximum} of $S_{\rm \star}$ and the gas kernel radius $H$ as the softening radius, in both directions of the pairwise interaction (thus conserving momentum). A natural choice for the fixed $S_{\rm \star}$ is to match it to the finest possible Jeans-resolved spatial MHD resolution, which we will show to be $\sim 20 \rm AU$ in \S\ref{sec:jeansresolution} for our fiducial mass resolution of $10^{-3}\msun$.


In all pairwise interactions, the tidal tensor $\mathbf{T}$ and jerk $\mathbf{j}$ (for sinks) are summed using spatial derivatives of the same softened force kernel that is used for $\mathbf{g}$, with the same symmetrization scheme used for that particular interaction.

\subsection{Timestepping}
\label{sec:timestepping}

Gas cells and sink particles are advanced in time in a hierarchical powers-of-two individual block-timestepping scheme \citepalias{Springel_2005_gadget}. To compute the timestep taken by an element, we compute numerous timestep criteria for capturing the various physical processes in the simulation, take the smallest of these, and round it down to the next step in the powers-of-two hierarchy. Individual timesteps are essential because the shortest timesteps required are typically on the order of of a few days, requiring $\sim  10^9$ timesteps over the $\sim 10\rm Myr$ lifetime of a GMC, but the vast majority of elements in the simulation require much less-frequent updates. 

\subsubsection{Timestep criteria}
\label{sec:timestep_criteria}
Gas cells obey all of the standard local, Galilean-invariant, MHD-specific timestep criteria given in \citetalias{hopkins_gizmo_mhd}, except that we neglect the gravitational component of the acceleration in the \citet{power_2003_timestep} acceleration criterion. Instead, both gas cells and sink particles obey the tidal timestep constraint \citep{tidaltimestep}:
\begin{equation}
    \Delta t_i < \Delta t_{\rm tidal} = \sqrt{\eta}\left(\frac{\|\mathbf{T}\|^2}{6}\right)^{-1/4},
    \label{eq:tidaltimestep}
\end{equation}
where $\|\mathbf{T}\|$ denotes the Frobenius norm of the tidal tensor and $\eta$ is a tolerance parameter controlling the overall accuracy of integration. In \citet{tidaltimestep} we showed that $\|\mathbf{T}\|$ encodes a reliable estimate of the local gravitational dynamical time $t_{\rm dyn} \sim \Omega^{-1} = \sqrt{r^3/GM}$, respecting the equivalence principle (invariance to the addition of a uniform external field $\mathbf{g}'$) and interpolating between appropriate limits for a continuous mass distribution and in the vicinity of a point mass (e.g. sinks) more accurately and robustly than the usual acceleration-based criterion.

Sink particles also obey their own special timestep criterion for ensuring orbital integration accuracy \citep{vonhoerner:1960}:
\begin{equation}
    \Delta t_{\rm s} < \Delta t_{\rm 2-body} = \sqrt{\eta} \left(t_{\rm c,min}^{-1} + t_{\rm dyn, min}^{-1}\right)^{-1},
    \label{eq:dt_2body}
\end{equation}
where 
\begin{equation}
    t_{\rm c,min} = \min_{s'\neq s}\frac{\sqrt{r^2_{ ss'}+\epsilon_{\star}^2}}{v_{ ss'}}
    \label{eq:tcross}
\end{equation}
and
\begin{equation}
    t_{\rm dyn,min} = \min_{s'\neq s} \sqrt{\frac{\left(r_{ ss'}^2 + \epsilon_{\rm \star}^2\right)^{3/2}}{G\left(m_s + m_s'\right)}},
    \label{eq:tdyn}
\end{equation}
where $j$ runs over all other sink particles, $\epsilon_{\rm \star}=h_{\rm \star}/2.8$ is the Plummer-equivalent sink softening radius, and $r_{ss'}$, $v_{ ss'}$, $m_{ s}$, $m_{ ss'}$ are the separation, relative velocity, and respective masses of sink particles $s$ and $s'$. Note that $\Delta t_{\rm 2-body}$ is simply the harmonic mean of a kinematic orbital crossing timescale $\sim r/v$ and an orbital dynamical timescale $t_{\rm dyn} = \Omega^{-1}\sim \sqrt{r^3/GM_{\rm tot}}$, but replacing $r$ with a softened version $\sqrt{r^2 + \epsilon_\star^2}$. We treat this as a single timestep criterion using the harmonic mean of the two timescales because the smooth interpolation in the regime $t_{\rm c,min} \sim t_{\rm dyn,min}$ yields slightly better integration accuracy for eccentric binary orbits. Unlike the tidal criterion (Eq. \ref{eq:tidaltimestep}), $\Delta t_{\rm 2-body}$ is symmetric between pairs of sinks, ensuring that binaries are updated synchronously when this is the dominant timestep constraint, which can give better conservation of orbital parameters. The global $\min$ operations can be evaluated efficiently in the pass through the gravity tree, combining stellar masses that exist within the same tree node if it is not opened (consistent with the force approximation).

Sink particles also observe various timestep constraints derived from local gas conditions, to ensure the stability of local gas interactions occuring within the hydrodynamic stencil, such as accretion and feedback injection. First, it cannot timestep more than $4\times$ the smallest timestep of a gas neighbor:

\begin{equation}
    \Delta t_{s} < \Delta t_{\rm ngb} = \min_{g} \,4 \Delta t_{g},
\end{equation}
where $g$ runs over all overlapping, potentially-interacting gas neighbors, i.e. $r_{ sg} < \max(H_s, H_g)$. This is analogous to the constraint imposed for neighboring gas cells in {\small GIZMO}, following \citet{saitoh_makino_timestep}. A sink particle's timestep is also constrained to anticipate the infall and/or orbital motion of surrounding gas, via a gas freefall time criterion:

\begin{equation}
    \Delta t_{s} < \Delta t_{\rm ff} = \sqrt{\frac{\eta \left(\max \left(\epsilon_{\rm \star}, \Delta x_s\right)\right)^3}{G m_{ s}}},
\end{equation}
where $\eta$ is the parameter controlling overall integration accuracy and $\Delta x_{\rm s}$ is the effective gas cell size in the vicinity of the sink. Sinks also obey a local Courant-Friedrichs-Lewy (CFL)-type timestep constraint:
\begin{equation}
\Delta t_{s} < \Delta t_{\rm CFL,\star} = C_{\rm CFL} \frac{\Delta x_{s}}{\mathrm{max}\left(\sqrt{c_{\rm s,s}^2 + v_{\rm A,s}^2  + |\vvector_{\rm s} - \vvector_{\rm gas,s}|^2},\tilde{c}, v_\mathrm{fb,s}\right)},
\label{eq:dt_CFL_sink}
\end{equation}
where $\mathbf{v}_{\rm s}$ is the velocity of the sink, $c_{\rm s,s}$, $v_{\rm A,s}$ and $\vvector_{\rm gas,s}$ are the local gas sound speed, Alfv\'{e}n speed, and and gas velocity (reconstructed using a simple kernel-weighted interpolation), $\tilde{c}$ is the (possibly reduced) speed of light (only included if radiative transfer is enabled), and $v_\mathrm{fb}$ is an estimate of the maximum velocity of gas emerging from the sink due to feedback:
\begin{equation}
    v_\mathrm{fb} = \max \left(v_\mathrm{SN}, \min\left(v_{\rm wind}, v_{\rm shell}\right)\right),
\end{equation}
where $v_\mathrm{SN}$ is the SN ejecta velocity given by Eq. \ref{eq:vSN} (or 0 if the sink is not currently going SN), $v_{\rm wind}$ is the stellar wind velocity (Eq. \ref{eq:vwind}), and $v_{\rm shell}$ is the greater of the velocity of an energy-conserving \citep{weaver_1977_winds} or momentum-conserving \citep{steigman_1975_momentum_similarity} shell as its radius reaches the resolution scale $\Delta x$:
\begin{equation}
    v_{\rm shell} = \max \left(0.38 \left(\frac{L_{\rm wind}}{\rho \Delta x_s^2}\right)^{1/3}, \left(\frac{0.053 \left(L/c + \dot{M}_{\rm wind} v_{\rm wind}\right)}{\rho \Delta x_s^2}\right)^{1/2}\right),
\end{equation}
where $\dot{M}_{\rm wind}$ is the sink's wind mass loss rate (Eq \ref{eq:mdot_star}), $v_{\rm wind}$ is the wind velocity (Eq. \ref{eq:vwind}), $L_{\rm wind}=\frac{1}{2}\dot{M}_{\rm wind} v_{\rm wind}^2$ is the mechanical luminosity of the wind, $L$ is the bolometric luminosity of the sink, $\rho$ is the local gas density. We have found that including something like the $v_{\rm fb}$ term in $\Delta t_{\rm CFL,\star}$ can be important to prevent the sink particle from ``overshooting" the amount of feedback it injects, i.e. injecting feedback over a timestep longer than the time required for the local gas cells to react to it, leading to an unstable solution.\footnote{This is only required to stabilize feedback mechanisms using weighted local injection within the hydrodynamic stencil: the component of radiation pressure due to unresolved absorption, and stellar winds with unresolved free expansion (\S\ref{sec:feedback}). Feedback mechanisms that resolve the ejecta self-consistently (resolved winds, jets, and SN) are stable because the high-velocity ejecta ``wake up" ambient gas cells as they approach, bringing them down to the necessary timestep automatically \citep{saitoh_makino_timestep}.}

Reciprocally to the $v_{\rm fb}$ term in Eq. \ref{eq:dt_CFL_sink}, gas cells also obey a timestep constraint to anticipate the arrival of feedback from a star:

\begin{equation}
    \Delta t_{ g} < \Delta t_{\rm fb} = C_{\rm CFL} \min_{s \in \rm sinks} \frac{\sqrt{r_{ gs}^2 + \max\left(\epsilon_{\rm \star},\Delta x_{ g}\right)^2}}{v_{\rm fb,s}},
\end{equation}
where $s$ runs over all sink particles and the $\min$ can be evaluated efficiently in the gravity tree pass. If a hyperbolic RT solver (e.g. M1) is enabled, we also enforce a local radiation CFL condition:
\begin{equation}
    \Delta t_{ g} < \Delta t_{\rm CFL,rad} = C_{\rm CFL} \frac{2h_{ g}}{\tilde{c}}.
\end{equation}
Likewise, we enforce appropriate local timestep criteria in the relevant methods papers for the various optional physics (e.g. non-ideal MHD, dust) described above.

\subsubsection{Time integration}
Given a choice of individual timestep as in \S\ref{sec:timestep_criteria}, we require a time integration scheme that achieves acceptable truncation error. The error budget of a multi-physics SF simulation is dominated by errors in MHD, radiative transfer, stellar evolution/feedback, and gravity, all of which are necessarily approximate and/or have large modeling uncertainties. Some errors may not even converge away in the limit of infinite resolution: e.g. moments-based radiative transfer methods will not converge to the exact radiative transfer solution in general. Hence, for gas, high-order integration schemes are unlikely beat down the leading-order error terms in the global GMC and star cluster evolution. For all gas cells, and as a robust fall-back option for stars in special circumstances, we use the standard second-order Kick-Drift-Kick (KDK) integrator \citepalias{Springel_2005_gadget}:
\begin{equation}
\begin{split}
    \mathbf{v}_{ i}&\mapsto \mathbf{v}_{ i} +  \frac{1}{2}\Delta t_{ i} \,\mathbf{a}_{ i},\\
    \mathbf{x}_{ i}&\mapsto \mathbf{x}_{ i} +  \Delta t_{ i}\,\mathbf{v}_{ i},\\
    \mathbf{v}_{ i}&\mapsto \mathbf{v}_{ i} +  \frac{1}{2}\Delta t_{ i}\,\mathbf{a}_{ i}, \\
\end{split}
\end{equation}
where $\mathbf{a}_{ i}$ is the total gravitational + MHD + radiative acceleration of cell/particle $i$, which is re-evaluated after every initial half-step kick.

\begin{figure}
    \centering
    \includegraphics[width=\columnwidth]{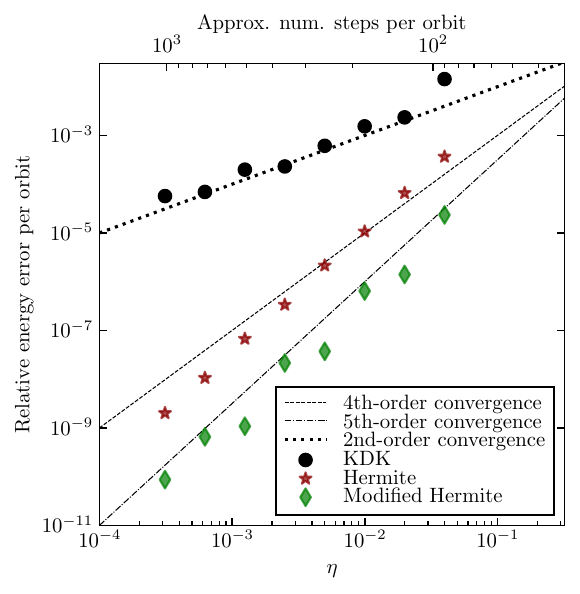}\vspace{-0.8cm}
    \caption{Relative energy error accumulated per orbit integrating $e=0.9$ binary motion with the second-order Kick-Drift-Kick (KDK) \citepalias{Springel_2005_gadget}, 4th-order Hermite \citep{makino:hermite} and our modified Hermite integrator (Eqs. \ref{eq:predictor}-\ref{eq:corrector}), as a function of the timestep tolerance parameter $\eta$, which controls the number of steps taken per orbit per our adaptive timestepping scheme (Eqs. \ref{eq:tidaltimestep} and \ref{eq:dt_2body} and powers-of-2 block scheduling, \S\ref{sec:timestepping}). Conserving binary properties in a $\sim 10\rm Myr$ GMC simulation ($\gtrsim 10^4-10^7$ binary orbits) is only practical with a higher-order scheme. Both Hermite schemes happen to converge at 5th order in this problem when using our timestep criteria, and our modified version performs better at fixed $\eta$, for an extra force and jerk evaluation per timestep.}
    \label{fig:integrator}
\end{figure}

Some additional control on orbital integration accuracy for stars is needed, to preserve the properties of binaries once formed. Any numerical integration scheme will incur a certain fractional energy error $\Delta \mathcal{E}/\mathcal{E}$ per orbit (as well as an angular momentum error $\Delta J/J$ and phase error $\Delta \phi/\phi$, but here we use $\Delta \mathcal{E}/\mathcal{E}$ as an overall proxy for integration error, as is standard). A true symplectic integrator such as the leapfrog with constant timesteps will preserve orbital energy and angular momentum on average, but true symplecticity is lost once the adaptive KDK version is adopted and errors accumulate over time, causing the semimajor axis to change with each orbit \citepalias{Springel_2005_gadget}. If a fairly typical $\sim 100 \rm yr$, $e=0.9$ binary is to survive the $\sim 10\rm Myr$ lifetime of its host GMC in a simulation, then we require $|\Delta \mathcal{E}/\mathcal{E}| << 1$ and hence an energy error per orbit of $<< 10^{-5}$. In Figure \ref{fig:integrator} we show that this would require $>> 2000$ timesteps per orbit with the KDK integrator (and would demand a minimum timestep at periastron of $\lesssim 1$ day), which we have found to demand an excessively-large overhead. Instead, we adopt a modified version of the 4th-order Hermite integrator \citep{makino:hermite} for stars. At the beginning of the timestep we evaluate the initial accelerations $\mathbf{a}_\mathrm{s,0}$ and jerk $\mathbf{j}_{ s,0}$ of all sinks in a special sinks-only gravity tree pass. We then perform the initial prediction step:
\begin{equation}
    \begin{split}
        \mathbf{x}_{ s}& \mapsto \mathbf{x}_{ s,0} + \Delta t\,\vvector_{ s,0} + \frac{1}{2}\Delta t^2 \mathbf{a}_{ s,0} + \frac{1}{6} \Delta t^3 \mathbf{j}_{ s,0},\\
        \vvector_{ s} &\mapsto \vvector_{ s,0} + \Delta t\,\mathbf{a}_{ s,0} + \frac{1}{2}\Delta t^2 \mathbf{j}_{ s,0},
    \end{split}
    \label{eq:predictor}
\end{equation}
re-evaluate $\mathbf{a}_{ s}$ and $\mathbf{j}_{,s}$ using the new positions and velocities, and then perform the correction step:
\begin{equation}
    \begin{split}
        \mathbf{v}_{ s} &\mapsto \mathbf{v}_{ s,0} + \frac{1}{2}\Delta t\left(\mathbf{a}_{ s} + \mathbf{a}_{ s,0}\right) + \frac{1}{12}\Delta t^2 \left(\mathbf{j}_{ s,0} - \mathbf{j}_{ s}\right), \\
        \xvect_{ s} &\mapsto \xvect_{ s,0} + \frac{1}{2}\Delta t \left(\mathbf{v}_{ s} + \mathbf{v}_{ s,0}\right) + \frac{1}{12} \Delta t^2 \left(\mathbf{a}_{ s,0} - \mathbf{a}_{ s}\right), \\
    \end{split}
    \label{eq:corrector}
\end{equation}
where the order here matters because the update to $\mathbf{x}_{ s}$ requires the updated version of $\mathbf{v}_{ s}$. This is subtly different from the original implementation in \citet{makino:hermite}, in that we perform {\it two} force/jerk evaluations per timestep, one at the beginning of the timestep and one after the prediction step, whereas the original Hermite scheme only reevaluates the force/jerk after the prediction step. We discovered serendipitously that this small modification gives a scheme that converges at the same order, but can give order-of-magnitude smaller energy errors at fixed timestep size in binary integration (Figure \ref{fig:integrator}). In a typical direct N-body application the entire cost of the simulation is force/jerk evaluation and there is not much parallelization overhead, so this advantage would be nullified by simply taking $2\times$ smaller timesteps at equal cost. In {\small GIZMO}, the force/jerk comes relatively cheaply, but there can be significant global overheads involved in taking smaller timesteps, so our modified Hermite scheme is more suitable. For our standard choice of $\eta=0.01$, this method achieves a relative energy error of $<10^{-6}$ per orbit for an $e=0.9$ binary (and this decreases steeply for smaller $e$).

In a given timestep, a sink particle is first provisionally timestepped according to the KDK scheme, co-evolving it alongside the gas update cycle so that the gas-star coupling seen by the gas is unaltered by the Hermite integration (but saving the initial state of the timestep). At the end of the timestep, the sink particle is eligible to accrete gas cells. If it does, low-order integration errors are introduced and $\mathbf{j}$ ceases to be well-defined, so we simply keep the more-robust KDK result for that timestep. If it does not accrete, it is eligible to take a Hermite timestep, updating via Eq. \ref{eq:predictor} using the saved values $\mathbf{x}_{s,0}$, $\mathbf{v}_{s,0}$, $\mathbf{a}_{s,0}$, and $\mathbf{j}_{s,0}$, and performing the subsequent force evaluation and correction step (Eq. \ref{eq:corrector}). Given the order of the MHD reconstruction, and the inability to define the jerk given e.g. shocks, there would be no gain from using this integrator for gas.

\begin{figure}
    \centering
    \includegraphics[width=\columnwidth]{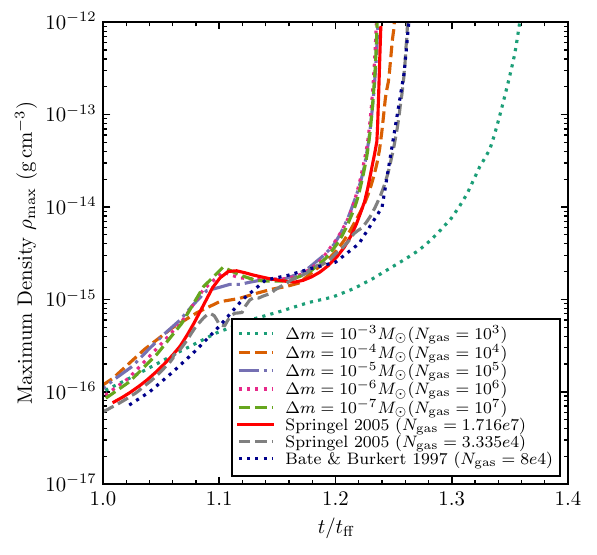}\vspace{-0.8cm}
    \caption{Evolution of the maximum density $\rho_{\rm max}$ in the standard isothermal test problem \citep{boss_bodenheimer_1979}, with time in units of the global cloud freefall time $t_{\rm ff}$. We plot MFM results for various mass resolutions $\Delta m=10^{-7}-10^{-3}\msun$, and compare with SPH results from \citet{bate_1997_resolution} and \citetalias{Springel_2005_gadget}.}
    \label{fig:bossbodenheimer}
\end{figure}
\begin{figure*}
    \centering
    \includegraphics[width=\textwidth]{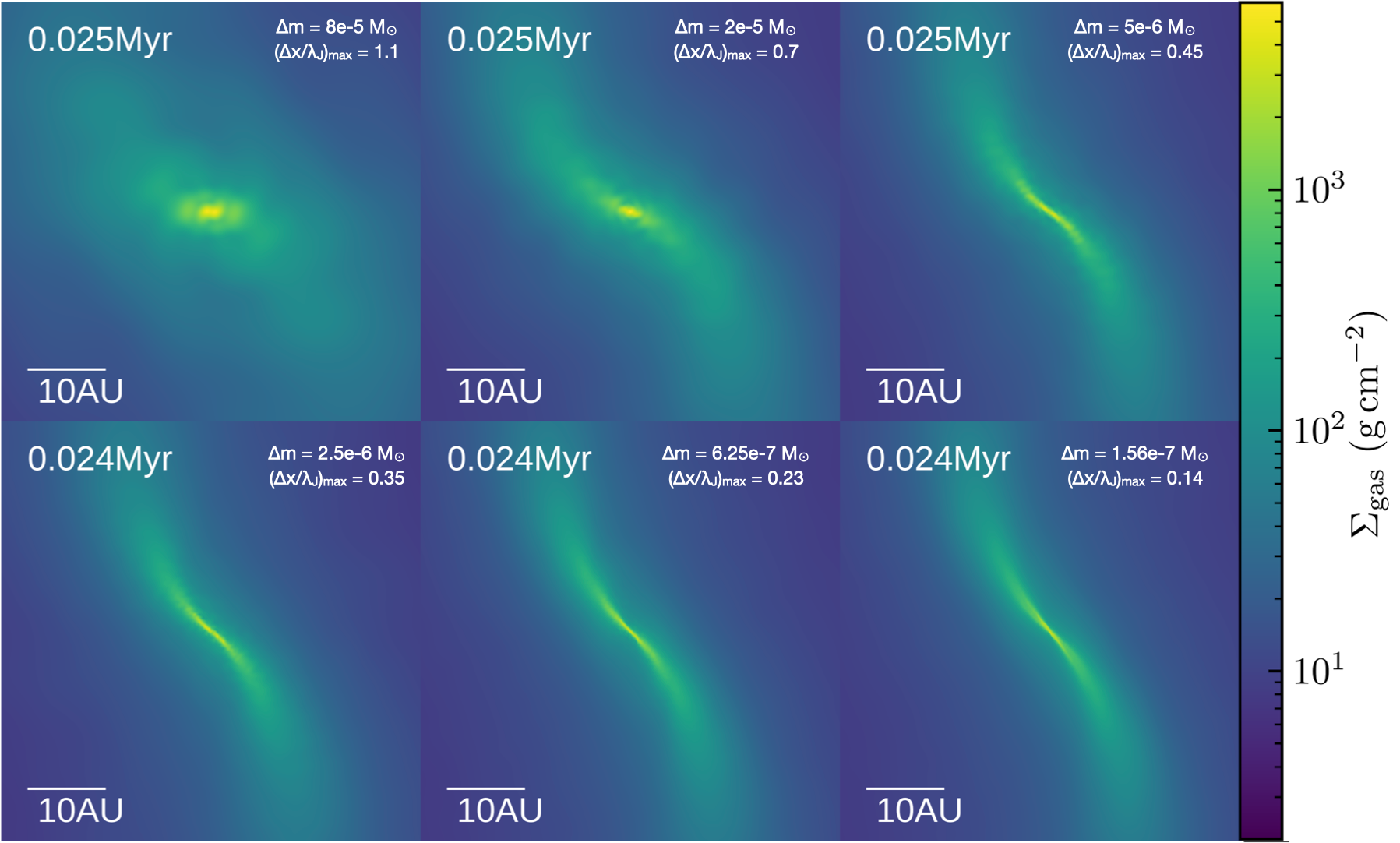}\vspace{-0.6cm}
    \caption{Effect of (under-)resolving the Jeans length in isothermal collapse with the Meshless Finite Mass (MFM) method with adaptive gravitational softening. We plot a surface density map of the filament formed in the \citetalias{truelove_1997_dens_condition} self-gravitating isothermal hydrodynamics test problem, at the time that the maximum density is $\rho_\mathrm{max}=10^{-9.5} \mathrm{g \,cm}^{-3}$ (cf. \citetalias{truelove_1997_dens_condition} Fig. 4). From top left to bottom right, we vary the mass resolution $\Delta m$ from $8\times 10^{-5}-1.6\times 10^{-7} M_\odot$, which varies the maximum number of Jeans wavelengths per cell $f_{\rm J} = \left(\Delta x/\lambda_{J}\right)_{\rm max} \approx \sqrt{G} c_s^{-1} \left(\Delta m\right)^{1/3} \rho_\mathrm{max}^{1/6}$ from 1.1-0.14. Failure to resolve the Jeans length simply coarse-grains the structure of the filament -- there is no evidence of artificial fragmentation when the Jeans length is poorly-resolved.}
    \label{fig:truelove}
\end{figure*}


\subsection{Isothermal hydro+gravity tests and resolution requirements}
\label{sec:jeansresolution}

Before discussing sink particles, it is useful to first examine the behaviour of our methods in test problems involving simple isothermal hydrodynamics and gravity.

\subsubsection{Existing tests}
The standard MHD and gravity algorithms in {\small GIZMO} have been extensively tested and applied to hundreds of different problems in the literature, so we will not repeat these. We do note these tests have demonstrated that our default implementation can simultaneously accurately evolve phenomena including gas in regular or warped Keplerian disks, strong interacting shocks, current sheets and flux tubes, supersonic {\em and} sub-sonic turbulence, fluid mixing instabilities (Kelvin-Helmholz, Rayleigh Taylor, etc.), multi-fluid dust-gas dynamics, collisional+collisionless gravitational dynamics, and reproduces the correct linear growth rates of the magneto-rotational instability (MRI) and non-ideal Hall MRI and anisotropic MHD instabilities (magneto-thermal, heat-flux-bouyancy) \citep{hopkins2015_gizmo,hopkins_gizmo_mhd,zhu:2016.sph.vs.gizmo.cosmo.sims.mw.mass.galaxy,lupi:2017.gizmo.galaxy.form.methods,deng:giant.impact.sim.mfm,deng:2019.mri.turb.sims.gizmo.methods,rennehan:turb.diff.implementation.fancy,moseley_2019_grains,panuelos:gizmo.kd.methods,hu:2020.subgrid.subsonic.turb.gizmo.methods}. Tests of idealized problems involving self-gravitating MHD including the \citet{evrard:1988.gas.collapse.problem} problem (spherical collapse of a self-gravitating polytrope),
the MHD \citet{zeldovich:1970.pancakes} pancake (self-gravitating collapse of an initially linear density perturbation along one dimension in a 3D Hubble flow) demonstrate that the MFM/MFV methods in {\small GIZMO} (as well as related moving-mesh methods) converge much more rapidly than popular AMR or SPH methods applied to the same problem \citep{hopkins2015_gizmo,hopkins_gizmo_mhd,hubber:gandalf.gizmo.methods}.

Several studies have argued that the most notable advantages of MFM compared to SPH or AMR methods may come in astrophysical disks, which are crucial for the physics of stellar accretion but are often marginally-resolved in our simulations (meaning that more-rapid convergence at fixed resolution is especially valuable). For example, (1) MFM accurately conserves angular momentum and prevents both unphysical disk ``spreading'' and/or clumping/fragmentation via artificial viscous instabilities in SPH or catastrophic angular momentum loss from spurious coordinate-alignment torques which are inescapable in AMR \citep{hopkins2015_gizmo,few:2016.disk.spiral.arm.sims.gizmo.vs.ramses.vs.sph,zhu:2016.sph.vs.gizmo.cosmo.sims.mw.mass.galaxy,lupi:2017.gizmo.galaxy.form.methods,panuelos:gizmo.kd.methods,deng:2020.parametric.instab.free.disks}. (2) \citet{few:2016.disk.spiral.arm.sims.gizmo.vs.ramses.vs.sph} found MFM more rapidly converges to correct linear growth rates for spiral arms and other disk instabilities, compared to AMR or SPH, while \citet{deng:2020.parametric.instab.free.disks} found a similar result for physical parametric instabilities of warped disks. (3) \citet{deng:gravito.turb.frag.convergence.gizmo.methods} showed MFM was the only method surveyed which exhibited convergence to exact solutions for gravito-turbulent fragmentation in cooling disks. (4) MFM, at fixed resolution, has been shown to more accurately capture boundary-layer mixing in disks (especially those formed via collisions), avoiding artificial suppression of sub-sonic turbulence and mixing common in SPH \citep{deng:giant.impact.sim.mfm,zhu:2016.sph.vs.gizmo.cosmo.sims.mw.mass.galaxy}. (5) \citet{hubber:gandalf.gizmo.methods} demonstrated that MFM simulations of ``gap opening'' via massive planets or binaries in disks converged more rapidly and maintained the gaps more accurately than equivalent SPH or AMR simulations (which tended to produce artificially-high torques and therefore stellar accretion rates in this regime). 

\subsubsection{Isothermal collapse tests}

Next, we consider a variant of the ``isothermal test case" from \citet{boss_bodenheimer_1979}: a uniform-density, un-magnetized, spherical solar-mass core with initial radius $5\times 10^{16}\rm cm$, in uniform rotation with $\Omega = 7.2 \times 10^{-13} \rm rad \,s^{-1}$ and a $10\%$ $m=2$ azimuthal density perturbation with an isothermal equation of state $P=c_{\rm s}^2 \rho$, $c_{\rm s}=0.166\rm km\,s^{-1}$ \citep{burkert_bodenheimer_1993, bate_1997_resolution, Springel_2005_gadget}. We use the MFM hydrodynamics solver with the default STARFORGE gravity and timestepping setup (\S\ref{sec:gravity}-\ref{sec:timestepping}), and initialize the cells in a glass configuration with the density perturbation imposed by rescaling cell masses. In Figure \ref{fig:bossbodenheimer} we plot the maximum density in the simulation as a function of time while varying the average cell mass $\Delta m$ from $10^{-3}-10^{-7} \msun$, and compare with SPH results from \citet{bate_1997_resolution} and \citetalias{Springel_2005_gadget}. The solution appears to converge to an answer in fair agreement with the highest-resolution SPH results in \citetalias{Springel_2005_gadget}. Moreover, our solution with $N_{\rm gas}=10^{4}$ cell resolution is closer to its respective converged limit than SPH simulations with $3.34\times 10^4$ and $8\times 10^4$ particles respectively. However, at low enough resolution numerical effects become apparent, as evidenced by the $\sim 10\%$ delay of the collapse of our lowest-resolution run with only $10^3$ gas cells.

 It is important to assess the effects of resolution upon SF simulations, as this will inform our sink particle prescription. A common convergence parameter for self-gravitating isothermal hydrodynamics simulations is the number of Jeans lengths per cell \citep{bate_1997_resolution,truelove_1997_dens_condition, Hubber_2006_sph_resolution_requirements}:
\begin{equation}
\begin{split}
f_{\rm J} \equiv \frac{\Delta x}{\lambda_{\rm J}} &= \sqrt{\frac{G}{\uppi c_{\rm s}^2}} \left(\Delta m\right)^{1/3} \rho^{1/6} \\
&\approx 0.03 \left(\frac{\Delta m}{10^{-3}\msun}\right)^{1/3} \left(\frac{n_{\rm H}}{10^{3} \rm cm^{-3}}\right)^{1/6} \left(\frac{c_{\rm s}}{0.2\rm km\,s^{-1}}\right)^{-1},
\end{split}
\label{eq:jeansnumber}
\end{equation}
using $\Delta x = \left(\frac{\Delta m}{\rho}\right)^{1/3}$ and $\lambda_{\rm J} = c_{\rm s}\sqrt{\frac{\uppi}{G \rho}}$. The consequences of under-resolving $\lambda_{\rm J}$ (i.e. allowing $\Delta x >> \lambda_{\rm J}$) vary from method to method, and have been the subject of extensive study. \citet{truelove_1997_dens_condition} (hereafter \citetalias{truelove_1997_dens_condition}) found that Eulerian grid simulations that do not enforce $f_{\rm J} < \frac{1}{4}$ are subject to artificial fragmentation (AF), wherein fragments of unphysical origin can form even in a smooth, symmetric collapse. A similar effect is seen in SPH simulations if care is not taken to match the gravitational resolution to the hydrodynamic resolution (\S\ref{sec:softening}), e.g. adopting a constant gas softening length that is much smaller than the particle spacing \citep{bate_1997_resolution}. Clearly AF is undesirable, so a variety of approaches have been developed to prevent it, e.g. by fine-tuning the sink particle formation, accretion, and merger criteria in conjunction with the refinement scheme \citep[e.g.][]{Krumholz_2004_sinks_in_eulerian,Haugbolle_Padoan_isot_IMF}. AF does {\it not} occur in SPH simulations that maintain consistency between gravitational and hydrodynamic resolution \citep{bate_1997_resolution,whitworth98a,Hubber_2006_sph_resolution_requirements}, and more recently it has been confirmed that this is true for MFM as well in the linear Jeans problem (\citealt{hubber:gandalf.gizmo.methods}, Yamamoto et al. in prep.). With these methods, fragments that should physically collapse but are insufficiently Jeans-resolved either do not collapse, or simply collapse more slowly.

Here we also check for AF in the exact test problem simulated in \citetalias{truelove_1997_dens_condition}, a variant of the \citet{boss_bodenheimer_1979} problem using an initial Gaussian density profile. With a $10\%$ $m=2$ initial density perturbation, \citetalias{truelove_1997_dens_condition} found that the converged solution is the formation of a single collapsing filament, but if the Jeans resolution criterion $\Delta x > \lambda_{\rm J}/4$ was violated then they would obtain an unphysical solution containing {\it two} filaments instead. In Figure \ref{fig:truelove} we plot the structure formed in the simulation at the time that the maximum density exceeds $10^{-9.5}\rm g\,cm^{-3}$, at a variety of mass resolutions such that the \citetalias{truelove_1997_dens_condition} criterion is strongly violated at our lowest resolution ($8\times 10^{-3}\msun$, $\Delta x \approx 1.1\lambda_{\rm J}$), and is well-satisfied at our highest ($1.56\times 10^{-7} M_\odot$, $\Delta x \approx 0.14 \lambda_{\rm J}$). No additional filament or fragment forms even when the \citetalias{truelove_1997_dens_condition} criterion is strongly violated -- the effect of poor resolution appears consistent with a simple spatial coarse-graining of the structure of the filament. \citetalias{truelove_1997_dens_condition} also found that the version of the problem with no initial density perturbation resulted in the formation of numerical fragments, unless $f_{\rm J} < 1/4$ was enforced. We have verified that this is not the case for MFM: axisymmetry is preserved accurately throughout the collapse, even when the Jeans resolution criterion is strongly violated. 

Our findings for MFM appear consistent with previous results in the linear Jeans problem (\citealt{hubber:gandalf.gizmo.methods}, Yamamoto et al. in prep.): unstable scales that are well-resolved ($f_{\rm J} << 1$) collapse as they should, and scales that should be stable are stable. Marginally-resolved ($f_{\rm J} \sim 1$) unstable wavelengths are either artificially stabilized, or collapse more slowly than is physical (e.g. the lowest-resolution run in Fig. \ref{fig:bossbodenheimer}), and these effects converge away with sufficient resolution. 

\subsubsection{Resolution criteria}
What density- and length-scales should then be considered ``resolved" in isothermal self-gravitating flows? This depends on what threshold value of $f_{\rm J}$ is considered acceptable for the question at hand, which is generally problem-dependent with no one straightforward answer. 
But assuming we do adopt a certain maximum $f_{\rm J,max}$ to demarcate the boundary of ``trusting" results in a certain problem, the maximum Jeans-resolved density is,
\begin{equation}
    \begin{split}
    \rho_{\rm J} &\equiv f_{\rm J,max}^6 \frac{\uppi^3 c_{\rm s}^6}{G^3 \Delta m^2} \\
    &\approx 3 \times 10^{-14} \mathrm{g\,cm^{-3}}\left(\frac{f_{\rm J,max}}{0.5}\right)^6\left(\frac{\Delta m}{10^{-3} \msun}\right)^{-2}\left(\frac{c_{\rm s}}{0.2\rm km\,s^{-1}}\right)^6,
    \end{split}
\label{eq:rhoJ}
\end{equation}
and the minimum Jeans-resolved cell length is
\begin{equation}
    \Delta x_{\rm J} \equiv f_{\rm J,max}^{-2} \frac{G \Delta m}{\uppi c_{\rm s}^2 } \approx 30\mathrm{AU} \left(\frac{f_{\rm J,max}}{0.5}\right)^{-2}\left(\frac{\Delta m}{10^{-3} \msun}\right)\left(\frac{c_{\rm s}}{0.2\rm km\,s^{-1}}\right)^{-2}.
    \label{eq:deltax_J}
\end{equation}
We caution that direct comparisons of the ``resolved" density- or length-scale between SF simulations should ideally be made at fixed $f_{\rm J,max}$ (i.e. correcting by appropriate $f_{\rm J,max}$ factors), and that even then there can be many confounding factors when comparing across different methods.

This discussion of Jeans resolution neglects magnetic fields, which can supplement thermal pressure as a source of support against gravitational collapse. For the purposes of gravitational stability analyses, it effectively adds the Alfv\'{e}n speed $v_{\rm A} = |\Bvector|/\sqrt{\mu_{\rm 0} \rho}$ to the thermal sound speed $c_{\rm s}$ in quadrature, i.e. $c_{\rm s} \mapsto \sqrt{c_{\rm s}^2 + v_{\rm A}^2} = \sqrt{1+\frac{2}{\beta}} c_{\rm s}$,
modulo some geometry-specific $\mathcal{O}\left(1\right)$ factors in the $\beta$ term \citep{chandrasekhar_grav_collapse, Mouschovias_Spitzer_1976_magnetic_collapse}. Assuming that the convergence parameter for isothermal, self-gravitating MHD is instead the {\it magnetic} Jeans number obtained by substituting the above into Eq. \ref{eq:jeansnumber}, as has been argued in various works \citep{Federrath_2010_sinks, Myers_2013_ORION_radiation_IMF}, our assessment of the resolving power of the simulations (Eqs. \ref{eq:rhoJ} and \ref{eq:deltax_J}) is overly conservative. However, the densest gas in isothermal MHD core collapse attracts toward $\beta \sim 1$ \citep{mocz_2017_core_sim, Wurster_2019_no_magnetic_break_catastrophe, guszejnov_isothermal_mhd}, so the corrections to our analysis from magnetic fields are expected to be modest for the present purposes. Even if not, this would merely make our effective $f_{\rm J}$ threshold more conservative, so e.g. our sink algorithm would not follow gas as far into the marginally-resolved regime, and our simulations are better-resolved than as quoted in Eqs. \ref{eq:rhoJ} and \ref{eq:deltax_J}. 

\subsection{Sink particles}
\label{section:sinks}

We use sink particles to model the accretion, dynamics, and feedback of individual stars and protostars \citep[e.g.][]{Bate_1995_accretion,Krumholz_2004_sinks_in_eulerian,Federrath_2010_sinks,hubber:2013.sinks,bleuler_teyssier_sinks}. A sink particle represents a designated region in the domain of the simulation in which physical processes are considered unresolved, and are relegated to sub-grid prescriptions. The general strategy is to put a sink in the centre of a collapsing core once the collapse process can no longer be followed self-consistently by the MHD scheme, and to allow this sink to accrete subsequent infalling material according to certain physically-motivated rules. 

Our sink implementation formally distinguishes between {\it resolved} accretion, i.e. the actual mass transfer from the gas in the simulation domain to the sink particle, and {\it unresolved} accretion: the transfer of mass from the sink's internal gas reservoir (comprising unresolved gas in the envelope or the protostellar disk) onto the protostar itself (and potentially into the protostellar outflow). Other works equate the two types of accretion, often assuming that gas removed from the simulation domain arrives at the protostar immediately \citep[e.g.][]{Krumholz_2004_sinks_in_eulerian}, or using a detailed subgrid model to decide how rapidly resolved accretion should occur \citep{hubber:2013.sinks}. For us it is important to model accretion onto the protostar distinctly from resolved accretion into the sink region, because we discretize resolved accretion into quanta -- the mass resolution $\Delta m$ -- but would like a continuous estimate of the protostellar accretion rate for modeling the protostellar evolution and accretion luminosity. \footnote{We have experimented with our own implementation of the algorithm of \citet{hubber:2013.sinks}, which uses an estimate of $\dot{M}_{\star}$ that interpolates between disk-like and Bondi-like regimes based on local gas kinematics, and uses that estimator to directly determine how much gas should be removed. However, we have found that in some problems the estimator of $\dot{M}_{\star}$ used to set the rate of resolved accretion can underestimate the actual accretion rate of the surrounding flow, so mass piles up within the softening radius of the sink particle and the actual accretion rate ends up being set by the need to remove gas cells on too small a timestep (circumventing the normal criteria), defeating the purpose of trying to estimate and enforce the proper accretion rate as determined by physical processes. One potential issue is that the $\alpha$-disk parameter used in the disk-like regime must be known {\it a priori}, otherwise the accretion rate will not match the boundary flow. This will generally vary with turbulent and numerical viscosity, magnetic torques, gravitational torques, etc, and cannot generally be fit by a single choice of $\alpha$.} Our algorithm is most similar to that of \citet{Bate_1995_accretion}, with some additional rules for formation and accretion, and some additional modeling of protostellar accretion and feedback. We sketch the flow of mass dictated by our algorithm in Figure \ref{fig:sinkdiagram}, and describe the algorithm in detail in this section.

\subsubsection{Formation}
A gas cell is eligible to turn into a sink particle if and only if it satisfies the following criteria:
\begin{enumerate}
    \item {\bf Density threshold:} The gas cell is denser than a density threshould $\rho_{\rm th}$, which we take to be the maximum density of marginal Jeans resolution, $\rho_{\rm J}$ (Eq. \ref{eq:rhoJ}), assuming $f_{\rm J}=1/2$.
    \label{criterion:density}
    \item {\bf Density maximum/no overlapping sink:} The gas cell is the densest of all neighboring gas cells or sink particles with overlapping kernel radii (with $r_{ gi} < \max \left(H_{ g}, H_{i}\right)$). For the purposes of this criterion we take sinks to have infinite density, i.e. overlapping with a pre-existing sink always prevents sink formation.
    \item {\bf Increasing density}: The gas cell's density is increasing: $\grad \cdot \vvector < 0$, according to the same least-squares matrix gradient estimator of $\grad \vvector$ used for reconstructing  fluid quantities for the MHD solver.
    \label{criterion:divv}
    
    \item {\bf Virial/Jeans criterion:} The gas cell is gravitationally unstable/self gravitating at the resolution scale, as determined by a local Jeans analysis including contributions from thermal pressure, magnetic fields, and velocity dispersion \citep{Federrath_2010_sinks,hopkins2013_sf_criterion}. We evaluate a local virial parameter for the gas cell:
    \begin{equation}
        \alpha_{g} = \frac{\frac{2\uppi^2}{\Delta x^2}\left(c_{\rm s}^2 + v_{\rm A}^2\right) + \|\grad \vvector\|^2}{4 \uppi G \rho},
        \label{eq:virialcriterion}
    \end{equation}
    where $\Delta x = \left(\Delta m/\rho\right)^{1/3}$ is the local cell length, and $\|\cdot \|$ denotes the Frobenius norm. We permit sink formation only if $\alpha_{g} < 2$. It is easy to verify that this reduces to the usual requirement that the cell is Jeans-unstable at the resolution limit where kinetic energy is negligible, and recovers the \citet{hopkins2013_sf_criterion} kinematic virial criterion when the $\|\grad \vvector\|$ term dominates (e.g. preventing sink formation in Toomre-stable flows stabilized by shear near a star).
    \label{criterion:virial}
    \item {\bf Tidal criterion:} The tidal tensor $\mathbf{T}$ at the position of the gas cell is fully compressive (possesses 3 negative eigenvalues). Note that the linearization of the gravitational field about a point $\mathbf{x}_{\rm i}$ is $\mathbf{g}\left(\xvect\right) - \mathbf{g}\left(\xvect_{\rm i}\right)\approx \mathbf{T}\cdot \left(\mathbf{x}-\mathbf{x}_{\rm i}\right)$, so if $\mathbf{T}$ has 3 negative eigenvalues then the gravitational force seen in the frame comoving with a ballistic trajectory starting at $\mathbf{x}_{\rm i}$ will always be directed back toward the origin. This is similar in motivation to the potential minimum requirement in \citet{Federrath_2010_sinks} and the Hill sphere criterion in \citet{hubber:2013.sinks}, intended to pick out actual centres of collapse from the shape of the gravitational landscape. Unlike a potential minimum criterion, the tidal criterion respects the equivalence principle, i.e. it is invariant to the transformation $\mathbf{g}\mapsto \mathbf{g}+\mathbf{g}'$ for a spatially-uniform $\mathbf{g}'$, which should not physically affect the internal dynamics of the simulation in any way, but would displace the location of a potential minmum. However, it is less strict than a potential minimum criterion, e.g. it is satisfied at every point in a uniform sphere (in which $\mathbf{T}$ is constant and negative-definite), whereas the potential minimum criterion is satisfied at one point, or none if the external field $\mathbf{g}'$ exceeds the internal field. 
    
    \item {\bf Can collapse before accretion}: The local gas freefall time $t_{\rm ff} =\sqrt{\frac{3 \uppi}{32 G \rho_{\rm g}}}$ is shorter than both the timescale for approaching a sink particle and the orbital timescale around that sink particle, as estimated by evaluating Eqs. \ref{eq:tcross} and \ref{eq:tdyn} for gas cells.

\end{enumerate}

When a gas cell is converted to a sink particle, it is removed from the simulation domain, and the volume it occupied is reassigned to surrounding cells when they re-compute their volume partitions.

\subsubsection{Accretion criteria}
\label{sec:sinkaccretion}

Gas cells are accreted by a sink particle if they satisfy the following criteria:

\begin{enumerate}
    \item {\bf  Sink radius:} A gas cell is only eligible for accretion if its centre approaches within sink radius $R_{\rm sink}$. We take $R_{\rm sink}$ to be the greater of the sink particle softening radius $S_{\rm \star}$ or volume-equivalent radius of a gas cell at the density of marginal Jeans resolution $\rho_{\rm J}$ (Eq. \ref{eq:rhoJ}, assuming $f_{\rm J}=1/2$):
\begin{equation}
\begin{split}
    R_{\rm Sink} &= \max\left(S_{\rm \star}, 0.79\frac{G \Delta m}{c_{\rm s}^2}\right) \\
     &= \max\left(S_{\rm \star}, 18\mathrm{AU}\left(\frac{\Delta m}{10^{-3} \msun}\right)\left(\frac{c_{\rm s}}{0.2\rm km\,s^{-1}}\right)\right),
\end{split}
    \label{eq:Rsink}
\end{equation}
where $c_{\rm s}$ denotes the isothermal sound speed {\it at sink formation} (i.e. it is set at formation, and kept constant thereafter). 
    \item {\bf Boundedness criterion:} The gas cell satisfies
    \begin{equation}
        2u_g + v_{\mathrm{A},g}^2 + |\vvector_{g}-\vvector_{ s}|^2 < v_{\rm esc}^2 = -2 \Phi\left(r_{ gs}\right),
        \label{eq:boundednesscheck}
    \end{equation}
    where $u_g$ is the specific internal energy of the gas, $v_{\mathrm{A},g}$ is its Alfv\'en speed, and $\Phi\left(r_{gs}\right)$ is the softened gravitational potential of the sink, evaluated at the separation between the gas and sink $r_{gs}$. This checks that the sink-gas system is gravitationally bound, and could not escape to infinity in isolation.
    \item {\bf Angular momentum criterion:} the gas cell possesses less angular momentum than a circular Keplerian orbit around the sink at $r_{ gs}$ \citep{Bate_1995_accretion}:
    \begin{equation}
        |\left(\xvect_{ g} - \xvect_{ s}\right) \times \left(\vvector_{ g}-\vvector_{ s}\right)|^2 < G m_{ s} r_{ gs}.
    \end{equation}
    In the limit of ballistic flow, this ensures that the orbit of the gas cell lies within $R_{\rm sink}$ (so we do not capture e.g. a gas cell that only makes a single close passage but then interacts outside $R_{\rm sink}$ and escapes).
    \item {\bf Size/density criterion}: The volume of the gas cell is less than the volume within $R_{\rm sink}$:
    \begin{equation}
        V_{ g} = \frac{m_{ g}}{\rho_{ g}} < \frac{4 \uppi}{3}R_{\rm sink}^3.
    \end{equation}
    This has the effect of ensuring that only gas having spatial resolution on the scale of $R_{\rm sink}$ can be accreted, which may be necessary for the other criteria to be reliable predictors of the gas's dynamics (and whether it is legitimately being accreted). In any true resolved accretion flow, gas will pile up around the sink until this criterion is eventually satisfied. It is analogous to maintaining the maximum refinement level in the vicinity of a sink in an AMR simulation \citep{Krumholz_2004_sinks_in_eulerian}.
    
\end{enumerate}
It is possible for a gas cell to satisfy all of these criteria for more than one sink. In such instances, the gas is accreted by the sink $s$ with which it has shortest mutual dynamical time $t_{\rm dyn} = \Omega^{-1} = \sqrt{\frac{r_{ gs}^3}{G\left(m_{ g} + m_{ s}\right)}}$.

The quantization of resolved accretion into parcels of mass $\Delta m$ has certain important limitations. Clearly, a sufficiently slow accretion flow with $\dot{M} << \Delta m / t$, where $t$ is some timescale of interest, cannot be captured. In the limit of a ballistic, Bondi-like flow, we can take $t = t_{\rm dyn} \left(<R\right) = \sqrt{R^3/GM_\star }$ at some radius of interest $R$. Then, assuming the physical accretion flow has a certain $\dot{M}_\star$, the most optimistic radius down to which the flow can be resolved is
\begin{equation}
    R_{\rm min} = 73 \mathrm{AU} \left(\frac{\Delta m}{10^{-3}M_\odot}\right)^{2/3} \left(\frac{\dot{M}_\star}{10^{-5} M_\odot \,\rm yr^{-1}}\right)^{-2/3}\left(\frac{M_\star}{1M_\odot}\right)^{1/3}, 
\end{equation}
where we insert typical values for $\Delta m$, $\dot{M}_\star$, and $M_\star$. Hence the accretion flow becomes less well-resolved for smaller accretion rates and greater stellar masses. This may impose some numerical bias toward higher accretion rates in the accretion histories of sink particles, and underestimate more extended periods Bondi-Hoyle accretion from low density gas. However, the effect does converge to the correct solution with sufficient mass resolution.

\subsubsection{Updating conserved quantities}
When a gas cell is accreted, it is deleted from the simulation domain and the volume partition of neighbouring gas cells is re-computed. Its mass, first mass moment $m_{\rm g}\xvect_{\rm g}$, momentum, and angular momentum are added to the sink:
\begin{equation}
    m_{\rm s} \mapsto m_{\rm s} + m_{\rm g},
\end{equation}
\begin{equation}
    \xvect_{\rm s} \mapsto \frac{m_{\rm s} \xvect_{\rm s} + m_{\rm g} \xvect_{\rm g}}{m_{\rm s} + m_{\rm g}} = \mathbf{x}_{\rm s}',
\end{equation}
\begin{equation}
    \mathbf{p}_{\rm s} \mapsto \mathbf{p}_{\rm s} + \mathbf{p}_{\rm g} = \mathbf{p}_{\rm s}',
\end{equation}
\begin{equation}
    \mathbf{J}_{\rm s} \mapsto \mathbf{J}_{\rm s} + \left(\mathbf{p}_{\rm s} \times \mathbf{x}_{\rm s} + \mathbf{p}_{\rm g} \times \xvect_{\rm g} - \mathbf{p}_{\rm s}' \times \mathbf{x}_{\rm s}'\right) ,
\end{equation}
conserving mass, centre of mass, momentum, and angular momentum, respectively. The stored value of $\mathbf{J}_s$ does not necessarily correspond to the physical angular momentum of the star, merely the angular momentum within the sink (consisting of the star and a presumed surrounding gas disk or envelope)\footnote{The raw accreted angular momentum of a sink particle is typically of order $\sqrt{G M_{s} R_{\rm sink}}$, which depends on the numerical parameter $R_{\rm sink}$, and is typically orders of magnitude greater than the angular momentum of a star \citep{hubber:2013.sinks}. To determine the actual stellar angular momentum evolution one must model unresolved AM transfer processes.}. Within the sink, the accreted mass is initially stored in the sink's accretion reservoir:
\begin{equation}
    M_{\rm acc,s} \mapsto M_{\rm acc,s} + m_{\rm g}.
\end{equation}
Note that our implementation does not address the long-standing issue of violating conservation of magnetic flux when a Lagrangian gas cell is deleted \citep[e.g.][]{price_bate_2007_mhd_sf}. The removal of a gas cell will also generally create a $\grad \cdot \Bvector$ error, and we rely upon our divergence cleaning scheme to damp it away. However, in \S\ref{sec:jettests} we show that the main quantities of interest that we wish to predict (star formation histories and the IMF) are in good agreement with results from a constrained-transport AMR code, which does not accrete magnetic flux and enforces $\grad \cdot \Bvector$ to machine precision\footnote{One possible solution for Lagrangian MHD codes (not explored here) would be to introduce a numerical resistivity $\eta_{\rm sink}$ that interpolates between $\sim 0$ when $r>>R_{\rm sink}$ and $\eta_{\rm sink} \sim \sqrt{G M_{\star} R_{\rm sink}}$ when $r \sim R_{\rm sink}$, which would diffuse flux away from the star as mass is carried into the sink, modeling the physical non-ideal processes that occur near protostars.}.

\begin{figure}
    \centering
    \includegraphics[width=\columnwidth]{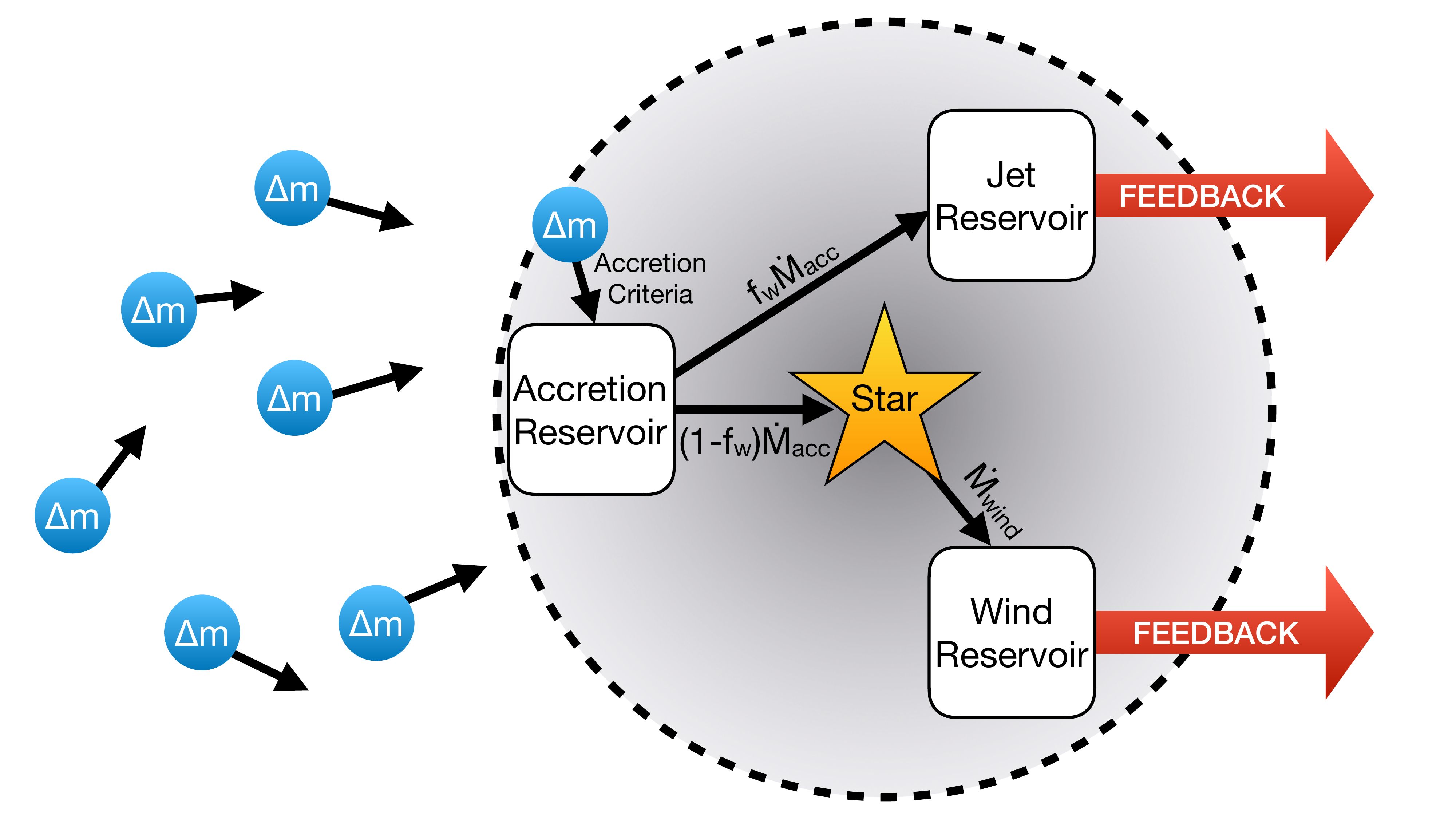}\vspace{-0.7cm}
    \caption{Diagram of the flow of mass due to accretion and feedback, as managed by our sink particle algorithm. We follow gas cells of mass $\Delta m$ until they satisfy all sink particle accretion criteria (\S\ref{sec:sinkaccretion}) and they are transferred to the sink's accretion reservoir representing the envelope or disk gas mass present on scales $<R_{\rm sink}$. Mass is accreted from the reservoir toward the protostar according to the smoothed accretion prescription (Eq. \ref{eq:mdot_star}), and if protostellar jets are enabled a fraction of this mass $f_{\rm w}$ is diverted to the jet reservoir. The rest arrives at the star, and mass is transferred from the star to the wind reservoir according to the wind mass loss rate (which is set to an extremely large value (with appropriate velocity) if the star goes SN, Eqs. \ref{eq:mdot_sn},\ref{eq:vSN}). The jet and wind reservoirs return gas to the simulation domain via their respective feedback channels (waiting until a sufficient mass is available to inject or spawn their respective mass quanta).}
    \label{fig:sinkdiagram}
\end{figure}

\subsubsection{Accretion from reservoir onto protostar}
To model the continuous accretion of the protostar for the purposes of modeling protostellar evolution and feedback, we use a simple prescription:
\begin{equation}
    \dot{M}_{\rm \star,s} = \left(1-f_{\rm w}\right)\frac{M_{\rm acc,s}}{t_{\rm acc}},
    \label{eq:mdot_star}
\end{equation}
where $\dot{M}_{\rm \star,s}$ is rate at which mass arrives at the protostar, $f_{\rm w}$ is the fraction of gas mass transferred into the protostellar outflows instead (\S\ref{sec:jets}), and $t_{\rm acc}$ is the accretion timescale. Both $f_{\rm w}$ and $t_{\rm acc}$ are variable, prescription-dependent quantities (we discuss $f_{\rm w}$ further in \S\ref{sec:jets} and in \citetalias{starforge_jets_imf}), but by default we take $t_{\rm acc}$ to be the mean time interval between the arrival of gas cells of mass $\Delta m$, assuming the accretion rate is $c_{\rm s}^3/G$:
\begin{equation}
    t_{\rm acc} = \frac{G \Delta m}{c_{\rm s}^3} = 530\mathrm{yr}\left(\frac{\Delta m}{10^{-3}\msun}\right)\left(\frac{c_{\rm s}}{0.2\mathrm{km\,s^{-1}}}\right)^{-3},
    \label{eq:tacc}
\end{equation}
which is dimensionally the same as the freefall time at the maximum Jeans-resolved density, $t_{\rm J} \sim \left(G \rho_{\rm J}\right)^{-1/2}$. This feeds the protostar with an exponentially-smoothed version of the discrete resolved accretion rate, with a $1/e$-folding time equal to $t_{\rm acc}$. For the smallest plausible continuous accretion rate in the initial core collapse, $\dot{M}_{\star} \sim c_{\rm s}^3/G$ \citep{shu_1977_isothermal_collapse}, our choice of $t_{\rm acc}$ is simply the mean time interval between the accretion of mass quanta $\Delta m$, which guarantees that it limits unphysical discreteness noise without ``over-smoothing" accretion.

Note that the prescription in Eq \ref{eq:mdot_star} is not meant to model the physical accretion rate at the protostellar surface in detail, and is merely a numerical scheme to obtain a continuous version of the resolved accretion rate with a smoothing timescale adapted to the mass resolution. If the accretion flow is a direct radial infall (e.g. Bondi accretion) then the relevant physical accretion timescale is the freefall time (generally shorter than $t_{\rm acc}$). In the regime where the gas hits an angular momentum barrier before reaching the protostar, accretion will generally take many orbits, and might be better described by e.g. a \citet{shakura_sunyaev} $\alpha$-disk type model (in which the dimensionless parameter $\alpha$ encodes the net effect of gravitational torques, magnetic fields, outflows, and viscosity upon angular momentum transport). In principle, our continuous accretion rate estimator could be fed into a physical model to obtain a more realistic estimate of the rate at which mass arrives at the protostar. However, protostellar accretion on sub-$10\rm AU$ scales is subject to a wide variety of poorly-understood complex microphysics (e.g. making the specific choice of $\alpha$ an open problem), so we do not attempt to model such processes here. 

\subsubsection{Stellar evolution}
\label{sec:stellarevol}

In simulations with feedback, it is necessary to model the evolution of the protostar or star in the sink particle to inform the emergent luminosity, spectral energy distribution (SED), mass loss rate, and wind/outflow velocity. We model the star or protostar according to a one-zone subgrid model whose sole input is the present protostellar mass and the accretion rate (Eq. \ref{eq:mdot_star}), originally following \citet{nakano_2000_protostellar} and based upon the particular implementation of \citet{Offner_2009_radiative_sim}. The model evolves the protostellar radius $R_\star$ explicitly, and is calibrated to recover the results of detailed numerical simulations of individual protostellar evolution. This model has been used in many subsequent works by different groups with different codes \citep[e.g.][]{myers2014_sim,Federrath_2017_IMF_converge_proceedings,murray_2018_jets}, so we describe it only briefly and refer the reader to \citet{Offner_2009_radiative_sim} for full details. The evolution is separated into distinct phases: 
\begin{enumerate}
    \item {\it Pre-collapse:} If $M_{\star} < 0.01\msun$ then the protostar is presumed to be a $\approx 4\rm AU$ first Larson core that has yet to undergo the second collapse phase \citep{larson_1969, masunaga1998}.
    \item {\it No burning:} Once $M_\star > 0.01 \msun$ the core undergoes the second collapse to protostellar density, but deuterium has yet to ignite.
    \item {\it Deuterium burning at fixed core temperature:} D burning has started, fixing the core of the protostar at $\approx 1.5 \times 10^6 \rm K$.
    \item {\it Core burning at variable core temperature:} The core temperature has begun to rise and D is convected to the core on short timescales (burning it roughly as rapidly as it arrives at the protostar).
    \item {\it Shell deuterium burning:} If D is still arriving rapidly enough, the protostar swells and forms an outer convective zone where the D ignites.
    \item {\it Main sequence:} The star has reached a central core temperature sufficient to ignite H.
\end{enumerate}

\begin{figure}
    \centering
    \includegraphics[width=\columnwidth]{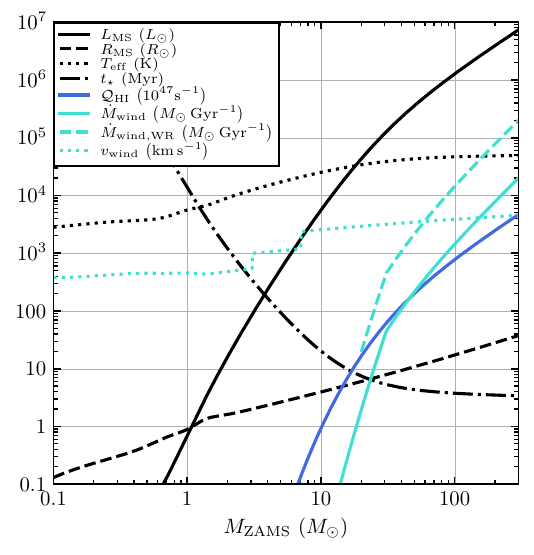}\vspace{-0.8cm}
    \caption{Stellar properties as a function of zero-age main-sequence mass $M_{\rm ZAMS}$ used to model feedback and stellar evolution in STARFORGE. We plot the main-sequence luminosity $L_{\rm MS}$, radius $R_{\rm MS}$, and resulting effective temperature $T_{\rm eff}$ from \citet{tout_1996_mass_lum}, the stellar lifetime $t_{\rm \star}$ per Eq \ref{eq:stellarlifetime}, the flux of H-ionizing $>13.6\rm eV$ photons $\mathcal{Q}_{\rm HI}$ (assuming a black-body spectrum of temperature $T_{\rm eff}$, \S \ref{sec:radiation}), the wind mass-loss rates $\dot{M}_{\rm wind}$ and $\dot{M}_{\rm wind,WR}$ for main-sequence and Wolf-Rayet stars, and the wind velocity $v_{\rm wind}$ (with wind quantities assuming solar metallicity, see \S\ref{sec:winds}).}
    \label{fig:stellarproperties}
\end{figure}

At each timestep, the state of the protostar is updated based upon the present mass, accretion rate, and evolutionary phase, and dictates the evolution of the stellar radius $R_{\rm \star}$ and the emergent luminosity $L_{\rm \star}$ (which includes terms from accretion, Kelvin-Helmholtz contraction, D burning, and H burning, as given in \citealt{Offner_2009_radiative_sim}). We use the \citet{tout_1996_mass_lum} fits for the mass-dependent zero-age main sequence luminosity $L_{\rm MS}$ and radius $R_{\rm MS}$, and neglect the effects of stellar evolution beyond the main sequence (apart from modeling a Wolf-Rayet phase for winds, \S\ref{sec:winds}, and an eventual supernova for $>8\msun$ stars, \S\ref{sec:SNe}). For the purposes of modeling SNe, stars $>8\msun$ have a mass-dependent stellar lifetime:
\begin{equation}
    t_{\rm \star} = 9600 \rm Myr \left(\frac{M_{\rm \star}}{M_\odot}\right)\left(\frac{L_{\rm \star}}{L_{\rm \odot}}\right)^{-1}+3 \rm Myr,
    \label{eq:stellarlifetime}
\end{equation}
which interpolates between $\sim 10\rm Gyr$ for solar-type stars and $\sim 40 \rm Myr$ for $8\msun$ stars, and asymptotes to $\sim 3\rm Myr$ for the most massive ($\gtrsim 100 \msun$) stars. In Figure \ref{fig:stellarproperties} we plot $L_{\rm MS}$, $R_{\rm MS}$, $t_{\rm \star}$, and various other useful derived quantities for stellar feedback (\S\ref{sec:feedback}) as a function of the zero-age main-sequence mass $M_{\rm ZAMS}$.

We do not model the end of life of stars less massive than $8 \msun$ (i.e. planetary nebulae), but this could be important for calculations that run for much longer than a GMC lifetime (e.g. \S\ref{sec:galsims}). We also presently neglect the formation of relic compact objects, but this would be a trivial modification to the inputs of the SN/wind module (simply reserving a certain relic mass), given a more-detailed stellar evolution prescription.

\subsubsection{Merging criteria}
In the code, sink particles are allowed to merge if they have a binary semimajor axis $<R_{\rm sink}$ and the secondary has a mass $<10\Delta m$. In theory this helps eliminate unphysical, spurious low-mass sinks that may form in proximity to legitimate sinks, or $\sim$few-AU clumps of mass $<0.01\msun$ that would physically be accreted by a protostar \citep{Offner_2009_radiative_sim}. In practice, this merger condition is not satisfied in most simulations, and generally only a few times (out of $\gtrsim 1000$ stars) if so. Hence our results are not sensitive to our sink particle merging strategy. It is possible that physical stellar mergers are a channel for the formation of very massive stars in the centres of dense star clusters \citep{portegies_zwart_1999_runaway_mergers, bonnell_2005_mergers,  shi_2020_mergers}, but we generally run stellar softenings significantly larger than physical stellar radii and hence cannot follow mergers self-consistently without some kind of sub-resolution modeling.\footnote{One possible approach to stellar mergers is to use the orbital energy and angular momentum (which are conserved absent perturbations) of stellar pairs passing within their respective softening radii to determine the physical, un-softened periastron radius, and hence whether the stars should physically merge.}

\subsubsection{Singular isothermal collapse test}
\begin{figure}
    \centering
    \includegraphics[width=\columnwidth]{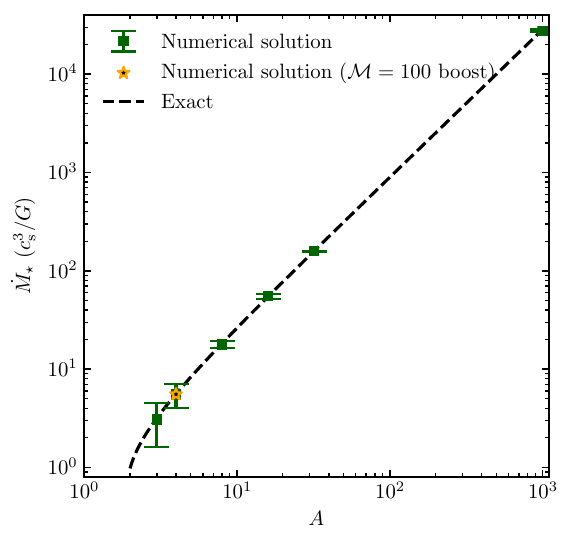}\vspace{-0.8cm}
    \caption{Simulated accretion rate in the \citet{shu_1977_isothermal_collapse} singular isothermal collapse problem, in units of $c_{\rm s}^3/G$, as a function of the instability parameter $A$, such that the initial density profile is $\rho = \frac{A c_{\rm s}^2}{4 \uppi G r^2}$ and $A=2$ is the threshold of stability. To the analytic solution (dashed) we compare results simulated with our default hydro, gravity, and sink particle algorithms simulated with the ICs at rest (squares) and with the initial gas moving at Mach 100 to verify Galilean invariance (star). Error bars indicate the $\pm \sigma$ quantiles of the accretion rate estimator used to feed the subgrid protostar (Eq. \ref{eq:mdot_star}) -- variance is driven by the discretization of resolved accretion into chunks of mass $\Delta m$. Agreement with the analytic solution is excellent, and in all instances exactly one sink particle is formed (replacing the central singularity).}
    \label{fig:shu_solution}
\end{figure}

We first validate the formation and resolved accretion criteria of sink algorithm in the \citet{shu_1977_isothermal_collapse} singular isothermal sphere problem, the collapse of a core with an initial density profile $\rho = \frac{A c_{\rm s}^2}{4\pi r^2}$, where $A$ parameterizes the family of solutions and collapse occurs for $A>2$. This problem possesses a single central singularity (to be represented by the sink), and admits a semi-analytic, spherically-symmetric solution for all fluid quantities (from the numerical solution of \citet{shu_1977_isothermal_collapse} Eqs. 11 and 12). This unambiguous reference solution allows it to quickly expose numerical quirks and bugs, whereas testing the sink particle algorithm on e.g. a full turbulent GMC collapse problem is both more expensive and less conclusive because the ``correct" solution (or whether it exists for a given physics setup) is unknown {\it a priori}. An insufficiently-strict sink formation prescription (or an overly-strict accretion prescription) can result in the formation of multiple spurious sinks when there should be a single singularity. Errors in momentum conservation or gravity can cause the sink to drift from the centre of collapse, causing subsequent gas to arrive off-centre and form spurious disks or sinks. Passing this test does not prove that a sink algorithm is valid for all problems, but failing this test is a strong indicator that the algorithm is flawed.

For reliable test results, the initial conditions should represent the analytic initial density field with equal-mass elements as we use in our simulations, but this is non-trivial. For MFM, we initialize 125,000 equal-mass gas cells on a uniform radial grid (producing the desired $r^{-2}$ density profile) with random initial angular positions, and relax the resulting Poisson sampling noise in the IC to a glass by reversing the sign of gravity and allowing cells to slide around on their respective initial radial shells, with an artificial drag force to damp out the motions toward equilibrium. We then rescale to survey various values of $A$. Exactly one sink forms in each test, and we plot its mass accretion rate in Figure \ref{fig:shu_solution} for $A$ values ranging from 3 to 1000. Agreement with the semi-analytic solution is excellent across the entire range of $A$ surveyed. We also verify Galilean invariance by running a version of the $A=4$ setup with a velocity boost of $100 c_{\rm s}$: even at this extreme bulk Mach number, the solution is preserved owing to the machine-precision Galilean invariance of the hydro, gravity, and sink particle algorithms. The error bars in Figure \ref{fig:shu_solution} plot the $\pm \sigma$ variations of our continuous accretion rate estimator (Eq. \ref{eq:mdot_star}), showing that its average error is at most a factor of $\sim 2$ for the lowest $A$ values and accretion rates, and generally much less for higher accretion rates.

\subsubsection{Effect of sink prescriptions}
\label{sec:sinktests}
Because we wish to use the properties of sink particles to predict the IMF that emerges from a given set of physics, it is important to ensure that the results of STARFORGE simulations are insensitive to the specific parameter choices made in our sink algorithm for e.g. the density threshold and sink radius, and ideally have some robustness to the specific choice of sink formation and accretion criteria as well. For this we re-run the ${\bf M2e4}$ GMC setup in \citetalias{guszejnov_isothermal_mhd}, a $2\times 10^4 \msun$ initially-spherical GMC of radius $10 \rm pc$ at $10^{-3}\msun$ resolution with numerous variations from the prescription described in this section, listed in full in Appendix \ref{appendix:sinktests}. By including only minimal physics (isothermal MHD and gravity), the incremental effects of sink numerics are expected to be more pronounced than in a more complex setup with realistic thermodynamics and feedback. 
In this sense this test could be considered a worst-case assessment of the sensitivity of STARFORGE results to sink prescriptions.

\begin{figure*}
    \centering
    \includegraphics[width=\textwidth]{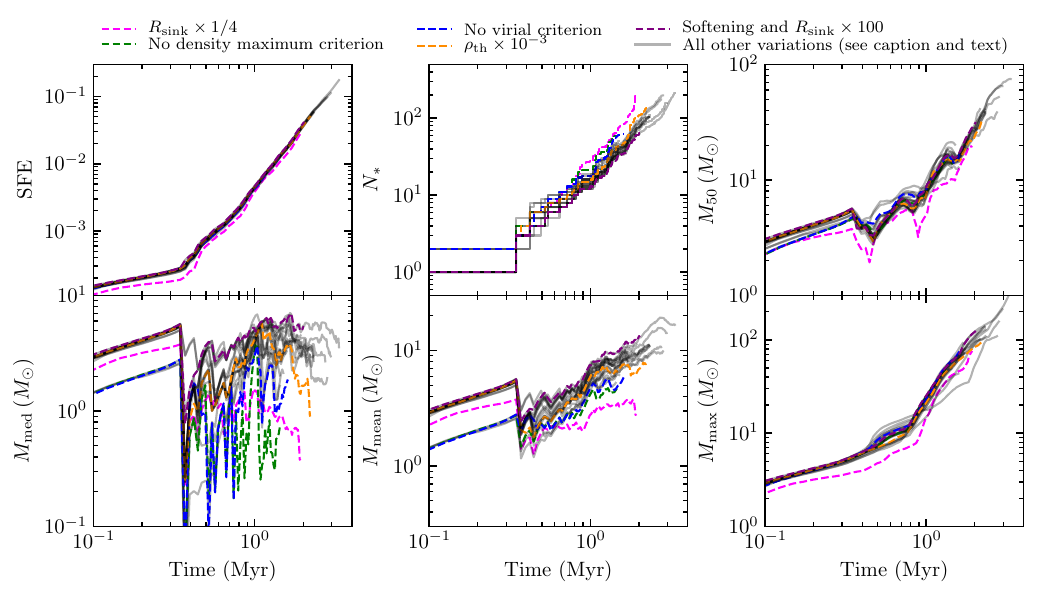}\vspace{-0.8cm}
    \caption{Effect of variations in sink particle formation and accretion prescriptions and properties upon the results of a simulation of a $2\times 10^4 \msun$ GMC of radius $10 \rm pc$, including just gravity and isothermal MHD (i.e. re-simulating ${\bf M2e4}$ from \citet{guszejnov_isothermal_mhd} at the same $10^{-3}\msun$ resolution). We plot the SFE (top left), number of sinks $N_\star$ (top centre), and the mass-weighted median ($M_{\rm 50}$), number-weighted median ($M_{\rm med})$, mean ($M_{\rm mean}$), and maximum ($M_{\rm max}$) mass statistics of the stellar mass function as a function of time from the beginning of SF. We highlight the most ``extreme" variations: neglecting the density maximum and virial sink formation criteria in turn, reducing $R_{\rm sink}$ by a factor of $1/4$ (to $\sim 5 \rm AU$) without reducing the softening, reducing the minimum density for sink formation $\rho_{\rm th}$ by a factor of $10^{-3}$, and increasing $R_{\rm sink}$ and the stellar softening $S_{\rm \star}$ by a factor of 100 (to $ 1800 \rm AU$). A variety of other overlapping sink parameter/prescription variations are plotted in grey, listed in full in Appendix \ref{appendix:sinktests}. Our predictions are fairly insensitive to most of these variations, provided they are within reasonable physical limits.}
    \label{fig:sinktests}
\end{figure*}

The reference numerical parameters in this setup are $\Delta m=10^{-3}\msun$, $R_{\rm sink}=S_{\rm \star}=18\rm AU$, and $\rho_{\rm J}= 2.6\times 10^{-14}\rm g\,cm^{-3}$, and some tests vary these quantities (Appendix \ref{appendix:sinktests}). We plot the results of this sink parameter study in Figure \ref{fig:sinktests}: the SFE, number of sinks $N_{\rm \ast}$, mass-weighted median sink mass $M_{\rm 50}$, median sink mass $M_{\rm med}$, and maximum sink mass $M_{\rm max}$. Our SFE and IMF results are remarkably robust to wide variations in the sink particle prescription and parameters, including $\rho_{\rm th}$ over a factor of $10^6$, and $R_{\rm sink}$ and $S_{\rm \star}$ over a factor of 200. The SFE is particularly robust, with very good agreement across all tests (the one outlier is consistent with a simple delay in accretion). The only setups that produced markedly different results in the IMF were ones with obvious flaws, such as ignoring the density maximum criterion (making the choice of which gas cell to turn into a sink generally non-unique), making $R_{\rm sink}$ much smaller than $S_{\rm \star}$ (making the accretion criteria unreasonably difficult to satisfy, because gravity is unresolved at the sink radius), and increasing $R_{\rm sink}$ and $S_{\rm \star}$ by a factor of $>100$ (a gross mismatch with the simulation resolution). Moreover our results do not hinge on any one particular ingredient or assumption -- neglecting each formation criterion in our prescription in turn had small effects (apart from the density maximum criterion). Stripping down our accretion criteria to simpler versions also made a negligible difference. Hence in practice there is a fair amount of redundancy between the different elements of our prescription. 

We conclude from this experiment that the results of STARFORGE simulations are unlikely to have any strong dependence upon the details of our sink implementation, at least within the space of \citet{Bate_1995_accretion}-like algorithms we have explored. Hence, to our knowledge, our sink implementation lacks parameter freedom for ``fine tuning" to ensure a particular desired result -- rather, simulation results are mainly sensitive to physical processes as desired. We generally recommend that $\rho_{\rm th}$ and $R_{\rm sink}$ be matched to the nominal density and spatial resolution limits of the simulations (e.g. Eqs \ref{eq:rhoJ},\ref{eq:deltax_J}), but e.g. the exact numerical prefactors we have adopted for these quantities are not important, within reasonable limits. Our results hint that simpler prescriptions can perform just as well as ours, but we err on the side of redundancy because the cost of evaluating the various sink formation and accretion criteria is small, and no criterion appears to be unreasonably strict (or else the sink algorithm would allow the simulation to crash due to a runaway gas pile-up).

\section{Thermodynamics}
\label{sec:thermo}
Although often idealized as such, GMCs are not truly isothermal, and many potentially-important effects in star formation require an explicit treatment of the thermal structure of the ISM, such as the dynamics of fragmentation \citep{bate2003, larson2005} and the evolution of bubbles driven by wind, radiative, and supernova feedback. We evolve the internal energy of the gas according to the MHD equations explicitly \citepalias{hopkins_gizmo_mhd}, accounting for all gravitational and MHD work terms with heating and cooling. We explicitly evolve the species Z, He, C, N,O, Ne, Mg, Si, S, Ca, and Fe, and by default assume initial solar abundances (Z, He, C, N,O, Ne, Mg, Si, S, Ca, Fe) = $(0.02,0.28, 3.26\times10^{-3}, 1.32\times10^{-3},8.65\times10^{-3}, 2.22\times10^{-3}, 9.31\times10^{-4}, 1.08\times10^{-3}, 6.44\times10^{-4},1.01\times10^{-4}, 1.73\times10^{-3})$, and re-scale abundances appropriately to the desired initial metallicity (but this can be freely varied).

We use a gas equation of state (EOS) with a variable adiabatic index $\gamma$, to account for variations in the equilibrium mixture of para- and ortho-hydrogen and the collisional dissociation of $\rm H_{\rm 2}$ above $\sim 2000 \rm K$ \citep{vaidya_2015_eos}. However we do not roll the heat of ionization into the EOS as in \citet{vaidya_2015_eos}, because this is handled separately by our cooling/chemistry solver. We fit to the values of $\gamma$ given in \citet{vaidya_2015_eos} (neglecting the feature corresponding to ionization) as a function of internal energy:
\begin{equation}
    \gamma = \frac{5}{3} + \sum_{k=1}^{5} \delta_k S\left(a_k(\log_{\rm 10} u- b_k)\right),
\end{equation}
where $S\left(x\right) = \frac{1}{2}\left(1+\frac{x}{\sqrt{1+x^2}}\right)$ is a sigmoid function, $\delta_k$=(-0.38,0.22,-0.068,-0.42,0.65), $a_k$=(5.95,6.18,10.26,7.71,98.87), $b_k$=(9.25,9.89,10.24,11.13,14.28), and $u$ is the specific internal energy in $\rm cm^{2}\,s^{-2}$.

We operator-split the adiabatic MHD evolution with a standard implicit cooling algorithm, which solves for equilibrium internal energy, temperature, net cooling/heating rate, mean molecular weight, and ionization state of the gas (treating the adiabatic heating rate from the MHD solver as an additional heating term). Our treatment of cooling and heating terms largely follows the FIRE-2 simulations (described fully in \citealt{hopkins2017_fire2} Appendix B), in accounting for free-free, photoionization/recombination, Compton, photoelectric,  metal-line, molecular, fine-structure, dust collisional, and cosmic ray heating and cooling processes. This cooling module has had various evolutionary updates since \citet{hopkins2017_fire2} that are not important for our results here (e.g. updating to the \citet{fauchergiguere_2020_uvb} UV background, which is similar to the previously-used \citet{fauchergiguere_2009_uvb} UVB at $z=0$), but will be detailed in full in an upcoming paper (Hopkins et al. 2021, in prep.). Note that our treatment of hot ($>10^6 \rm K$) gas considers the dominant radiative cooling mechanisms (i.e. free-free emission and metal lines) as described in \citet{hopkins2017_fire2}, but neglects heat conduction by thermal electrons by default, which may moderate expansion of wind and supernova bubbles.

\subsection{Background radiation}
In the intermediate-density ($\sim 100-10^4\rm cm^{-3}$) gas that makes up the bulk of the mass of GMCs, the thermal structure is set mainly by the balance of photoelectric heating and molecular or fine-structure cooling \citep{glover:2011.molecules.not.needed.for.sf}, necessitating some treatment of this background. When modeling solar circle conditions, we assume an isotropic \citet{draine_1978_isrf} background $e_{\rm FUV} = 9 \times 10^{-14} \rm erg\,cm^{-3}$ for purposes of photoelectric heating, i.e. 1.7 times the \citet{habing1968} flux of photons in the range $6-13.6\rm eV$. For each gas cell, we evaluate the optical depth to the FUV background on-the-fly using the {\texttt TreeCol} algorithm \citep{treecol}, i.e. summing the optical depths of tree nodes grouped into angular bins during the pass through the gravity tree. We default to a simple 6-bin angular binning of the sky, and assume an opacity of $\kappa_{\rm FUV}=500\rm cm^{-2} Z/Z_\odot$.

We also model the background radiation due to galactic dust emission as a black-body spectrum with energy density $0.31 \rm eV\,cm^{-3}$ and an effective temperature of $20 \rm K$. When we evolve this radiation component with explicit RHD, we simply implement this radiation energy density and temperature as the initial conditions, and allow both to evolve freely (\S\ref{sec:radiation}). Without RHD, it is simply held fixed.

The background radiation components quoted here are as measured in the Solar neighbourhood, and can be re-scaled to appropriate values for other environments.

\subsection{Dust cooling and heating}
Dust cooling and heating are dominant at high ($\gtrsim 10^6 \rm cm^{-3}$) ISM densities \citep{goldsmith_1978_cooling}. The dust heating/cooling term $\Lambda_{\rm dust}$ is \citep{meijerink.spaans:xray.cooling.models}:
\begin{equation}\
\begin{split}
    \Lambda_{\rm dust} &= 1.12\times10^{-32}\,{\rm erg\,s^{-1}\,cm^{3}}\,\left(T-T_{\rm dust}\right)\,T^{1/2}\,\\
    &\times \left(1-0.8\,\exp\left(\frac{-75}{T}\right)\right)\,\left(\frac{f_{\rm d}}{0.01}\right),
    \label{eq:lambdadust}
\end{split}
\end{equation}
where $T$ is the gas temperature in $K$, $T_{\rm dust}$ is the dust temperature, and $f_{\rm d}$ is the local dust-to-gas ratio, which we take to be $f_{\rm d}=0.01 Z/Z_\odot$ (ie. assume a constant dust-to-metals ratio equal to the local value) in simulations which do not explicitly follow dust dynamics (otherwise this is the actual local value $\rho_{\rm dust}/\rho_{\rm gas}$). How $T_{\rm dust}$ is determined depends on whether or not we are using an explicit RHD solver.

\subsubsection{Simulations with explicit RHD}

If explicit RHD is enabled, we co-evolve the gas, dust, and radiation temperature self-consistently as in \citet{FIRE_RT}, including the stellar luminosity in various bands accounting for photon transport, absorption and emission using dust opacity fits from \citet{semenov_2003}. Dust cooling is handled by including Eq. \ref{eq:lambdadust} as a radiation source term for the IR band, so that energy lost to dust cooling is transported away by the RHD solver. This automatically handles the trapping of cooling radiation in the optically-thick limit (setting e.g. the ``opacity limit" for fragmentation, \citealt{rees1976}). We explain our RHD treatment fully in \S\ref{sec:radiation}. 

\subsubsection{Simulations without explicit RHD}
If an explicit RHD solver is not enabled, we either make the minimal assumption that $T_{\rm dust}$ is constant, or use a simple, inexpensive RT approximation similar to \citet{guszejnov_feedback_necessity} and \citet{Federrath_2017_IMF_converge_proceedings}. This approximation uses the {\small LEBRON} radiative transfer algorithm \citep{hopkins2017_fire2} to estimate the IR radiation energy density from local sources at the position of a gas cell in the gravity tree pass, summing over contributions from all stars:
\begin{equation}
    e_{\mathrm{IR},g} = \sum_s \frac{L_s}{4 \pi r_{gs}^2 c}
\end{equation}
where $L_s$ are the respective bolometric luminosities of the sink particles, $r_{gs}$ are the distances from the gas cell to the sinks. We then solve for $T_{\rm dust}$ assuming local equilibrium between absorption and emission according to a  $\beta=1$ opacity law (i.e. $\kappa\left(\nu\right) \propto \nu$, e.g. \citealt{draine:2006.dust.opacity}). $T_{\rm dust}$ is then the solution to the quintic equation
\begin{equation}
    \left( T_{\rm dust}/2.92\,{\rm K}\right)^{5}  = \sum_{k} \left(T_{{\rm rad},\,k}/{\rm K}\right)\,\left(e_{k}/{\rm eV\,cm^{-3}}\right)
\end{equation}
where the index $k$ runs over three radiation field components with respective energy densities and effective temperatures: the above component from local sources, which is assumed to have $T_{\rm rad,IR}=T_{\rm dust,IR}$, the CMB with $e_{\rm CMB} = 0.262\left(1+z\right)^4 \rm eV\,cm^{-3}$ and $T_{\rm rad,CMB}=2.73 \mathrm{K} \left(1+z\right)$, and the dust-reprocessed component of the interstellar radiation field (ISRF) with $e_{\rm ISRF} = 0.31\rm eV\,cm^{-3}$ and $T_{\rm rad,ISRF}=\rm \max\left(20 K,T_{\rm rad,CMB}\right)$, with fiducial values appropriate for solar neighborhood conditions, and adjustable depending upon the simulated environment. Note that this differs slightly from \citet{guszejnov_feedback_necessity} and \citet{Federrath_2017_IMF_converge_proceedings}, who adopted the optically-thick grey-opacity radiative transfer solution in the static diffusion limit $T \propto e_r^{1/4}$, as was found in \citet{Offner_2009_radiative_sim}.

\subsection{Optically-thick cooling and the opacity limit}
If the trapping of cooling radiation is not being solved self-consistently by our RHD solver, we also adopt a simple prescription for interpolating the cooling rate between the optically-thin and -thick regimes. If the net absolute heating/cooling rate is $|\Lambda_{\rm Net}|$, then we enforce:
\begin{align}
|\Lambda_{\rm Net}| <& \Lambda_{\rm BB} \\ 
\Lambda_{\rm BB} & \equiv 5.67\times10^{-5}\,T^{4}\,\left(\frac{\mu}{a \Sigma_{\rm eff}} \right)\frac{1}{1 +  \kappa_{\rm eff}\, a \Sigma_{\rm eff}}\,n_{\rm H}^{-1},
\label{eq:rafikov}
\end{align}
where $\mu \approx 2.4$ is the mean molecular weight and $\Sigma_{\rm eff}$ is estimated via the {\rm TreeCol} algorithm, $\kappa_{\rm eff}$ is the effective opacity detailed in \citet{hopkins2017_fire2}, and $a=0.2$ is an uncertain geometric factor chosen to reproduce detailed RHD protostellar collapse calculations \citep{masunaga1998}. This limits the cooling (or heating) rate to the bound for a slab geometry derived in \citet{rafikov2007}. This is still approximate, but is more realistic than an ``effective equation of state" that transitions from isothermal to adiabatic \citep[e.g.][]{bate2003,bate09a}: we are still always allowing for heating and cooling, with radiation terms that explicitly account for optical depth, which is the physically-relevant quantity for radiative cooling. In a single isolated collapsing clump in a spherical geometry, an equation of state may be justified because $\Sigma_{\rm eff} \approx \rho \lambda_{\rm J}$, but this does not hold in general, and particularly not in systems that are optically-thick to dust {\it globally} ($\Sigma_{\rm eff} \gtrsim 1\rm g\,cm^{-3}$).
\subsection{Tests}
\begin{figure}
    \centering
    \includegraphics[width=\columnwidth]{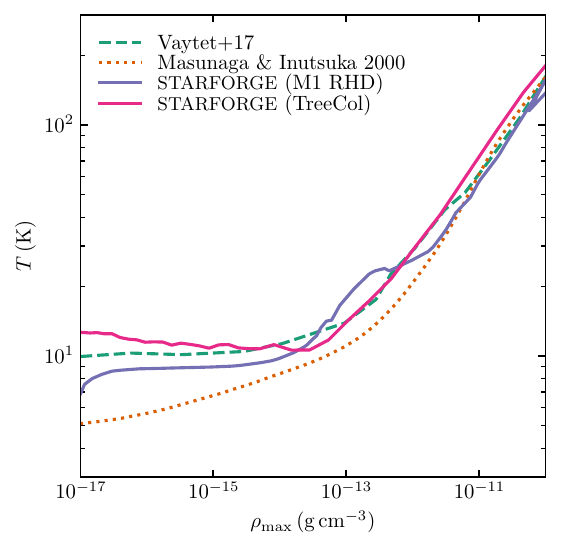}\vspace{-0.8cm}
    \caption{Thermal evolution of the densest gas cell in the centre of a collapsing $\msun$ core as a function of its density $\rho_{\rm max}$, replicating the test problem in \citet{masunaga1998} with our default physics using our self-consistent M1 RT solver with gas-dust-radiation coupling (\S\ref{sec:radiation}) and our simple optically-thick cooling approximation based upon the {\small TreeCol} algorithm (Eq. \ref{eq:rafikov}).} 
    \label{fig:masunaga}
\end{figure}

To test the code's ability to capture the transition from isothermal to adiabatic behaviours as the ISM gets optically-thick to cooling radiation (important e.g. for the opacity limit for fragmentation), we simulate collapse of a $1 \msun$, uniform-density Jeans-unstable core with both methods described here: explict RHD with the M1 solver and the simpler {\small TreeCol}-based prescription (Eq. \ref{eq:rafikov}). We initialize the cloud with a mass resolution of $10^{-5}\msun$ (sufficient to mariginally resolve the first Larson core, \citealt{bate2003}), arranging the cells in a uniform-density glass configuration with density $5.3 \times 10^{-18}\rm g\,cm^{-3}$, and assume solar metallicity with a dust-to-gas ratio of $0.01$, as has been simulated by many other RHD studies \citep{larson_1969,masunaga1998,vaytet_2017_protostellar_collapse}. In Figure \ref{fig:masunaga} we plot the thermal evolution of the center of the core as a function of its density, showing good agreement with previous calculations: in all instances, the transition from isothermal ($T \sim const.$) to adiabatic ($T \propto \rho^{2/5}$) evolution occurs at $\sim 10^{-13}\rm g\,cm^{-3}$. Small residual differences between the four calculations at the level seen here are expected, due to varied assumptions about radiation initial and boundary conditions, dust properties and opacities, molecular equation of state, etc.



\section{Feedback}
\label{sec:feedback}
We now describe the respective physical model and numerical implementations of each feedback mechanism in turn. For those with well-understood or analytic solutions (winds, SN, and radiation) we will verify that each module performs accurately. Where such solutions are not available (e.g. jets), we will do the next best thing: test for robustness to numerical details and agreement with other codes.

\subsection{Generic injection algorithms}

\begin{figure}
    \centering
    \includegraphics[width=\columnwidth]{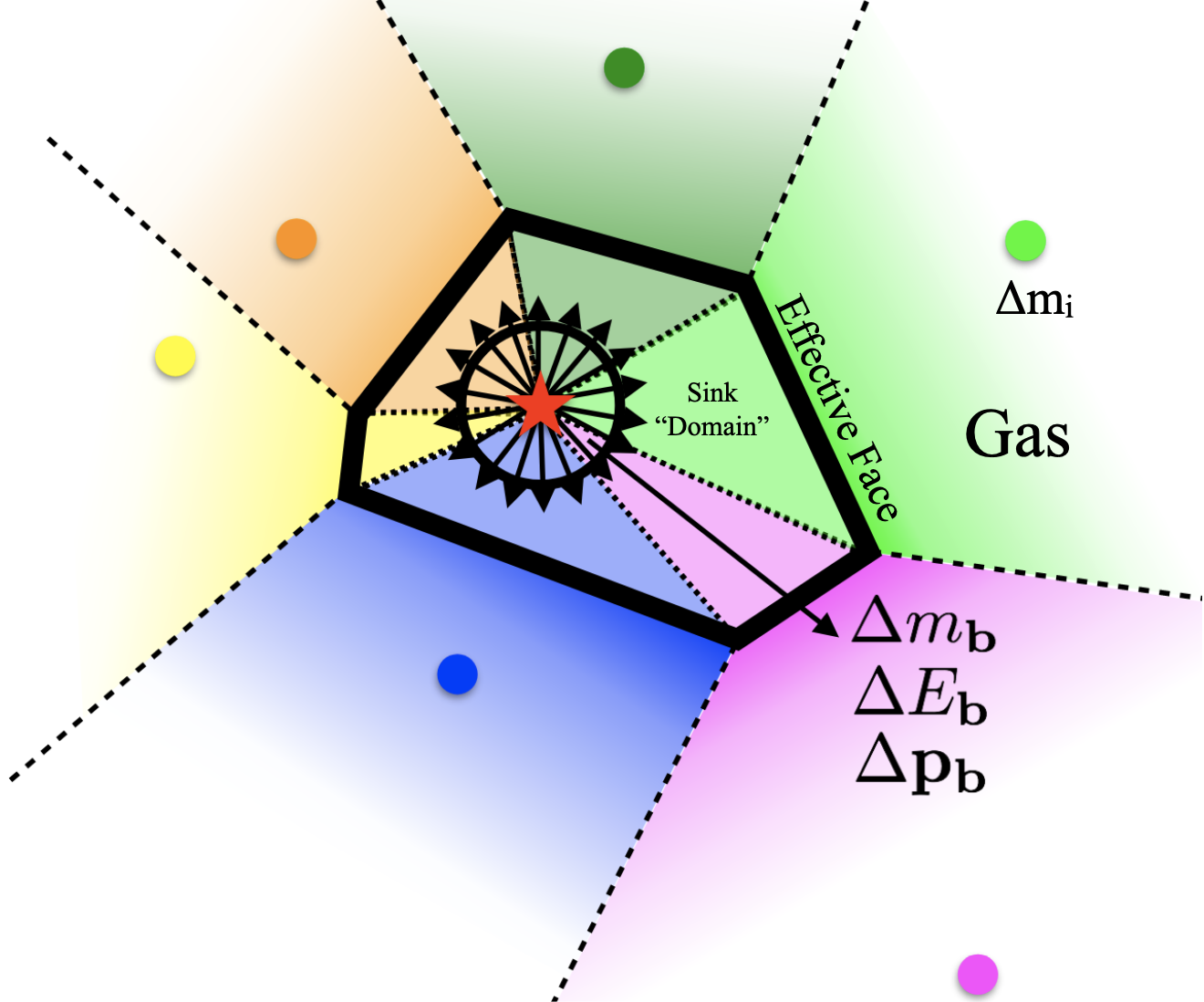}\vspace{-0.7cm}
    \caption{Illustration of the mesh-free local injection procedure used for coupling photons and stellar winds with unresolved free expansion to gas cells in STARFORGE (adapted from \citealt{Hopkins_2018_sne_feedback}). Local injection adds momentum, energy, and either mass or photons to pre-existing interacting gas cells (coloured domains with circles representing the mesh-generating points) around the sink particle (red star), in proportion to the solid angle subtended by each cell according to the ``effective" faces constructed around each source from the neighboring mesh-generating points (thick black lines).  We use a weighting scheme that ensures statistical isotropy and exact momentum and energy conservation, and in the case of photons accurately accounts for unresolved extinction between the star and cell centre (\S\ref{sec:radiation}).}
    \label{fig:localinjection}
\end{figure}
\begin{figure}
    \centering
    \includegraphics[width=0.7\columnwidth]{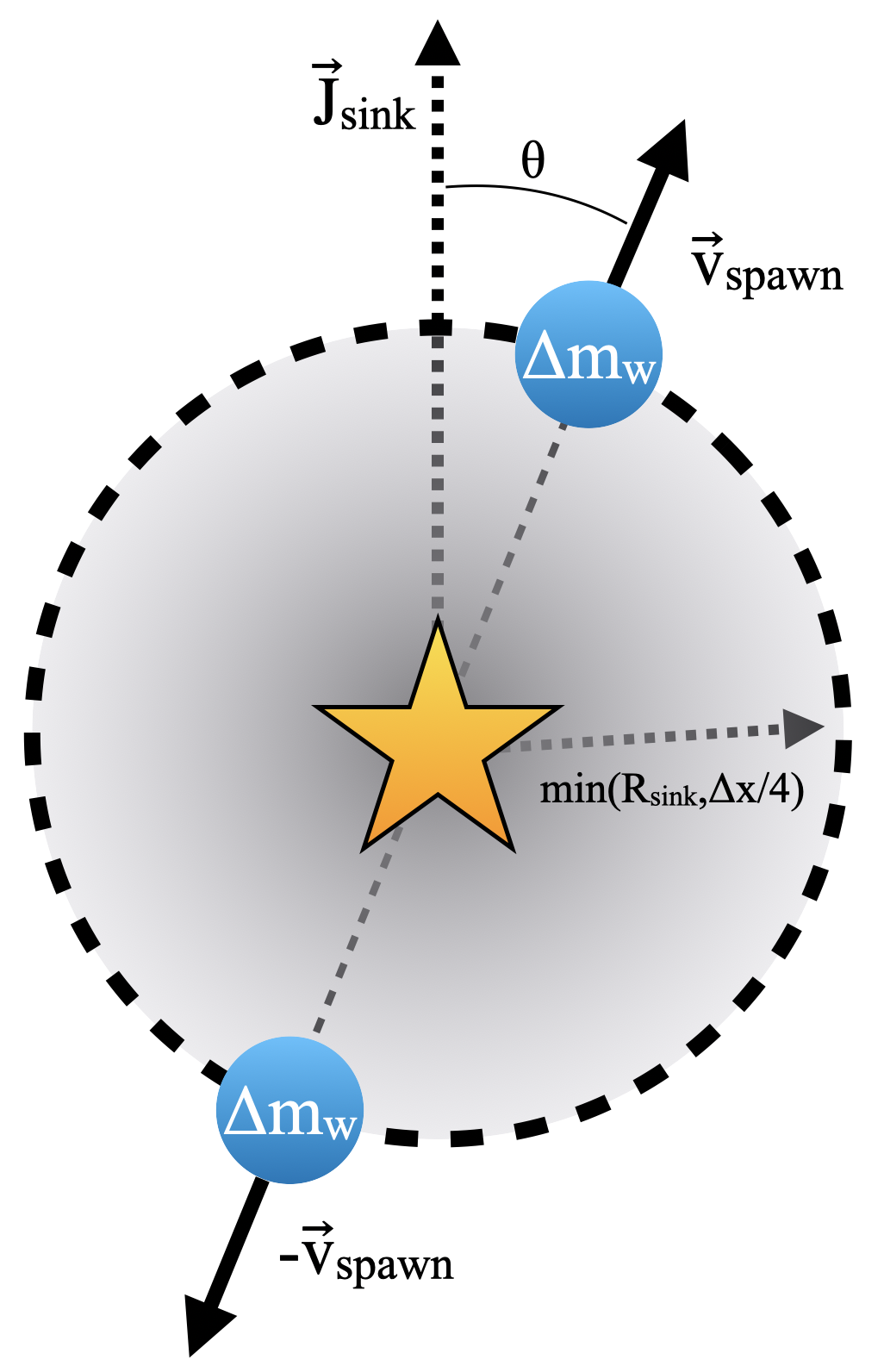}\vspace{-0.4cm}
    \caption{Illustration of the cell spawning technique for injecting mass return from sinks (winds, jets, and supernovae, \S\ref{sec:spawning}). We create new gas cells of mass $\Delta m_{\rm w}$ (blue circles) at the sink radius (or at a fraction of the local inter-cell spacing, whichever is smaller), at antipodal positions and velocities to ensure conservation of center of mass and momentum. The angle $\theta$ from the sink angular momentum vector $\mathbf{J}_{\rm sink}$ can be controlled to allow arbitrarily-collimated injection. $\Delta m_{\rm w}$ can be set smaller than $\Delta m$ (the ``normal" mass resolution of ambient gas) to improve resolution in diffuse feedback-driven cavities. Note that this depicts the launching of a pair of gas cells (as for jets and winds), but for SN we launch 24 at once in an angular grid (\S\ref{sec:SNe}), designed to ensure exact conservation and isotropy.}
    \label{fig:spawncartoon}
\end{figure}
\subsubsection{Local injection}
\label{sec:injection}
When coupling feedback in the form of mass, momentum, and energy from stars, a natural approach is analogous to the manner in which the MHD equations are solved: simply determine the fluxes of those respective quantities at the faces of surrounding gas cells, or more generally distribute those quantities in some weighted fashion in the local hydrodynamic stencil -- we refer to this technique as ``local injection" (illustrated for a meshless/unstructured mesh code in Figure \ref{fig:localinjection}). This is what is done in virtually all grid codes and is the technique typically used in {\small GIZMO} simulations for local radiative feedback, SNe, and stellar winds \citep{Hopkins_2018_sne_feedback,hopkins2017_fire2, FIRE_RT}. We adopt this algorithm for feedback coupling (described in full in \citealt{Hopkins_2018_sne_feedback}) where appropriate: for stellar winds when the free-expansion radius is unresolved (\S\ref{sec:winds}) and for photon injection (\S\ref{sec:radiation}). This method ensures machine-precision conservation of mass, momentum, and energy, and ensures that feedback is injected isotropically (or according to the desired angular weighting) even when the local spatial arrangement of cells is anisotropic (unlike e.g. a simple kernel weighting).

\subsubsection{Cell spawning}
\label{sec:spawning}
Local injection with Lagrangian methods can run into a major challenge: the resolution is concentrated where gas dens, but feedback structures such as jet cavities, supernova remnants, and wind bubbles can be very diffuse. Moreover, if feedback is driving mass away then this problem grows worse over time. 
And if the inter-cell spacing becomes sufficiently large, it may cease to be a good approximation to instantaneously dump mass, momentum, and energy, because the gas has finite travel time. Moreover, if we restrict injection to the nearest neighbor cells we cannot inject outflows more collimated than the solid angle subtended by a neighbouring cell (which can be large). So where local injection is not feasible or appropriate, we instead {\it create} new gas cells, in a procedure we refer to as ``cell spawning" (Figure \ref{fig:spawncartoon}). A similar technique has been used previously in SPH simulations of stellar winds \citep{Dale_Bonnell_2008_winds_IMF} and protostellar outflows \citep{rhode_2019_sph_jets}, and recently in {\small GIZMO} to simulate AGN jets \citep{torrey_2020_agn_jets}. We adapt it here to simulate protostellar jets, stellar winds (when resolvable), and supernova ejecta. 

 Cell spawning can be viewed as the inverse of the gas cell deletion operation that occurs during sink accretion: a new cell is created at a certain position with a certain mass, velocity, and internal energy, and the volume partition in its vicinity is re-computed to accommodate it in the next density iteration. We take the distance between the centre-of-mass mesh-generating point of the spaned cell and the sink to be
 \begin{equation}
     R_{\rm spawn} = \min \left(R_{\rm sink}, \Delta x_s/2\right),
 \end{equation}
 where $\Delta x_{\rm s}$ is the average inter-cell spacing in the vicinity of the sink. We prescribe the initial radial direction and velocity according to the desired angular pattern of the feedback mechanism being realized (\S\ref{sec:jets}-\ref{sec:SNe}). We assign purely radial initial velocities, but non-radial velocities could potentially be used to model angular momentum transport \citep{Federrath_2010_sinks}.  Spawned cells are assigned an initial temperature of $10^4 \rm K$, and a very small, random initial magnetic field scaled such that the initial plasma $\beta$ is $10^{6}$.
  
 Spawning is allowed to occur when the sink's internal respective feedback reservoir (Fig. \ref{fig:sinkdiagram}) contains enough mass to produce at least $N_{\rm spawn}\times \Delta m_{\rm w}$ cells, where the number of cells spawned at a time $N_{\rm spawn}$ and spawned cell mass resolution  $\Delta m_{w}$ are specified for the respective feedback channel. Note that cells do not necessarily have to be spawned with a mass resolution equal to the nominal average ``ambient" gas cell mass $\Delta m$ : rather $\Delta m_{\rm w}$ can be chosen to be smaller to achieve better time resolution of feedback and to improve spatial resolution within diffuse feedback cavities. For jets and winds, we have generally found the choice $\Delta m_{w} = 0.1\Delta m$ to be a good compromise between computational cost and resolution. For supernovae, we simply take $\Delta m_{\rm w} = \Delta m$. With these choices, we note the spatial resolution in diffuse feedback bubbles will generally be fairly coarse ($\sim 1 \rm pc$) for a typical $\Delta m \approx 10^{-3}M_\odot$, possibly making it challenging to resolve channels and leakage of hot gas \citep[e.g.][]{rogers_pittard_winds}.
 
 Care must be taken when handling MHD interactions between cells of greatly differing masses and sizes in MFM, particularly for new cells which change the local volume partition and can perturb $\nabla \cdot \mathbf{B}$ (interacting with out $\nabla \cdot \mathbf{B}$ cleaning algorithm). Spawned cells with $\Delta m_{\rm w} < \Delta m/2$ default to a lower-order but more-robust reconstruction in the Riemann problem. We also limit the magnitude of the oriented effective face area $\mathbf{A}_{g g'}$ between cells (all interacting cells, wind or not) to the lesser of their geometric areas, $A_{\rm max}= \min\left(\uppi h_g^2, \uppi h_{g'}^2\right)$. A spawned cell $g$ is merged into a normal cell $g'$ if they are hydrodynamically-interacting neighbors, they are moving toward each other, and $|\vvector_g - \vvector_{g'}|<\min\left(c_{\rm s,g}, c_{\rm s,g'}\right)$. We have found that this is a good indicator that the spawned cell has merged with the surrounding ISM, and hence its additional resolution is no longer required.

Lastly, to ensure physical and convergent results, it is important for any feedback algorithm to ensure conservation of momentum and centre of mass. We achieve this by always spawning cells in multiples of 2, such that each cell has an antipodal counterpart in the opposite direction, giving machine-precision conservation.

\subsection{Jets}

\begin{figure*}
    \centering
    \includegraphics[width=\columnwidth]{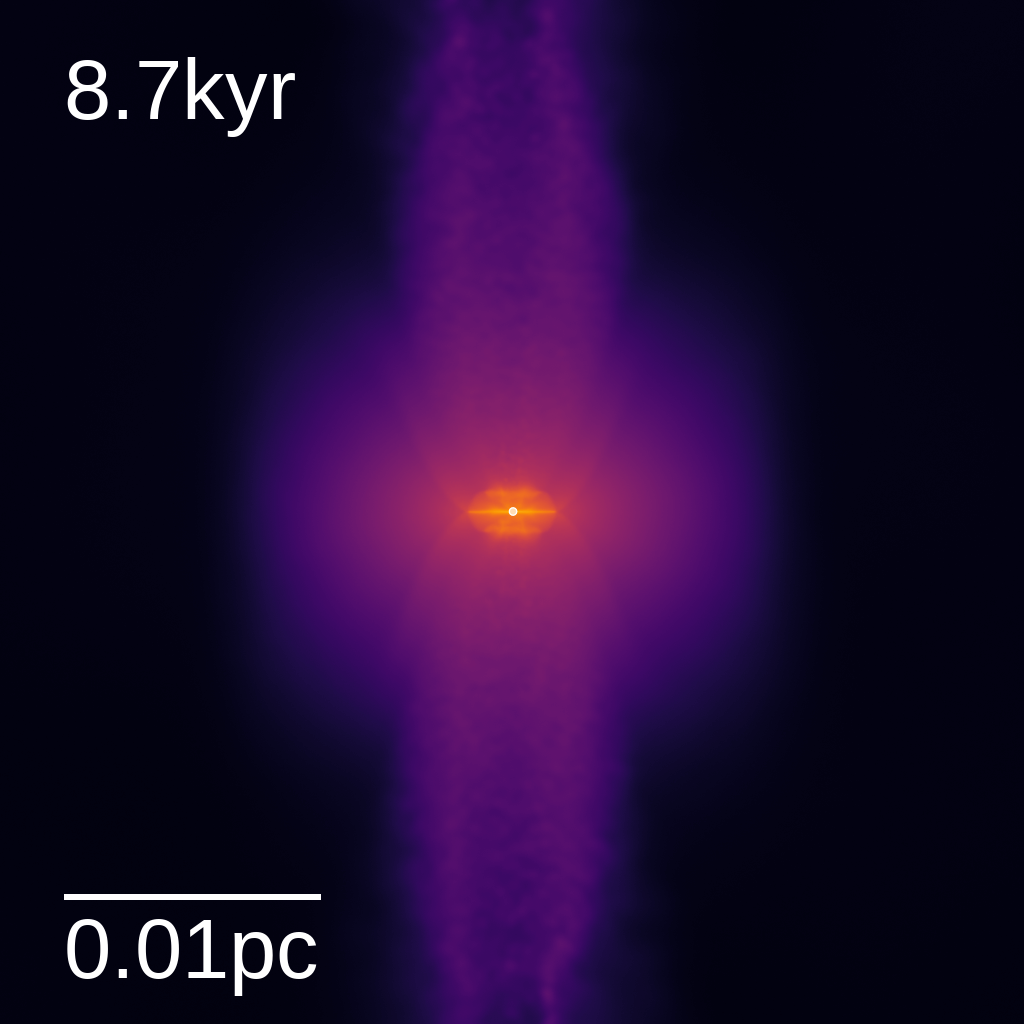}\includegraphics[width=\columnwidth]{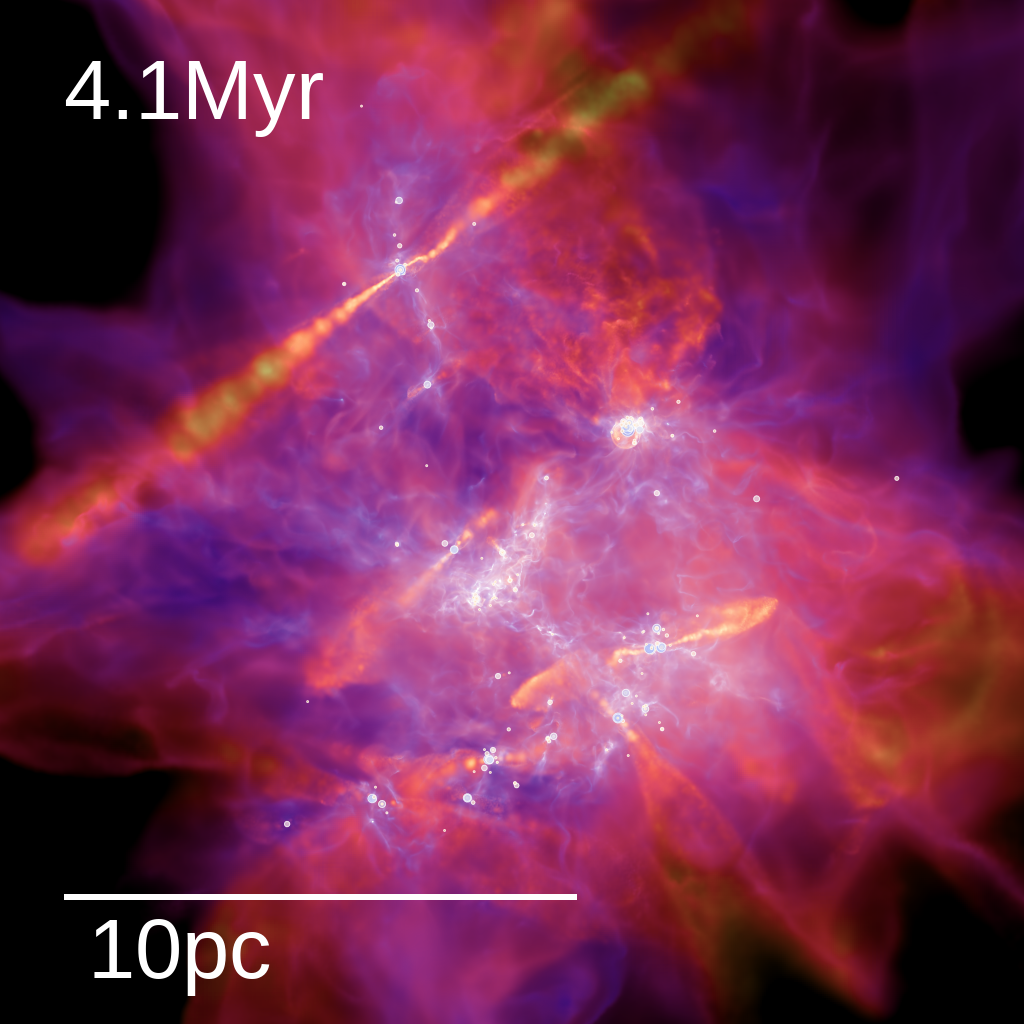}
    \caption{Examples of the protostellar jet module (\S\ref{sec:jets}) in action in SF simulations. {\it Left}: Idealized laminar rotating core collapse problem forming a single star, run at high ($10^{-5}\msun$) resolution. As the star accretes from a disk, jets clear out high-velocity diffuse cavities along the poles, entraining material away from the core. {\it Right:} Bipolar outflows (higlighted in orange) permeate a $2\times 10^4 \msun$ GMC run at $10^{-3}\msun$ resolution (typical of STARFORGE runs), with the largest penetrating out to $\sim 10 \rm pc$ scales before merging with the ISM. This map colors by 1D line-of-sight velocity dispersion (purple is $\sim 0.1 \rm km\,s^{-1}$, orange is $\sim 10 \rm km\,s^{-1}$ and modulates the lightness to encode surface density information (lighter is denser).}
    \label{fig:jets_pictures}
\end{figure*}

\begin{figure*}
    \centering
    \includegraphics[width=\textwidth]{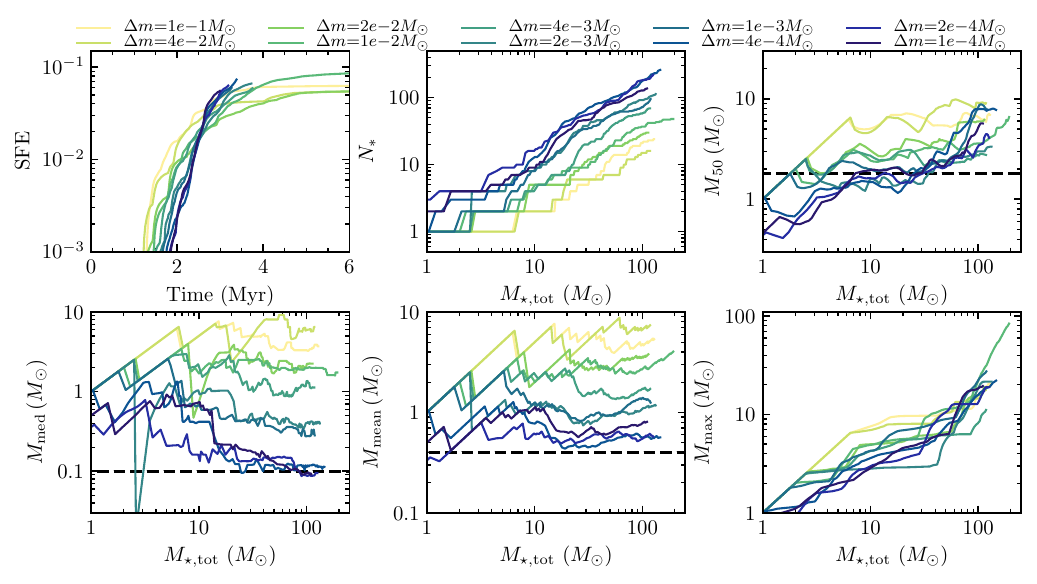}\vspace{-0.8cm}
    \caption{Resolution study of a simulation of a $2000\msun$ GMC with initial radius $3\rm pc$, with MHD, gravity, cooling physics, and protostellar jets (\S\ref{sec:jets}). We vary the mass resolution $\Delta m$ from $10^{-4}-0.1\msun$ (darker is finer), scaling sink accretion and softening radii $\propto \Delta m$ from $2-2000\rm AU$ (Eq. \ref{eq:Rsink}) and scaling the sink density threshold $\propto \Delta m^{-2}$ (Eq \ref{eq:rhoJ}), so that there is no characteristic numerical scale that could cause false convergence. We plot the resulting star formation history and the mass-weighted median, mean, and median stellar masses $M_{\rm 50}$, $M_{\rm mean}$, and $M_{\rm med}$, and black dashed lines correspond to the value for a \citet{kroupa_imf} IMF. SFE and IMF statistics cease to scale systematically with resolution past a certain threshold.}
    \label{fig:jets_convergence}
\end{figure*}

\label{sec:jets}
\subsubsection{Physics prescription}
Collimated, bipolar protostellar outflows (jets) have special importance as a feedback mechanism. Many works have demonstrated their potential importance both in setting the IMF and SFE, because they are the only channel can be both prompt and omnipresent, immediately regulating the growth of individual stars without requiring e.g. massive stars with dynamically-relevant wind or radiative fluxes to be present \citep{matzner_1999_clusters, krumholz_2019_cluster_review}. We find that they are likely to be an important ingredient for the IMF in  \citetalias{starforge_jets_imf}, consistent with previous studies \citep{krumholz_2012_orion_sims,myers2014_sim, Federrath_2014_jets, li_2018_mhd_jets_sf,Cunningham_2018_feedback}. The details of the MHD jet launching mechanism (e.g. whether it is better-described by the canonical ``X-wind" or ``D-wind" models, \citealt{pudritz_1983_dwind, shu_1994_xwind}) are the subject of active research, and depend upon a variety of complex microphysics operating at sub-$\rm AU$ scales that are not practical to resolve in our simulations (although the simulations may provide important context for subsequent ``zoom-in" studies of individual stars). Thus we model jets according to a simple phenomenological prescription following \citet{Cunningham_2011_outflow_sim}, parametrizing the jets' properties in three parameters: the fraction $f_{\rm w}$ of mass accreted by the envelope-disk-star system that is diverted to the jet (see Fig. \ref{fig:sinkdiagram}), the fraction $f_{\rm K}$ of the Keplerian velocity at the protostellar radius $R_{\rm \star}$ at which jets are launched, such that 
\begin{equation}
    v_{\rm jet} = f_{\rm K}\sqrt{\frac{G M_\star}{R_\star}},
    \label{eq:vjet}
\end{equation}
and the collimation angle $\theta_{\rm 0}$, such that the angular distribution of injected wind momentum is given by \citep{matzner_1999a_jets}:
\begin{equation}
    \xi\left(\theta,\theta_{\rm 0}\right) = \left(\ln \left(\frac{2}{\theta_0}\right) \sin^2\theta + \theta_0^2\right)^{-1},
    \label{eq:jet_angular_dist}
\end{equation}
where $\theta$ is the angle with respect to the angular momentum axis of the sink $\mathbf{J}_s$ of the sink. This concentrates injection in a narrow cone of angular size $\approx \theta_{\rm 0}$ about the angular momentum axis of the star (see Fig. \ref{fig:jets_pictures}). Our default choices are $f_{\rm w}=f_{\rm K}=0.3$ and $\theta_0=0.01$, following \citet{Cunningham_2011_outflow_sim} and other authors. These choices put the momentum loading $f_{\rm w} f_{\rm K}$ roughly in the middle of the observed range (see \citealt{Cunningham_2011_outflow_sim} \S2.4, \citealt{Federrath_2014_jets} \S3.5 and references therein). The values of $f_{\rm w}$ and $f_{\rm K}$ do matter: in \citetalias{starforge_jets_imf} we find that the product $f_{\rm w} f_{\rm K}$ affects the IMF peak. The specific value of $\theta_0$ is likely to be unimportant because in any realistic turbulent accretion scenario $\mathbf{J}_s$ will generally tend to precess during accretion over an angular region much larger than $\theta_0$ \citep[][]{rosen_2020_jets_radiation}, and even without precession jet cavities will expand in the perpendicular direction, opening up an ever-increasing solid angle \citep{Arce_2006,Offner_2011_outflows}. In our tests we will show that our results are insensitive to variations in $\theta_{\rm 0}$ of at least a factor of 10, which is consistent with prior hydrodynamic outflow simulations carried out by \citet{Offner_Arce_2014}.

\subsubsection{Numerical methods}
We always couple jets via cell spawning (\S\ref{sec:spawning}), waiting until sufficient mass is available in the jet reservoir to spawn 2 cells of mass $\Delta m_{\rm w}$. The angular direction with respect to the sink angular momentum $\mathbf{J}_s$ is sampled randomly for the first cell from Eq. \ref{eq:jet_angular_dist}, and the second cell is pointed in the opposite direction, conserving momentum and centre of mass. Our prescription ignores both the angular momentum and magnetic flux content of the jet material. Although outflows can be the dominant mechanism of angular momentum transport within the disk (i.e. on scales smaller than the sink radius), the material in the disk already had to have very low angular momentum to get to the base of the jet, so it is unlikely that this angular momentum has important effects on larger scales once transported back out in the jet (however it is considered by other jet feedback models, such as by \citealt{Federrath_2014_jets} and  \citealt{rhode_2019_sph_jets}). The effects of the magnetic field in the jet are less readily dismissed, however this would be nontrivial model in the present numerical framework due to the problem of determining what initial $\mathbf{B}$ should be assigned to the spawned cells while observing conservation laws.

We plot some examples of the effects of the jet module in simulations in Figure \ref{fig:jets_pictures}. We generally observe realistic-looking structures in the simulations, with broad bipolar cavities penetrated by a narrow jet, surrounded by a disk (if angular momentum support is important) or a pseudo-disk (of infalling material funneled by the bipolar cavities). $v_{\rm jet}$ tends to increase as the star accretes (because $R_{\star}$ varies only weakly in Eq. \ref{eq:vjet}), so the jet tends to catch up with itself, piling up and cooling in a plume-like region reminiscent of Herbig-Haro objects \citep[e.g.][]{bally_2016_jets}.

\begin{figure*}
    \centering
    \includegraphics[width=\textwidth]{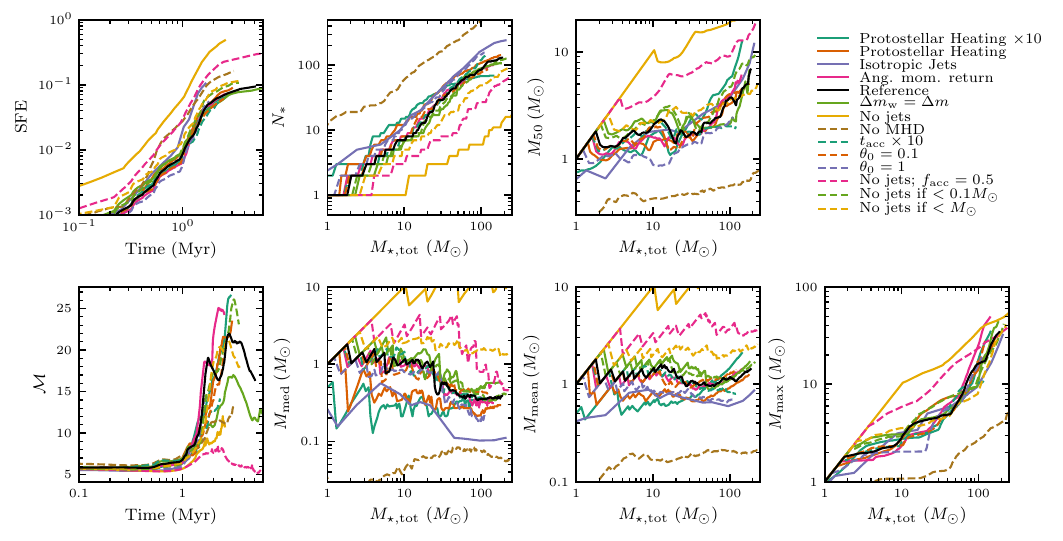}\vspace{-0.8cm}
    \caption{Effects of various numerical and physical variations upon the star formation history, kinematics, and IMF evolution of the same $2000\msun$ GMC simulated in Fig. \ref{fig:jets_convergence}, at $10^{-3}\msun$ resolution. Quantities are as described in Fig. \ref{fig:jets_convergence}, plus the evolution of the 3D mass-weighted RMS Mach number $\mathcal{M}$ (bottom left). {\it Reference}: Baseline settings, with cooling, MHD, and protostellar jets enabled. {\it No jets}: Jet module disabled. {\it No jets; $f_{\rm acc}=0.5$}: Jet module disabled, and sinks delete half the mass that they accrete. {\it Protostellar heating}: Including the approximate protostellar heating prescription described in \S\ref{sec:spawning}, with physical and artificially-large ($\times 10$) luminosities. {\it Isotropic jets}: spawning jet cells in random (vs. highly-collimated, Eq \ref{eq:jet_angular_dist}) directions. {\it $\theta_{\rm 0} = X$}: Changing the jet collimation angle $\theta_{\rm 0}$ (Eq. \ref{eq:jet_angular_dist}) from the standard value of 0.01. {\it No MHD}: Setting the magnetic field to 0. {\it $t_{\rm acc}\times 10$}: Scaling the accretion smoothing timescale ($\ref{eq:tacc}$) $\times 10$. {\it No jets if <X}: Jets are disabled for stars with mass $<X$. {\it $\Delta m_{\rm w}=\Delta m$}: Setting the jet cell mass resolution to the nominal mass resolution $\Delta m$, as opposed to the standard $\Delta m/10$. {\it Ang. mom return:} returning accreted angular momentum to surrounding gas as in \citet{hubber:2013.sinks}.}
    \label{fig:jettests}
\end{figure*}

Lacking a test problem for our jet module that admits an analytic or universally agreed-upon solution, we resort to heuristic methods to validate it: checking for numerical convergence, checking for robustness to uncertain or arbitrary numerical parameters, and finally comparing with another published solution from a different code.

\subsubsection{Resolution tests}
\label{sec:jettests}
Here we consider the effects of numerical resolution upon a simulation of the ${\bf M2e3}$ GMC model introduced in \citetalias{guszejnov_isothermal_mhd} with gravity, MHD, the cooling module without explicit RT or protostellar radiation (\S\ref{sec:thermo}), and the jet module enabled. The initial condition is a spherical, uniform-density GMC with mass $2\times 10^3 \msun$, radius $3\rm pc$, and virial parameter $2$ (for an initial turbulent Mach number of $\sim 9$). The GMC is surrounded by a diffuse medium with $1/1000$ the density filling a $30\rm pc$ box, and the initial magnetic field is uniform throughout the box with a strength of $2.3 \mu G$. The normal mass resolution $\Delta m$ varies from $0.1-10^{-4}\msun$, and the jet mass resolution $\Delta m_{\rm w}$ varies from $0.01-10^{-5}\msun$. The sink formation density threshold, sink radius, and sink softening scale are varied according to the mass resolution (\S\ref{section:sinks}), scaling $\rho_{\rm th} \propto \Delta m^2$ over a factor of $10^6$ between $3\times 10^{-17}-3\times 10^{-11} \rm g\,cm^{-3}$, and scaling $R_{\rm sink} = S_{\rm \star} \propto \Delta m$ from $1800-1.8 \rm AU$. Hence there is no purely-numerical, resolution-related quantity that is held fixed in our resolution study, so the possibility of inferring ``false" convergence is ruled out\footnote{Scaling all purely-numerical, dimensional sink-related quantities with resolution is important for resolution studies in SF simulations. Otherwise, it is possible that the constant value of e.g. the density threshold or sink radius imprints a characteristic Jeans mass or length, leading one to falsely infer convergence. In near-isothermal problems with Lagrangian codes, this entails scaling $\rho_{\rm th}\propto \Delta m^2$ and $R_{\rm sink} \propto \Delta m$. For AMR codes, one must scale $\rho_{\rm th} \propto \Delta x_{\rm min}^{-1/2}$ and $R_{\rm sink} \propto \Delta x_{\rm min}$, where $\Delta x_{\rm min}$ is the spatial resolution at the finest refinement level (and assume a \citet{truelove_1997_dens_condition} Jeans refinement scheme). AMR resolution studies may also need to scale the based-level grid resolution to resolve turbulent fragmentation \citep{Haugbolle_Padoan_isot_IMF}, i.e. fixing the number of refinement levels.}.

In the top left panel of Figure \ref{fig:jets_convergence} we plot the evolution of the SFE for the 10 simulations in our resolution study. Star formation tends to start sooner at lower resolution, opposite to what is seen in the collapse of Jeans-mass clumps (Fig. \ref{fig:bossbodenheimer}), possibly owing to increased numerical dissipation of turbulence at low resolution. The star formation history ceases to appear sensitive to resolution below $\approx 0.01\msun$. This corresponds to the resolution criterion we derived in \citetalias{guszejnov_isothermal_mhd}, that the sonic mass $M_{\rm sonic} \approx M_{\rm 0}\mathcal{M}^{-4} \approx 0.3\msun$ be resolved in at least $\approx 30$ gas cells.

We examine the resolution dependence of the stellar mass spectrum at fixed total stellar mass in panels 2-6 of Figure \ref{fig:jets_convergence}: the number of stars $N_{\rm \star}$,  the mass-weighted median stellar mass $M_{\rm 50}$, and the median, mean, and maximum stellar masses. All IMF statistics, whether mass- or number-weighted, eventually cease to change systematically with resolution. As in the isothermal case, the resolution threshold for $M_{\rm 50}$ and $M_{\rm max}$ to stabilize is $\approx 0.01\msun$. Number-weighted statistics require somewhat higher resolution, with $\approx 10^{-3}\msun$ being the marginal value for accurately predicting the mean stellar mass and $4\times 10^{-4}\msun$ for the median. We also plot the effects of resolution upon statistics taken over different stellar mass cuts in Appendix \ref{sec:masscut}, finding that the resolution required depends upon the mass cut (consistent with a simple resolution-dependent low-mass incompleteness effect, with incompletess starting below $< 100 \Delta m$).

By design, there is no purely-numerical dimensional quantity that could imprint a characteristic stellar mass here, so it is likely that the predicted IMF is shaped largely by the physical processes modelled in the simulation, i.e. there may exist a well-defined, physical IMF that emerges from the combined physics of cooling, MHD, gravity, stellar dynamics, and protostellar outflows, and this IMF resembles the observed one. We explore this IMF prediction across a wide parameter space in \citetalias{starforge_jets_imf}. Assuming that other feedback processes do not demand further resolution requirements, this experiment gives some idea of the mass resolution needed to predict e.g. the mean stellar mass in STARFORGE simulations. Note that some incompleteness in the IMF may persist to higher resolution -- to obtain a {\it complete} IMF, we may require a resolution of $\sim 10^{-5}\msun$ to fully resolve the collapse of clumps at the opacity limit, forming the smallest brown dwarfs \citep{bate2003}. But brown dwarfs contain only a small fraction of the total number and mass in stars and are not expected to exert significant feedback. Hence many major questions involving cluster formation, feedback, and the physics underlying the {\it typical} mass of stars can be addressed at much lower resolution. We adopt $10^{-3}\msun$ as our standard resolution for these purposes, but note that the coincidence of our results here with the \citetalias{guszejnov_isothermal_mhd} sonic mass resolution criterion $\Delta m \lesssim 0.03 M_{\rm sonic} \sim 0.03 M_{\rm 0} \mathcal{M}^{-4} \approx 0.01 \msun \left(T/10\rm K\right)^2 \left(\Sigma/100\rm \msun pc^{-2}\right)^{-1}$ suggests that this is the more-general convergence criterion. This resolution criterion can be more demanding for e.g. high-surface density GMCs found in the Galactic centre \citep{2001ApJ...562..348O, Longmore_2012_the_Brick} but not necessarily because such clouds can also be warmer.

We take this opportunity to comment on the computational cost and scaling of these simulations with feedback. Our fiducial resolution ($10^{-3} M_\odot$, $2 \times 10^6$ cells) run cost roughly 10,000 core-hours run on the Frontera supercomputer at the Texas Advanced Computing Center equipped with 2.7 GHz Intel Xeon ``Cascade Lake" processors (56 cores per node), and the simulations in the following section at the same resolution all had comparable cost. The largest simulation shown in Fig. \ref{fig:jets_convergence} (mass resolution $10^{-4}M_\odot$, $2 \times 10^7$ gas cells) required roughly 100,000 core-hours, running for 17 wall-clock days on 4 Frontera nodes. The largest STARFORGE simulation with jet feedback run so far ($M_{\rm GMC}=2\times 10^5 M_\odot$ with $2\times 10^8$ cells, \citetalias{starforge_jets_imf}) required roughly 4.8M core-hours in 70 wall-clock days on the Stampede-2 machine at the Texas Advanced Computing Center with 2.1Ghz Intel Xeon ``Skylake" processors (24 cores per node). Note that these are numbers for simulations without explicit RHD; we anticipate that our full RHD STARFORGE simulations currently in progress will be a factor $\sim 5-10$ more expensive than their non-RHD counterparts, mainly due to their more stringent timestep constraints.

\subsubsection{Effect of physics variations and numerical details}

In Figure \ref{fig:jettests} we explore the effects of 13 other variations on this setup (both numerical and physical) upon the same SFH and IMF statistics, as well as the GMC kinematics. Neglecting jet feedback altogether results in much higher terminal SFE, even if we delete half the accreted mass from the simulation, a simple prescription used by previous works to model jet feedback, deleting either on-the-fly or in post-processing \citep{padoan_2012_sf_law, federrath_klessen_2012, Haugbolle_Padoan_isot_IMF}. Models that do not explicitly treat feedback also seriously underestimate the level of turbulence in the GMC, and predict a much more top-heavy IMF (e.g. greater values of $M_{\rm 50}$). \citet{Haugbolle_Padoan_isot_IMF} found that deleting half the accreted mass gave a good fit to the observed IMF, but simulated a somewhat different GMC setup, and it can easily be reasoned on dimensional grounds that the accretion efficiency one needs to emulate the effect of jets could generally be problem-dependent. Overall, runs with jet feedback behave dramatically different to the baseline run with feedback: both the total stellar mass formed and average individual stellar masses are roughly an order of magnitude smaller, because the momentum content of the jets disrupts both local protostellar accretion flows and eventually the cloud itself (see \citetalias{starforge_jets_imf}).

All results are fairly insensitive to variations in the collimation angle $\theta_{\rm 0}$ from 0.01-1, the jet mass resolution $\Delta m_{\rm w}$ from $0.1-1\Delta m$, and the accretion smoothing timescale $t_{\rm acc}$ from $1-10\times \Delta m c_{\rm s}^3 / G$ (Eq. \ref{eq:tacc}). Results are also insensitive to whether we allow jets from stars $<0.1\msun$, whether we include protostellar heating, and whether we return accreted angular momentum to the surrounding gas as in \citet{hubber:2013.sinks} (which would influence the stellar angular momenta and jet directions in turn). Results only differ significantly for variations that are ruled out observationally and/or unphysical: neglecting magnetic fields (resulting in a much more bottom-heavy IMF, \citealt{guszejnov_isothermal_collapse}), assuming jets are emitted isotropically (reducing the typical stellar masses and increasing the overall SFE), artificially increasing the coupled protostellar heating luminosity by a factor of 10 (which made the IMF noticeably more top-heavy, as in \citealt{krumholz_2012_orion_sims}), and disabling jets for all stars $<1\msun$ (which also made the IMF more top-heavy). 

\subsubsection{Comparison with AMR simulations}
\label{sec:orion2}
To conclude our testing of the jet module, we test its results against a code that differs from ours in all regards except for the physical equations solved and the physical assumptions underlying our feedback models. Our objective here is to verify that the results of simulations with jets are robust to such details.

We have reproduced the setup described in \citet{Cunningham_2018_feedback}, who simulated low-mass cluster formation in a periodic box of side length $0.65 \rm pc$ containing $185 \msun$ with the {\small ORION2} AMR constrained-transport MHD code with radiative transfer and protostellar jets \citep{Cunningham_2011_outflow_sim,li_2012_orion2_mhd}. We re-run their driven turbulence simulation with an initial mass-to-flux ratio $\mu=2.17$, initially driving turbulence at $\mathcal{M}\sim  6.6$ for $\sim 1\rm Myr$ and then switching on gravity. We did not use exactly the same turbulent driving pattern, so our results should only be compared statistically, hence we ran an ensemble of simulations from 3 different initial turbulence realizations. We adopted the same modified prescription for the jet speed as \citet{Cunningham_2018_feedback}, $v_{\rm jet} = \min \left(\sqrt{GM_{\rm \star}/R_{\rm \star}}, 60\mathrm{km\,s}^{-1}\right)$, and the two codes' respective protostellar evolution modules setting $R_{\rm \star}$ both follow \citet{Offner_2009_radiative_sim}. \citet{Cunningham_2018_feedback} account for protostellar radiation with a flux-limited diffusion solver, while we use the inexpensive tree-based approximation described in \S\ref{sec:thermo}. Our simulations adopt a mass resolution of $10^{-3}M_\odot$ and cost roughly 200 core-hours each when run on a single Frontera CLX node. 

\begin{figure}
    \centering
    \includegraphics[width=\columnwidth]{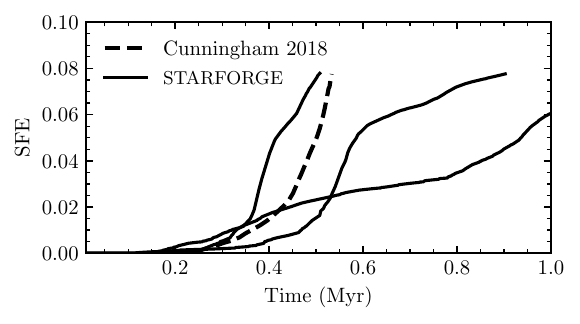}\vspace{-0.2cm}
    \includegraphics[width=\columnwidth]{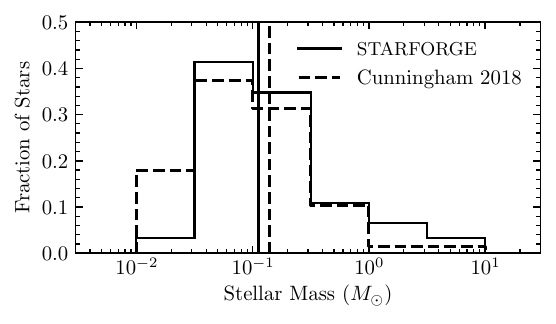}\vspace{-0.8cm}
    \caption{Comparison between the results of the driven $\mu=2.17$ simulation in \citet{Cunningham_2018_feedback} with MHD and protostellar heating and jets, and 3 different STARFORGE replications at standard $10^{-3}M_\odot$ resolution, using our simple approximate RT treatment of dust heating and our protostellar jet module (modified so that $v_{\rm jet} = \min \left(\sqrt{GM_{\rm \star}/R_{\rm \star}}, 60\mathrm{km\,s}^{-1}\right)$, as in \citealt{Cunningham_2018_feedback}). {\it Top:} Star formation efficiency versus time. {\it Bottom:} Stellar mass function at the final SFE of $\sim 8\%$, comparing the stacked mass function from the 3 STARFORGE runs with \citet{Cunningham_2018_feedback}, showing the respective median stellar masses as vertical lines.}
    \label{fig:cunningham}
\end{figure}

We plot the resulting star formation history and IMF at $8\%$ SFE in Figure \ref{fig:cunningham}. The specific shape of the SF history does appear to be a function of the initial turbulent driving, but we had two seeds that matched that of \citet{Cunningham_2018_feedback} for an appreciable fraction of the SF history, both modulo a small time difference due to the different intial turbulent states. Therefore predictions regarding the regulation of SF due to protostellar feedback appear similar for the two codes.

The final IMFs are also in fair agreement: the median stellar mass predicted by both codes is $\sim 0.15 \msun$ (shown as vertical lines). The only statistically-significant difference is between the predicted numbers of $0.01-0.03\msun$ brown dwarfs, with STARFORGE runs finding only $\sim 1/4$ as many. This may be due any number of details in the cooling and RT modules that differ (with more higher temperatures or more radiative heating suppressing brown dwarfs, \citealt{bate_2009_rad_importance,Offner_2009_radiative_sim}), or a mere resolution effect (as we expect some numerical IMF incompleteness at masses $\lesssim 30 \Delta m$).

In summary, the respective implementations of STARFORGE and {\small ORION2} find very similar results for the regulation of SF and the stellar mass range of the IMF in low-mass star cluster formation, despite these two codes' detailed numerical implementations differing in every regard. We scale a setup similar to this to GMCs as much as $>1000\times$ more massive in \citetalias{starforge_jets_imf}, exploring the broader implications of protostellar feedback in massive GMCs.

\subsection{Stellar Winds}
\label{sec:winds}
\subsubsection{Physics prescription}
\begin{figure}
    \centering
    \includegraphics[width=\columnwidth]{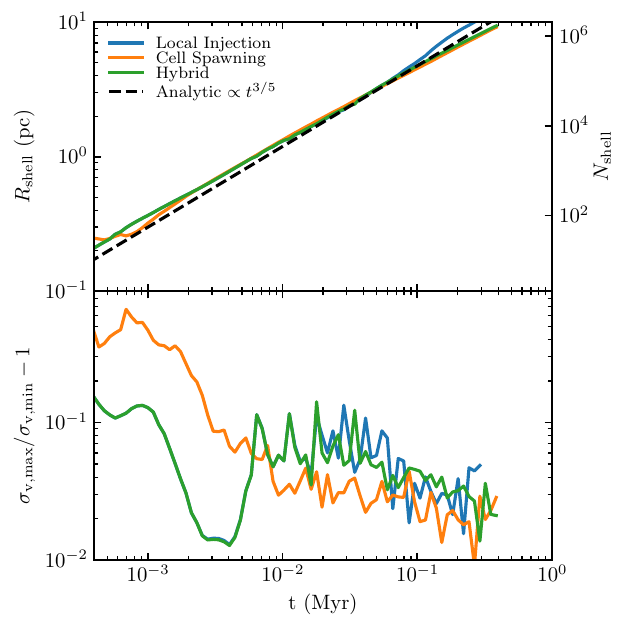}\vspace{-0.8cm}
    \caption{Self-similar expansion of a stellar wind bubble with an adiabatic interior and radiative outer shell, as simulated with our stellar wind module (\S\ref{sec:winds}) for a star with $\dot{M}_{\rm wind}=10^{-5}\msun$ and $v_{\rm wind}=3000\rm km s^{-1}$ in a $16\rm pc$ box containing $2\times 10^4 \msun$, using different numerical methods (local injection (\S\ref{sec:injection}), cell spawning (\S\ref{sec:spawning}), and a hybrid method that switches between them adaptively (\S\ref{sec:winds}). {\it Top:} Comparison of the evolution of the radius of the swept-up shell $R_{\rm shell}$ (corresponding to the number of swept-up gas cells $N_{\rm shell}$) to the known similarity solution, $R_{\rm shell} = 0.763\left(\frac{L_{\rm wind}}{\rho}\right)^{1/5} t^{3/5}$ \citep{weaver_1977_winds}. {\it Bottom}: Numerical velocity anistropy of the bubble (=0 in the similarity solution), generally $\lesssim 10\%$ except in very early phases when the bubble is not well-resolved (contains just a few cells). The switch between different coupling methods at $0.02\rm Myr$ is apparent when the ``Hybrid" curve deviates from the ``Local Injection" curve.}
    \label{fig:winds}
\end{figure}
    

We allow main-sequence stars more massive than $2 M_\odot$ to inject stellar winds. Stellar mass loss rates are subject to considerable theoretical and observational uncertainties, with various unresolved discrepancies between theory and observations \citep{smith_2014_winds}, so we default to a simple phenomenological prescription. Wind-emitting stars feed their wind reservoir from the stellar mass at a base rate of
\begin{equation}
   \frac{\dot{M}_{\rm wind}}{M_\odot\rm yr^{-1}} = \min \left(10^{-15} L_{\rm MS}^{1.5}, 10^{-22.2} L_{\rm MS}^{2.9} \right)  Z_{\rm \star}^{0.7},
   \label{eq:mdotwind}
\end{equation}
where the main-sequence luminosity $L_{\rm MS}$ and the metallicity $Z_{\rm \star}$ are in solar units. This is a fit to the the envelope of the ``de Jager / 3" and ``weak wind problem" scalings given in \citet{smith_2014_winds}, hence it is a conservative model accounting for the widely-acknowledged overestimation of $\dot{M}$ by theoretical line-driven stellar wind models (i.e. it is generally weaker than widely-used models such as \citet{vink_2001_winds}). The velocity of the winds is
\begin{equation}
    v_{\rm wind} = \sqrt{\frac{2 G M_{\rm \star}}{R_{\rm \star}}} \times \begin{cases} 
0.7 &  T_{\rm eff} < 12500\rm K \\

  1.3  & 12500\rm K  < T_{\rm eff} < 21000 \rm K \\ 
  2.6  & T_{\rm eff} \geq 21000K \\
   \end{cases},
   \label{eq:vwind}
\end{equation}
following \citet{lamers_1995_wind_vesc}.

Much of the energy and momentum in stellar winds from a stellar population originates in Wolf-Rayet stars. We use a simple model for the Wolf-Rayet phase for $>20\,\msun$ stars, multiplying $\dot{M}_{\rm wind}$ by a factor of 10 at the end of its lifetime. The time spent in the Wolf-Rayet phase is given by
\begin{equation}
    t_{\rm WR} = {1.5 \rm Myr} \min \left(1,\frac{M_{\rm \star}/\msun-20}{80}\right) \left(\frac{Z_{\rm \star}}{Z_\mathrm{\odot}}\right)^{0.5},
\end{equation}
an approximate fit to results from \citet{meynet_2005_wolfrayet}.
\subsubsection{Numerical methods}
Our numerical method for coupling winds uses either local injection or cell spawning, where appropriate. In the regime where the wind's free-expansion radius $R_{\rm free} = \sqrt{\dot{M}_{\rm wind} / v_{\rm wind}\rho}$ is much less than the size of a wind cell $\Delta x_{\rm w} = \left(\Delta m_{\rm w}/\rho\right)^{1/3}$, a spawned cell will generally stop within a single cell length, collide with the ISM, and thermalize its kinetic energy, so it is more efficient and accurate to instantaneously inject that mass, momentum, and energy isotropically into the neighbouring cells. But when the free-expansion radius is well resolved, the simulations resolve the travel time of the wind before it merges with the ISM, so cell spawning is more appropriate. Hence we switch between the two modules adaptively, based on whether $R_{\rm free}$ is resolved by at least 1 wind cell length. As with jets we spawn cells 2 at a time, but with an isotropic angular distribution instead of collimated.
\subsubsection{Tests}
We test this module on the problem of the self-similar expansion of a wind bubble propagating into a uniform medium with an adiabatic interior and radiative exterior with negligible exterior pressure, which has the analytic solution $R_{\rm shell} = 0.763\left(\frac{L_{\rm wind}}{\rho}\right)^{1/5} t^{3/5}$ \citep{weaver_1977_winds}. We place a star with $\dot{M}_{\rm wind} = 10^{-5} M_\odot$ and $v_{\rm wind}=3000 \rm km s^{-1}$ in a $16\rm pc$ box containing $2\times 10^4 \msun$, with $\Delta m = 0.01\msun$ and $\Delta m_{\rm w}=10^{-3} \msun$, with initial temperature $10\rm K$. In Figure \ref{fig:winds} we plot the bubble expansion and find that agreement with the similarity solution is good whether we use pure local injection, pure cell spawning, or our hybrid method \footnote{We have also found negligible differences between the different wind methods in full star cluster formation simulations including winds as the only feedback mechanism.}. We also examine the velocity anisotropy $\sigma_{\rm v,max}/\sigma_{\rm v,min}-1$ where $\sigma_{\rm v,max}$ and $\sigma_{\rm v,min}$ are the maximum and minimum gas velocity dispersions along the principal axes of the gas momentum distribution, which is 0 in the exact solution. This is typically $\lesssim 10\%$, except in the very early phase of the run with pure cell spawning (because the free expansion radius is not yet well-resolved, so shot noise from individual injection steps is still apparent). For the hybrid method, the transition between methods occurs smoothly, with no clear spurious numerical artifacts.

\subsection{Supernovae}
\label{sec:SNe}
\subsubsection{Physics prescription}
\begin{figure}
    \centering
    \includegraphics[width=\columnwidth]{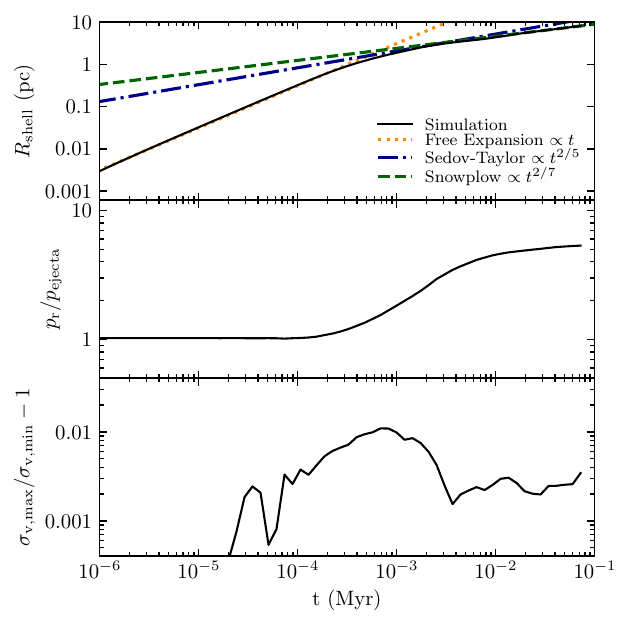}\vspace{-0.8cm}
    \caption{Evolution of the SN remnant from a $10 \msun$ progenitor in a $16 \rm pc$ box containing $2\times10^4 \msun$ ($n\sim 200 \rm cm^{-3}$, $Z =Z_\odot$), with cooling physics enabled (\S\ref{sec:thermo}) and with the SN modeled via cell-spawning as described in \ref{sec:SNe},  SNR evolution starting at the free-expansion phase. {\it Top:} Radius of the swept-up shell of dense gas as a function of time, which interpolates between the initial free-expansion ($\propto t$), Sedov-Taylor ($\propto t^{2/5}$), and pressure-driven snowplow phases ($\propto t^{2/7}$) as shown. {\it Middle:} radial momentum in units of the initial $p_{\rm ejecta} = M_{\rm ejecta} v_{\rm ejecta}$, initially conserved in the free-expansion phase but boosted by a factor of $\sim 6$ due to PdV work performed in the Sedov-Taylor phase. {\it Bottom}: Numerical velocity anisotropy, which is generally $\lesssim 1\%$.}
    \label{fig:sn}
\end{figure}
\begin{figure}
    \centering
    \includegraphics[width=\columnwidth]{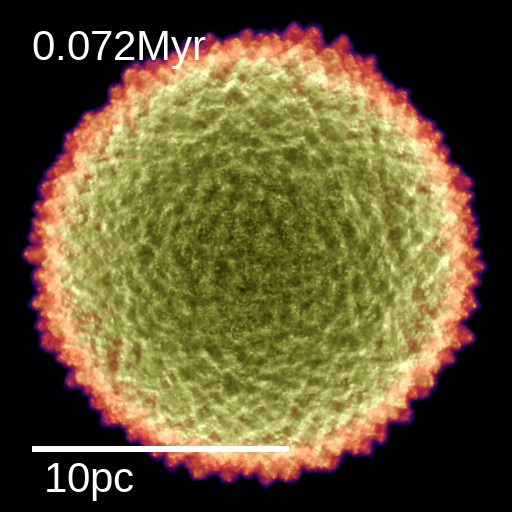}
    \caption{Morphology of the supernova remnant at the end of the supernova test in \S\ref{sec:SNe}, when the supernova remnant has entered the pressure-driven snowplow phase.}
    \label{fig:snr}
\end{figure}
We assume that all stars more massive than $8\msun$ go supernova at the end of their lifetime, with the lifetime given by Equation \ref{eq:stellarlifetime} (from $\approx 40\rm Myr$ for $8\msun$ to $\approx 3\rm Myr$ at $100\msun$). When flagged as a supernova, the star ceases all other forms of feedback, and rapidly expels its mass isotropically with velocity
\begin{equation}
    v_{\rm SN} = \sqrt{\frac{2 E_{\rm SN}}{M_{\rm ejecta}}} = 3200 \mathrm{km\,s^{-1}} \left(\frac{E_\mathrm{SN}}{10^{51} \rm erg}\right)^{1/2} \left(\frac{M_{\rm ejecta}}{10 \msun}\right)^{-1/2} ,
    \label{eq:vSN}
\end{equation}
where we assume $E_\mathrm{SN}=10^{51}\rm erg$ by default. We assume the entire star is destroyed, but we can in principle allow for finite-mass relic compact objects by reserving a certain final mass. We assume that the SN ejecta have IMF-averaged yields according to \citet{nomoto2006:sne.yields}, with mass fractions (He, C, N, O, Ne, Mg, Si, S, Ca, Fe) = (3.87, 0.133, 0.0479\,{MAX}[$Z/Z_{\odot},\,1.65$], 1.17, 0.30, 0.0987, 0.0933, 0.0397, 0.00458, 0.0741).

\subsubsection{Numerical methods}

SNe are realized numerically by the same cell spawning strategy as winds, except that 1) the spawned cells have the standard mass resolution $\Delta m_{\rm w} = \Delta m$ and 2) cells are spawned in shells of $N_{\rm spawn}=24$ cells at once until the progenitor mass is exhausted. Mass is transferred from the star to the wind reservoir at a rate of
\begin{equation}
    \dot{M}_{\mathrm{SN}} = \frac{N_{\rm spawn} v_{\rm SN} \Delta m}{R_{\rm sink}},
    \label{eq:mdot_sn}
\end{equation}
 which in our default simulations is $\approx 1M_{\rm\odot} \rm yr^{-1}$ . Hence a typical progenitor will actually take several years to eject all its mass, but this does not affect the solution on scales $\gtrsim 0.01 \rm pc$, where we actually resolve the dynamics. We impose this finite duration because a very massive star could, in a single timestep, spawn $\sim 10^5$ new cells, which would make operations like load-balancing and controlling $\nabla \cdot \mathbf{B}$ computationally challenging. Gas cells within a given shell are arranged in a regular angular grid pattern following \citet{bruls1999_angular_grid} to avoid pathological cell arrangements and ensure statistical isotropy, and the orientation of the grid is randomized between each shell to reduce grid alignment effects.
\subsubsection{Tests}
In Figure \ref{fig:sn} we test this algorithm by detonating a $10\msun$ progenitor star in the manner described here. The star is initially placed in a $16\pc$ box containing $2\times 10^4\msun$ in gas (with $0.01\msun$ mass resolution in the box and the ejecta), and we follow the evolution of the remnant from the free expansion through Sedov-Taylor through snowplow phases. The radius of the swept-up shell matches the expected similarity solutions in the different phases, and the terminal momentum boost factor from $P\mathrm{d}V$ work in the Sedov-Taylor phase is $\sim 6$, consistent with more-detailed SN simualtion studies \citep{martizzi_2015_sne, walch_naab_2015_sne, gentry_2016_snr,haid_2016_sne, Hopkins_2018_sne_feedback} given the ambient density $n\sim 200 \rm cm^{-3}$ and metallicity ($Z =Z_\odot$). Numerical momentum anisotropy is always small, peaking at $\sim 1\%$ during the Sedov-Taylor phase. We visualize the final morphology of the supernova remnant in Figure \ref{fig:snr}.

\subsection{Radiation}
\label{sec:radiation}
Following \citet{FIRE_RT}, STARFORGE simulations with the radiative transfer module enabled follow the emission, transport, and absorption of photons in 5 different bands in wavelength $\lambda$:
\begin{enumerate}
\item{{\bf Hydrogen ionizing} ($\lambda<912\,$\AA): Ionizing photons emitted by stars and responsible for the dynamics of HII regions, which are widely theorized to be the most important feedback effect from massive stars in typical Galactic conditions on global GMC scales \citep{mckee_1984_photoionization, dale_ion_feedback, krumholz_matzner_2009_hii_rp,geen_2017_sfe, kim_2018_gmc_raytrace, grudic_2018_mwg_gmc, olivier_2020_hii}.}
\item{{\bf Far-UV/photoelectric} ($912\,$\AA$<\lambda<1550\,$\AA): Responsible for heating the ISM via the photoelectric effect on dust grains, and likely an important component of the thermal balance of the cold and warm neutral media in the outer parts of galaxies \citep{wolfire_1995_fuv, ostriker_2010_fuv}.}
\item{{\bf Near-UV} ($1550$\AA$\,<\lambda<3600\,$\AA): Contains most of the photon energy and momentum emitted by a young stellar population, and hence is the most important term for direct stellar radiation pressure \citep{fall2010,Murray_2010_GMC_disruption,raskutti_2016,kim_2018_gmc_raytrace,hopkins_grudic_2019}.}
\item{{\bf Optical/near-IR} ($3600\,$\AA$\,<\lambda < 3\,\mu \rm m$): Contains most of the light from old stellar populations, and carries a non-negligible fraction of photon momentum that can potentially couple on larger scales in a GMC due to reduced dust opacity compared to NUV.}
\item{{\bf Mid/far-IR} (mainly $\lambda > 3\,\mu \rm m$): Radiation absorbed and re-radiated by dust, which is the primary cooling mechanism in the densest gas and can dominate the radiation pressure near massive protostars or in ULIRGs \citep{krumholz_2009_massive_sf,kuiper_2011_massive_sf,davis_2014_ir, rosen_2016_massive_sf, tsang_2018_ir}. This is treated specially, as a component of the radiation field having a black body SED with a local effective temperature $T_{\rm rad}$, which is evolved self-consistently. }
\end{enumerate}

We can optionally further ``fine grain'' these bands into narrower bins: for example, splitting ionizing and FUV photons into photo-electric, Lyman-Werner, H ionizing, He-ionizing, He-secondary-ionizing, soft X-ray, etc, as described in \citet{FIRE_RT}, but this generally produces second-order effects.  
Note that, unlike most RHD SF simulations, we {\em independently} and explicitly evolve the dust temperature $T_{\rm dust}$, radiation temperature $T_{\rm rad}$ (of the IR band)\footnote{Note that the IR radiation temperature $T_{\rm rad,\,IR}$ parameterizes the spectral shape (or equivalently wavelength of the IR SED peak or mean energy per photon). We therefore allow the IR radiation energy density $u_{\rm \gamma,\,IR}$ and spectral shape or $T_{\rm rad,\,IR}$ to evolve independently, rather than imposing the blackbody assumption $u_{\rm gamma,\,IR}\propto T_{\rm rad,\,IR}^{4}$ (which is only valid in the infinite optical-depth, tight-coupling limit).}, and gas temperature $T_{\rm gas}$.

All sinks/stars are treated as potential sources for all bands above. In our default simulations, we calculate the emitted flux in each band by treating each sink/star as a blackbody with effective temperature $T_{\rm eff,\star}\approx 5780{\rm K}\,(L_{\star}/L_{\odot})^{1/4}\,(R_{\star}/R_{\odot})^{-1/2}$ with $L_{\star}$ and $R_{\star}$ given by our stellar evolution module (\S\ref{sec:stellarevol}), integrated over the relevant wavelengths. We ignore ``primary'' gas emission at other wavelengths, as this is generally negligible in the problems of interest. Secondary gas/dust (re)-emission is treated as follows: recombination emission from absorbed ionizing photons are re-emitted into the optical/near-IR (OIR) band; the absorbed radiation energy in other bands (where the opacity is dust-dominated) is re-emitted by dust in the IR band with the evolved dust temperature $T_{\rm dust}$.

The limitation of this band-integrated treatment of radiation is that line emission and absorption are not followed explicitly. As such, it does not capture molecular line cooling explicitly (which we treat using approximate formulae, see \S\ref{sec:thermo}). It is also not applicable to dust-free conditions where multiply-scattered Ly-$\alpha$ photons are dynamically important \citep[e.g.][]{2017MNRAS.464.2963S}.

Absorption and scattering cross-sections/opacities and coupling to our thermo-chemistry (gas heating/cooling) routines largely follow \citet{FIRE_RT}. For ionizing bands, this employs standard photo-ionizing absorption cross sections which scale with the neutral H and neutral or partially-ionized fractions for He, and absorbed ionizing photons directly couple to our detailed photo-ionization heating rates (see \citealt{hopkins2017_fire2}; note these are always calculated, our RHD simulations simply include local sources in addition to the UVB and ISRF). The opacities in FUV, NUV, OIR are given by the grey expressions: $(\kappa_{\rm FUV},\,\kappa_{\rm NUV},\,\kappa_{\rm OIR}) = (0.2+2000\,f_{d}^{\prime},\,1800\,f_{d}^{\prime},\,180\,f_{d}^{\prime})\,{\rm cm^{2}\,g^{-1}}$ where $f_{d}^{\prime}=f_{d}/0.01$ is the local dust-to-gas ratio relative to solar, defined as in \S~\ref{sec:thermo}. The FUV intensity directly enters the photo-electric heating rate in our thermochemistry calculation (\citealt{hopkins2017_fire2}; Appendix~B); absorbed radiation in FUV+NUV+OIR+IR bands contributes to determine the dust temperature $T_{\rm dust}$ which interacts via dust-gas collisions (\S\ref{sec:thermo}). For the IR band, we calculate the opacity as a function of ionized fraction, local dust-to-gas ratio, dust temperature, and radiation temperature, as $\kappa_{\rm IR}=\kappa_{\rm gas}+\kappa^{0}_{\rm dust}\,f_{d}^{\prime}$ where $\kappa_{\rm gas}\approx 0.35\,x_{e}\,{\rm cm^{2}\,g^{-1}}$ from Thompson scattering with $x_{e}$ the free electron fraction, and $\kappa^{0}_{\rm dust}(T_{\rm dust},\,T_{\rm rad})$ calculated from the tables of \citet{semenov_2003}. Specifically we take the `standard' model with the `porous 5-layed sphere' composition in \citet{semenov_2003} and for each {\em dust} temperature $T_{\rm dust}$ (which gives a different dust composition) we explicitly calculate the Rosseland-mean opacity for each radiation temperature $T_{\rm rad}$ assuming a blackbody-like spectral shape. We provide detailed fits in Appendix \ref{appendix:dustopacity}. We assume a sublimation temperature of $T_{\rm dust}^{\rm sub}=1500\,$K. Finally because our RHD methods account for both absorption and scattering we must define the albedo $A$. For simplicity we assume $A_{\rm ion}=0$ (pure absorption) for ionizing bands; $A_{\rm FUV,\,NUV,\,OIR}=1/2$, i.e.\ equal absorption and scattering opacities which is roughly appropriate for dust grains in FUV through OIR bands \citep[see e.g.][]{draine.weingartner.2001}; and for IR we assume the Thompson portion of the opacity is pure-scattering while for the dust albedo we can interpolate reasonably accurately between the short-wavelength ($A\sim1/2$) and long-wavelength (Rayleigh scattering, $A\rightarrow 1$) regimes by taking $A_{\rm IR} \approx (\tilde{T}_{r}/2 + 1)/(\tilde{T}_{r} + 1)$ with $\tilde{T}_{r} \equiv (T_{\rm rad,\,IR}/725\,{\rm K})^{2}$.

Photon momentum (radiation pressure) is always transferred appropriately to the gas+dust when radiation is absorbed or scattered (from any band).

\subsubsection{Photon injection}
Photons from sinks must be injected into the simulation domain before they are propagated by the RT solver. We do this via local injection: constructing effective oriented faces ${\bf A}_{sg}$ between the sink particle and overlapping gas cells, and injecting photons conservatively with a weighting given by the solid angle subtended by the face (e.g. Fig. \ref{fig:localinjection}). A full description of the original algorithm for photons is given in \citet{hopkins_grudic_2019} Appendix A, but we make a small extension here. 


In the original algorithm, an extinction factor $f_{\rm abs}=\exp\left(-r_{sg}/\lambda_{\rm mfp}\right)$ (for photon mean free path $\lambda_{\rm mfp}=\left(\kappa \rho\right)^{-1}$) was applied to the injected photon energy and momentum of a cell, and the appropriate absorbed photon momentum was imparted. This models sub-resolution extinction, which is crucial for capturing radiation pressure effects when $\lambda_{\rm mfp}$ is unresolved -- which is very often the case in SF problems at practical resolutions \citep{krumholz_2018_rad_pressure,hopkins_grudic_2019}. Here we also take the photon {\it energy} absorbed on unresolved scales and ``downgrade" (re-emit) it to the appropriate band as defined above. The downgraded photons are then injected into their respective bands in addition to the photons originally in that band in the stellar SED. Hence, in practice a star in a highly optically-thick accretion flow will usually end up injecting most of its luminosity to the mid/far IR band, because $\lambda_{\rm mfp}$ is not resolved.


\subsubsection{Photon transport}
{\small GIZMO} employs modular RHD solvers, so in principle we can adopt and compare various methods for photon transport. But in our default explicit-RHD simulations we adopt the first-moment or M1 \citep{levermore_1984_M1} method, which has the advantages of being computationally efficient (well-adapted to hierarchical timestepping and multi-physics simulations), manifestly momentum and energy conserving in finite-volume form, able interpolate between optically-thick and optically-thin limits, and well-tested in simulations of star cluster formation \citep{geen_2015_rt, geen_2017_sfe, Gavagnin_2017_SF_feedback, he_2019_gmc_fb}. In particular, for questions involving radiation pressure forces on gas, shadowing in an inhomogeneous medium, and the transition between optically thin-thick regimes, it is (by construction) able to capture phenomena which cannot appear in the 0th-order flux-limited-diffusion (FLD) method \citep[see references above and e.g.][]{davis_2014_ir,Rosdahl_2015_M1_galaxy,zhang:2017.rhd.dusty.winds,kannan:2018.arepo.rhd}. For each band $i$ we explicitly evolve the first two moments of the intensity equation in the usual mixed-frame approximation keeping all terms to $\mathcal{O}(v^{2}/c^{2})$, with all terms appropriately integrated over the relevant bands \citep{mihalas:1984oup..book.....M,lowrie:1999.radiation.hydro.coupling}. This gives \begin{align}
\label{eqn:rad.egy} \frac{\partial e_{r}^{i}}{\partial t} + \nabla \cdot {\bf f_{r}^{i}} &= \left( \dot{e}_{\rm em}^{i} - \dot{e}_{\rm abs}^{i} \right) + (\psi_{a}^{i} - \psi_{s}^{i})\,{\bf u} \cdot {\bf g}^{i}_{r} \\
\label{eqn:flux} \frac{1}{\tilde{c}^{2}}\frac{\partial {\bf f}^{i}_{r}}{\partial t} + \nabla \cdot \mathbb{P}^{i}_{r} &= - (\psi_{a}^{i}+\psi_{s}^{i})\,{\bf g}^{i}_{r} + \frac{\bf u}{{c}^{2}}\,\left( \dot{e}^{i}_{\rm em} - \dot{e}^{i}_{\rm abs}  \right)
\end{align}
where ${\bf u}$ is the local gas/dust velocity, $\dot{e}_{\rm abs}\equiv\tilde{c}^{2}\,\psi_{a}\,e_{r}$ and $\dot{e}_{\rm em}$ are the volumetric absorption and emission rates, ${\bf g}_{r} \equiv {\bf f}_{r} - {\bf u}\cdot(e_{r}\,\mathbb{I} + \mathbb{P}_{r})$ with $e_{r}$ and ${\bf f}_{r}$ the radiation energy and flux densities, $\psi_{a,\,s}\equiv \rho\,\kappa_{a,\,s}/\tilde{c}$ are the absorption+scattering coefficients, $\tilde{c}$ and $c$ the reduced (RSOL) and true speed of light, and $\mathbb{P}_{r}\equiv e_{r}\,\mathbb{D}_{r}$ the radiation pressure tensor (Eddington tensor $\mathbb{D}_{r}$).\footnote{We adopt the common M1 closure for $\mathbb{P}_{r}$, taking $\mathbb{D}_{r} \rightarrow (1/2)\,[(1-\chi)\,\mathbb{I} + (3\,\chi-1)\,\hat{\bf f}_{r}\,\hat{\bf f}_{r}]$, $\chi \equiv (3+4\xi^{2})/(5+2\,[4-3\,\xi^{2}]^{1/2})$ and $\xi\equiv |{\bf f}_{r}|/(\tilde{c}\,e_{r})$, which interpolates between thin and thick regimes \citep{levermore_1984_M1}.}
These are discretized and inter-cell fluxes are integrated in the same finite-volume (conservative) form as the MHD equations \citep{FIRE_RT}.

Note that Eqs.~\ref{eqn:rad.egy}-\ref{eqn:flux} include terms up to formal $\mathcal{O}(v^{2}/c^{2})$ such as the ${\bf u}\cdot(e_{r}\,\mathbb{I}+\mathbb{P}_{r})$ term differentiating ${\bf g}_{r}$ and ${\bf f}_{r}$ or the $(\psi_{a}-\psi_{s})\,{\bf u}\cdot g_{r}$ term representing the work done by radiation pressure which are often dropped in SF simulations where typical speeds $v\ll c$. However, as many authors have pointed out, these terms actually dominate the behavior in the infinite-optical-depth, tight-coupling or photon-trapped limit, where their actual order scales closer as $\mathcal{O}(v/v_{\rm rad,\,diff})$ (with $v_{\rm rad,\,diff}$ the effective bulk speed of photon diffusion). Without this terms, the RHD equations in the trapped limit will give unphysical solutions (photons will not properly be advected and arbitrarily large radiation-pressure forces can arise). The terms in $(\dot{e}^{i}_{\rm abs}-\dot{e}^{i}_{\rm em})\,{\bf u}/c^{2}$, on the other hand, represent true relativistic beaming, and are negligible for our problems of interest.

In the gas/dust momentum equation we add the terms $\sum_{i}\,(\psi^{i}_{a}+\psi^{i}_{s})\,{\bf g}^{i}_{r} + (\dot{e}^{i}_{\rm abs}-\dot{e}^{i}_{\rm em})\,{\bf u}/c^{2}$, the  former representing the normal radiation pressure acceleration and the latter accounting for beaming.
In the gas/dust total energy equation we add the terms $\sum_{i}\,(\dot{e}^{i}_{\rm abs}-\dot{e}^{i}_{\rm em}) + (\psi_{s}^{i}-\psi^{i}_{a})\,{\bf u}\cdot{\bf g}^{i}_{r}$, the former representing the energy of absorption+emission (handled with our thermochemistry as described above) and the latter work terms. The gas temperature is then evolved according to the thermodynamics modules in \S~\ref{sec:thermo}, coupled also to the dust temperature by way of dust-gas collisions (Eq.~\ref{eq:lambdadust}). The dust temperature is set assuming grain absorption-emission equilibrium, giving $T_{\rm dust}^{4} = (\langle Q_{\rm abs} \rangle/\langle Q_{\rm em} \rangle)\,e^{\rm tot}_{r}\,c/(4\,\sigma_{T})$ where $e^{\rm tot}_{r}=\sum_{i} e_{r}^{i}$, and $Q_{\rm abs,\,em}$ are the appropriate absorption and emission efficiencies\footnote{For IR-IR band interactions, we assume $Q_{\rm abs}\approx Q_{\rm em}$, though this could lead to small differences in $T_{\rm dust}$.}.

As noted above, the IR radiation field is treated as a blackbody shape with local effective temperature $T_{\rm rad,\,IR}$ and total energy integrated over the cell domain $E_{\rm gamma,\,IR}$, which evolves as new photons are emitted or when radiation is exchanged between cells of different $T_{\rm rad}$. In emission, sinks emit with $T_{\rm rad,\,IR}^{\rm em}=T_{\rm eff}$ and dust has $T_{\rm rad,\,IR}^{\rm em}=T_{\rm dust}$; given a total emitted radiation energy $\Delta E_{\rm em}$ in the cell timestep, the effective $T_{\rm rad,\,IR}$ is then updated to guarantee {\em both} radiation energy and photon number conservation, giving $T_{\rm rad,\,IR}(t+\Delta t) = [E_{\rm \gamma,\,IR}(t) + \Delta E_{\rm em}] / [E_{\rm \gamma,\,IR}(t)/T_{\rm rad,\,IR}(t) + \Delta E_{\rm em}/T_{\rm rad,\,IR}^{\rm em}]$. The same update scheme is used when cells exchange radiation energy.

Finally, we follow common practice and adopt a reduced speed of light (RSOL) with $\tilde{c}<c$ to enable larger timesteps. In general, $\tilde{c}$ should be larger than the bulk speeds of radiative diffusion or ionizing front expansion to capture the dynamics; \citet{geen_2015_rt} argue this is satisfied for $\tilde{c} \gtrsim 30\,{\rm km\,s^{-1}}$ in our problems of interest, and we verify this below.

\subsubsection{Tests}
First, we remind the reader of the test in \S\ref{sec:thermo}, which demonstrates the accuracy of our IR RHD+thermochemistry models evolving $T_{\rm dust}$, $T_{\rm rad}$, and $T_{\rm gas}$ appropriately in the collapse of a Jeans-unstable core. To test other aspects of our RHD models, we next repeat the test setup of an O star in a box described in the previous sections (\S\ref{sec:winds}-\ref{sec:SNe}) for radiation. First, we examine the ability of the photon injection, transport, and absorption schemes to capture radiation pressure in two regimes: where the photon mean free path is well-resolved ($\Delta x << \lambda_{\rm mfp}$) and totally unresolved ($\Delta x > \lambda_{\rm mfp}$). We inject radiation in a single band with opacity scaled so that the cell opacity $\tau_{\rm cell}=\Delta x/\lambda_{\rm mfp}$ ranges from 0.015 to 1.5, and the global optical depth $\tau_{\rm box} = L_{\rm box}/\lambda_{\rm mfp}$ ranges from 1.9 to 190. Because the box is always optically-thick and thermal pressure is negligible, in all cases we expect the expanding shell solution to approach the momentum-conserving similarity solution $R_{\rm shell} = \left(\frac{L_{\rm \star}}{6\uppi \rho_{\rm 0} c}\right)^{1/4}t^{1/2}$  at late times with radial momentum approaching the total emitted photon momentum $L_{\rm \star} t/c$, and the solution should be spherically symmetric.

\begin{figure}
    \centering
    \includegraphics[width=\columnwidth]{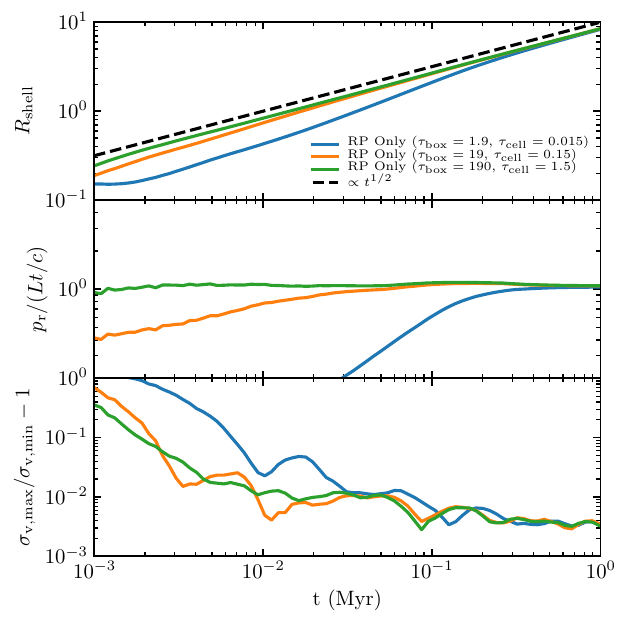}\vspace{-0.8cm}
    \caption{Radiation pressure-only test with an O star in a homogeneous box, accounting only for radiation pressure feedback with a range of opacities, varying the global and cell optical depths $\tau_{\rm box}$ and $\tau_{\rm cell}$ and injecting and transporting photons as described in \S\ref{sec:radiation}. {\it Top}: position of the spherical shell swept up by the radiatively-driven bubble. All solutions approach the analytic similarity solution for momentum-driven bubbles $\propto t^{1/2}$, with the time lag determined by the absorption timescale as expected. {\it Middle}: Radial gas momentum in units of the total emitted photon momentum $Lt/c$. The correct momentum is always coupled even when $\tau_{\rm cell}>1$, by accounting for unresolved local extinction in the injection phase \citep{hopkins_grudic_2019}. {\it Bottom}: Numerical velocity anisotropy, which is $<1\%$ in all but the earliest (i.e. worst-resolved, few-cell) phases of the expansion.}
    \label{fig:rt_momentum}
\end{figure}

Figure \ref{fig:rt_momentum} shows that the dynamics of a radiation pressure-driven shell are captured accurately: all three solutions eventually approach the similarity solution (with the least optically-thick run having a time delay $\lambda_{\rm mfp}/\tilde{c} \approx 0.3\rm Myr$, owing to the finite light travel time before absorption). The correct radial momentum is imparted whether or not $\lambda_{\rm mfp}$ is well-resolved (due to the face-integrated injection method, detailed in \citealt{hopkins_grudic_2019}), and numerical anisotropy falls rapidly below $\sim 1\%$ once the bubble becomes well-resolved.

We repeat this experiment with our ionizing radiation band enabled, without radiation pressure, to test the ability of the code to follow the dynamics of expanding HII regions (and also to determine an appropriate value for $\tilde{c}$ to accomplish this). We compare with the approximate \citet{hosokawa_2006_hii} solution for the position of the ionization front:
\begin{equation}
    R_{\rm HII} = R_{\rm St} \left(1 + \frac{7}{4}\sqrt{\frac{4}{3}}\frac{c_{\rm i}t}{R_{\rm St}}\right)^{4/7},
    \label{eq:hii}
\end{equation}
where $c_{\rm i} \approx 11 \rm km\,s^{-1}$ is the isothermal sound speed in ionized gas and the $R_{\rm St}$ is the Str\"{o}mgren radius:
\begin{equation}
    R_{\rm St} = \left(\frac{3\dot{N}_{\rm LyC} m_{\rm p}^2}{ \alpha_{\rm B} \rho_{\rm 0}^2}\right)^{1/3},
\end{equation}
where $\dot{N}_{\rm LyC}$ is the emission rate of H ionizing photons and $\alpha_{\rm B}$ is the case-B HII recombination coefficient (invoking the ``on-the-spot" approximation, \citealt{osterbrock_1989_onthespot}). We fix the ratio $\dot{N}_{\rm LyC}/\alpha_{\rm B}$ so that $R_{\rm St}$ is initially resolved in only 2 cell lengths, $R_{\rm St} = 2 \left(\Delta m/\rho_{\rm 0}\right)^{1/3}$, and survey $\tilde{c}$ values between $5\rm km\,s^{-1}$ and $300\rm km\,s^{-1}$. We also run a version with $\tilde{c}=30\rm km\,s^{-1}$ with the gas fixed in place, to compare with the static Str\"{o}mgren sphere solution and isolate artifacts of the RSOL approximation.

\begin{figure}
    \centering
    \includegraphics[width=\columnwidth]{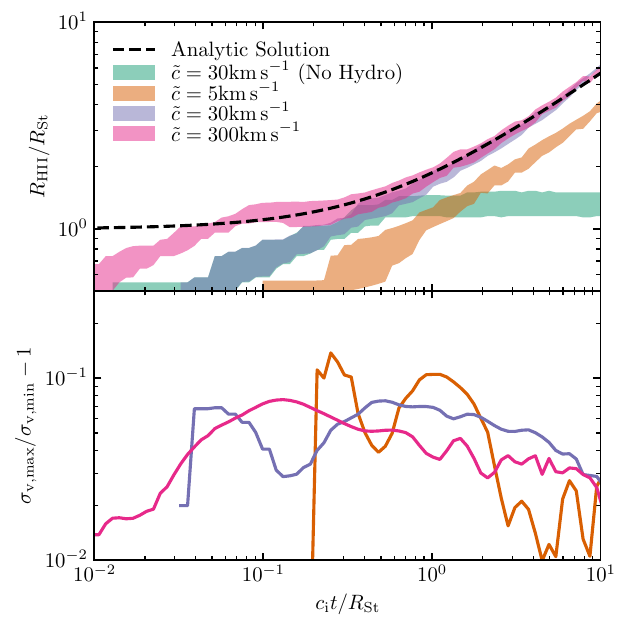}\vspace{-0.8cm}
    \caption{Evolution of an HII region surrounding a single O star in an initially-static, homogeneous box, with only ionizing radiation feedback and no photon momentum. {\it Top}: Evolution of the ionization front, plotting the interval over which the ionization fraction falls from $99\%$ to $1\%$, comparing results for a range of reduced speed of light values to the analytic solution \citep{hosokawa_2006_hii}. The ``No Hydro" solution freezes the gas in place, simulating a static Str\"{o}mgren sphere. {\it Bottom}: Numerical velocity anisotropy as a function of time (0 in the exact solution).}
    \label{fig:hiiregion}
\end{figure}

We plot the expansion of the ionization front for the different runs in Figure \ref{fig:hiiregion}. The ``No Hydro" frozen solution relaxes to the static Str\"{o}mgren sphere solution after a time $\sim t_{\rm St}= R_{\rm St}/\tilde{c}$ as expected, and remains statistically (but not exactly) static thereafter. The solutions with $\tilde{c}>30\rm km\,s^{-1}$ relax to the Str\"{o}mgren solution similarly, but then start to expand after a time $\sim R_{\rm St}/c_{\rm i}$ and agree well with the Eq. \ref{eq:hii} solution. But for $\tilde{c}=5 \rm km\,s^{-1}$, $t_{\rm St}$ is longer than the physical sound crossing time of the bubble, so the bubble expansion is delayed artificially, confirming the finding of \citet{geen_2015_rt} that $\tilde{c} \sim 30 \rm km\,s^{-1}$ is roughly the marginal RSOL value for following the dynamics of HII regions accurately. Numerical anisotropy is again small (a few per cent) once the bubble actually expands and becomes well-resolved.

\section{Discussion}
\label{sec:discussion}
We have presented, demonstrated, and tested the methods used for STARFORGE simulations, and we refer the reader to \citetalias{starforge_jets_imf} for the preliminary science results of the STARFORGE project. We now discuss some further applications of the methods presented here and enumerate several caveats, limitations, and possible extensions to our setup.

\subsection{Applications}
The particular suite of physics and numerical methods developed here is optimized and intended for GMC and star cluster formation simulations, but the methods described in this work are potentially suitable for wider applications in astrophysical simulations involving stars and feedback.

\subsubsection{Dedicated feedback simulations}
The methods we have presented for coupling feedback from individual stars do not necessarily need to be combined with star formation simulations -- in principle our feedback implementation is suitable for any problem involving stellar winds, jets, radiation, or SNe from individual stars. Notably, our methods can be used to capture multi-scale flows in complicated geometries, such as the evolution of a supernova remnant from the free-expansion phase at sub-$\rm AU$ scales onward in an inhomogeneous ISM, or following interacting binary stellar winds from the scale of the binary separation all the way to interaction with the ISM. These geometries are historically challenging for AMR methods owing to high-velocity, non-grid-aligned motion.

\subsubsection{Local and global galaxy simulations}
\label{sec:galsims}
Stratified and/or shearing-box simulations have been used to simulate the evolution of a patch of the ISM within a galaxy at a resolution that is generally higher than what is attainable in global galaxy simulations \citep[e.g.][]{hennebelle_iffrig_2014_box,Walch_2015_SILCC_ISM_SN,tigress, martizzi_2016_stratified_sne}. These can be used to follow the formation and dispersal of GMCs self-consistently. All the algorithms presented here translate directly to this type of setup -- the only differences are the initial conditions, boundary conditions, and additional inertial forces (all currently implemented in {\small GIZMO}). Resolved SF simulations could also be performed in a galactic context via a ``zoom-in" re-simulation of a GMC, taking a coarsely-resolved GMC in a simulated galaxy \citep[e.g][]{guszejnov_GMC_cosmic_evol} and up-sampling it to higher resolution \citep{reyraposo_2015_gmc_zooms}. 

The total stellar masses we form in the largest simulations in \citetalias{starforge_jets_imf} ($\gtrsim 10^4 \msun$) is within an order of magnitude of the total stellar mass in the faintest known dwarf galaxies \citep{wheeler_2019_fire_ufds, simon_2019_ufd_review}, so it may even be possible to perform a {\it global} cosmological galactic zoom-in simulation of an ultra-faint dwarf (UFD) with individually-resolved star formation. State-of-the-art simulations like \citet{wheeler_2019_fire_ufds} simulate UFD formation at a mass resolution of $\sim 30 \msun$, so much higher resolution would be required, but this could potentially be achieved by using a \citetalias{truelove_1997_dens_condition}-like refinement criterion to reach the required resolution in the dense gas, while the more diffuse gas not engaged in star formation could be kept at coarser resolution.

\subsubsection{Stellar zoom-in simulations}
Our sink particle method creates an open boundary condition for gas to flow into a stellar system, but we do not follow physics on scales smaller than $R_{\rm sink}$, for lack of resolution. Meanwhile, detailed simulations of individual protostellar systems can capture processes on smaller scales that we cannot, but lack the broader context of star cluster formation that should inform the accretion history of the system, as well as environmental effects like ionizing radiation and close encounters \citep{concharamirez_close_encounters,concharamirez_disk_ionization}. Simulations like those here, however, can be used to inform the initial conditions for simulating the formation of individual star systems at much higher ($<10^{-6}\msun$) resolution, sufficient to resolve the structure of the inner envelope and disk, and follow the dynamics of dust and non-ideal MHD with a suitable adaptive refinement scheme \citep[e.g.][]{tomida_2015_nimhd_disks,mocz_2017_core_sim, anglesalcazar_zoomin_agn}.

\subsection{Caveats and room for improvement}
\label{sec:caveats}
\subsubsection{Gravity and full N-body dynamics}

In \S\ref{sec:timestepping} we introduced a treatment of stellar dynamics using an integrator that gives superior accuracy to the usual second-order integrators used in multi-physics simulations (Fig. \ref{fig:integrator}), allowing us to preserve the properties of binaries over the GMC lifetime. However the accuracy and efficiency of our algorithms in pure N-body applications still pales in comparison to dedicated N-body codes \citep[e.g.][]{aarseth_nbody,wang_nbody6}. Standard N-body treatments do not generally require a minimum softening length, using a variety of techniques to optimize binary motion and close encounters (e.g. regularization). The gravitational force is also generally exact to machine precision in pure N-body applications, or approximate to a specified very fine, dynamically-controlled tolerance \citep{mcmillan_1993_treecode}. The approximate tree-force is not necessarily an issue, because as discussed in \S\ref{sec:timestepping} the error budget in SF simulations is dominated by errors and uncertainties in RMHD algorithms and feedback, but it is not presently clear what physics relevant to SF may be missed when softening is introduced (but we do not find qualitatively-different results with softenings as small as $1.8 \rm AU$, \S\ref{sec:jettests}). A regularization scheme would improve the efficiency and accuracy of binary integration, potentially allowing larger timesteps and optimizing the simulations. However, such methods are non-trivial to couple to multi-physics simulations (see however recent successes with the {\small AMUSE} framework, \citealt{wall_2020_amuse_torch_fb}).

\subsubsection{Radiative transfer}
Although we have validated our implementation of the M1 radiative transfer method in various simple problems relevant to SF (\S\ref{sec:radiation}), it is by no means clear that M1 can capture all important radiation phenomena in SF. As a moments method, it does not capture the collisionless nature of photons (colliding streams will shock, not pass through each other), so the radiation field streaming within a cluster will not generally be particularly accurate (but should still be reasonable if a single source is dominant).

Unfortunately the idiosyncrasies of different RT methods often only reveal themselves in complex, nonlinear problems with nontrivial geometries (such as SF), where exact solutions are unknown (as opposed to the simple problems considered here). This motivates an empirical approach to studying the behaviours of different RT methods in SF, i.e. comparing their results in the full SF problem, and determining which best reproduces observations. This will be important to do but is beyond the scope this work.

\subsubsection{Resolving disk evolution}
Our use of the sink particle method (which identifies each sink with an individual star) effectively ignores any possibility of fragmentation on scales $<R_{\rm sink}$ ($\sim 20 \rm AU$ in a typical STARFORGE application), and does not model the detailed accretion flow onto the protostar on scales smaller than the sink radius. As discussed in \S\ref{sec:sinkaccretion}, the rate at which mass arrives at the protostar need not be the same as the rate at which it enters the sink boundary, and this difference in accretion rate would ultimately influence the evolution of feedback rates from the protostar, and the surrounding environment in turn.

If fragmentation occurs frequently on these smaller scales as may happen in gravitationally unstable disks \citep{kratter10a,kratter2016}, then our predicted IMF will be incomplete for a given set of physical assumptions. Our resolution study (\S\ref{sec:jettests}) does not show any hint of IMF incompleteness as $\sim \rm AU$ scales start to become resolved, but our simulation assumes ideal MHD and hence may overestimate magnetic braking \citep{li_2011_mhd_braking}. This likely exaggerates accretion onto the central star and reduces its disk mass, suppressing any possible fragmentation. The impact of disk properties and evolution on the IMF should be investigated further with higher-resolution simulations accounting for non-ideal MHD and radiative transfer \citep[e.g.][]{Wurster_2019_no_magnetic_break_catastrophe}.

\subsubsection{Subgrid accretion and feedback modeling}

As previously discussed in \S\ref{sec:sinkaccretion}, another caveat of not following sub-AU physics is that the rate at which mass arrives at the protostar, and is launched in an outflow, must be assumed. We show that our results are insensitive to our assumed accretion rate at at least the factor of $10$ level in protostellar jet simulations (\ref{sec:jettests}), but if the flow is angular momentum-supported at unresolved scales then accretion may proceed slower still, regulated by the rate of angular momentum transport. If accretion proceeds much more slowly, then the rate of accretion-powered protostellar radiation and outflows will be reduced in turn.

Another potential issue is our assumptions about the power and collimation of protostellar outflows. We have assumed a simple parametrized model following \citet{Cunningham_2011_outflow_sim}, with parameters chosen to roughly match observations, but in reality these parameters may exhibit systematic scalings according to e.g. stellar type and accretion rate, non-ideal MHD processes and the dust grain distribution \citep{pudritz_2019_jets_review}, and the magnetic field geometry \citep{gerrard_2019_outflow_mhd_geometry}. Because protostellar outflows can have such powerful effects upon SF, efforts should be made to constrain sub-grid prescriptions. 

\subsubsection{Cooling and chemistry}
Our treatment of cooling and chemistry assumes equilibrium abundances, i.e. we do not explicitly follow the formation and destruction of the various molecular species that can serve as coolants or useful observational tracers. In principle this could affect the dynamics of the simulations, if the molecular cooling rate was severely over- or underestimated, but in practice the cold, dense initial conditions we simulate can safely be assumed to be fully molecular, and even if not the cooling rate is actually fairly insensitive to the specific species into which e.g. C and O are locked \citep{glover:2011.molecules.not.needed.for.sf}. The presence of molecules is a {\it consequence} of gas collapse, not a prerequisite \citep{orr_fire_ks}. Rather, the main utility of self-consistent chemistry in simulations is to enable the simulations to predict molecular emission self-consistently, e.g. to help determine which physical processes or which regions are being probed by different lines. In future work will explore simulations adopting detailed molecular networks such as {\small CHIMES} \citep{richings_2014a_chimes,richings_2014b_chimes}, which has been implemented in {\small GIZMO} (Richings et al., in prep).

\section{Conclusion}
\label{sec:conclusion}
We have presented and demonstrated the methods of the STARFORGE, combining the physics of MHD, gas self-gravity, stellar dynamics, thermodynamics, and all major dynamically-important (proto-) stellar feedback mechanisms into a detailed numerical model of star formation. We have shown that the respective techniques for each mechanism give satisfactory results in test problems with known solutions. We also discovered a remarkable degree of robustness in the sink particle prescription (\S \ref{sec:sinktests}), and found good agreement when comparing with results in similar problems from a code that implements the same physics with completely different numerical methods (\S \ref{sec:jettests}).

We found stable numerical results for the IMF down to a completeness limit of $\sim 0.1\rm \msun$ at the modest (by SF simulation standards) mass resolution of $\approx 10^{-3}\msun$ (\S\ref{sec:jettests}, Figs \ref{fig:jets_convergence}), and in \citetalias{starforge_jets_imf} we scale this setup up to GMCs as massive as $2 \times 10^5 \msun$, mapping out exploring the effects of protostellar jets upon the IMF at this scale for the first time. In subsequent works we will present the full results of radiation MHD simulations of those same massive GMC models, with all feedback modules described in this work acting in concert.

We anticipate that STARFORGE will be a useful theoretical laboratory for disentangling the many physical mechanisms at work in GMCs. By starting with a realistic picture and switching different physics on and off in controlled experiments, it can help distil the essential elements of a working theory of star formation. It can also be used to calibrate sub-resolution prescriptions for effects such as stellar kinematics and protostellar feedback for use in lower-resolution star cluster and galaxy formation simulations, increasing the predictive power of such simulations in the densest gas and star clusters, where the details of prescriptions become important \citep[e.g.][]{hopkins_2013_dense_gas,elephant, li_2020_smuggle_gmcs}. It should also be useful as interpretive tool for observations, mapping out the effects of different physics upon the relations between observed gas tracer properties and star formation.

An important goal of this project is to reduce the dependence of SF simulations upon sub-grid prescriptions, which must make highly-uncertain assumptions about how individual stars form. Our setup helps to accomplish this, but it only peels back one layer. To simulate feedback and the emergence of the IMF, we must make assumptions about how various sub-AU physics (accretion, stellar winds, jet launching, protostellar and stellar evolution/death) proceed, and we list various ways in which incorrect assumptions about these processes could affect our results (\S\ref{sec:caveats}). Hence it is crucial to continue to advance our understanding of the processes governing the formation and internal evolution of individual stars and star systems.

\section{Data availability}
The data supporting the plots within this article and the initial conditions used for the numerical tests are available by request to the corresponding authors. A public version of the {\small GIZMO} code is available at \url{http://www.tapir.caltech.edu/~phopkins/Site/GIZMO.html}.

\section*{Acknowledgements}
We warmly thank the theoretical and observational star formation communities for the innumerable enlightening exchanges that informed and motivated this work over the past several years. We are especially grateful to fellow starsmith Anna Rosen and to the referee Chris Matzner, whose careful readings helped improve the manuscript. MYG is supported by a CIERA Postdoctoral Fellowship. DG is supported by the Harlan J. Smith McDonald Observatory Postdoctoral Fellowship.  Support for PFH was provided by NSF Collaborative Research Grants 1715847 \&\ 1911233, NSF CAREER grant 1455342, and NASA grants 80NSSC18K0562 \&\ JPL 1589742. SSRO acknowledges support by NSF CAREER Award AST-1748571, NSF grant AST-1812747, NASA ATP grant 80NSSC20K0507  and a Cottrell Scholar Award from the Research Corporation for Science Advancement. CAFG is supported by NSF through grants AST-1715216, and CAREER award AST-1652522; by NASA through grant 17-ATP17-0067; and by a Cottrell Scholar Award from the Research Corporation for Science Advancement. 
This work used computational resources provided by XSEDE allocation AST-190018, the Frontera allocation AST-20019, and additional resources provided by the University of Texas at Austin and the Texas Advanced Computing Center (TACC; http://www.tacc.utexas.edu).




\bibliographystyle{mnras}
\bibliography{bibliography} 




\appendix

\section{List of sink particle tests}
\label{appendix:sinktests}

In \S\ref{sec:sinktests} we re-run the {\bf M2e4} GMC simulation from \citetalias{guszejnov_isothermal_mhd} with a large space of parameters and prescriptions for our sink particle algorithm, with results shown in Figure \ref{fig:sinktests}. These tests include:
\begin{itemize}
    \item Fiducial parameters and prescriptions, simply re-running the simulation in \citetalias{guszejnov_isothermal_mhd} with exactly the same code version as the other tests, with the methods described in \S\ref{sec:corephysics}.
    \item Simultaneously increasing $R_{\rm sink}$ and $S_{\rm \star}$ by $\times 10$ and $\times 100$ (from $18\rm AU$ to $180-1800\rm AU$). These tests generally formed the upper envelope of the predicted IMF mass scale statistics because accretion is made much easier and IMF incompleteness is introduced by accreting independently-collapsing cores within the sink radius. However, these effects are small.
    \item Increasing $R_{\rm sink}$ and $\times 10$ and $\times 100$ without rescaling $S_{\rm \star}$. These had similar results to tests in which we scaled $S_{\rm \star}$ as well.
    \item Decreasing $S_{\rm \star}$ from $18\rm AU$ to $1.8\rm AU$. This had negligible effects.
    \item Changing the sink formation density threshold $\rho_{\rm th}$ by a factor of $10^{-3}$ and $10^{3}$. The $\rho_{\rm th} \times 10^{-3}$ test also neglected the thermal term in the virial criterion (Eq. \ref{eq:virialcriterion}), which would otherwise have effectively imposed a density threshold of its own. This has the effect of rescaling the threshold number of Jeans wavelengths per cell at the sink formation threshold from $0.5$ to $1.58$ and $0.15$ respectively (i.e. strongly violating vs. satisfying the \citet{truelove_1997_dens_condition} criterion). All results of the $\times 10^3$ run were difficult to distinguish from the fiducial run, possibly because following collapse far beyond $\rho_{\rm J}$ is unlikely to reveal any new fragments (\S\ref{sec:jeansresolution}). The $\rho_{\rm th} \times 10^{-3}$ run formed a slightly larger number of sinks than the fiducial run, because the other sink formation criteria are not perfect predictors of runaway collapse when looking at gas of modest density, but effects were still quite small.
    \item Rescaling $R_{\rm sink}$ from $18\rm AU$ to $9\rm AU$ and $4.5 \rm AU$ respectively, while keeping the softening radius fixed at $18\rm AU$. The $R_{\rm sink}=9\rm AU$ run had no clear systematic difference from the fiducial run. The $4.5 \rm AU$ run (highlighted in Figure \ref{fig:sinktests}) was the largest outlier of our survey, forming stars with a slight delay with respect to the modal SFE, and producing a noticeably larger ($\times 2$) number of sinks, affecting the median and mean sink masses in turn. Mass-weighted statistics such as $M_{\rm 50}$ and $M_{\rm max}$ were affected more modestly, but were also systematically lower than the modal solution. Note that this is not a particularly reasonable prescription: it forces sink particles to have a volume $1/64$ the volume of a gas cell at the resolution limit, resulting in a gross mismatch between $R_{\rm sink}$, $S_{\rm \star}$, and the gas resolution scale at the time of sink formation, making it very difficult to satisfy all accretion criteria.
    \item A minimal accretion prescription requiring only $r<R_{\rm sink}$. This had negligible effects.
    \item A simple accretion prescription requiring boundedness (Eq. \ref{eq:boundednesscheck}) and $r<R_{\rm sink}$. This had negligible effects.
    \item Rescaling $R_{\rm sink}$ by a factor of $1/4$ while {\it also} rescaling $S_{\rm \star}$ by the same factor and  $\rho_{\rm th}$ by a factor of 64 so that the cell spacing at sink formation matched $R_{\rm sink}$. This was much closer to the reference run than reducing $R_{\rm sink}$ alone. 
    \item Running our fiducial sink formation prescription in conjunction with the simpler sink accretion prescription given in \citet{Bate_1995_accretion} (except neglecting the correction terms to the hydro force). This amounts to neglecting thermal and magnetic pressure in the boundedness calculation (Eq \ref{eq:boundednesscheck}), and not requiring that gas cells physically fit inside the sink volume. This had negligible effects because there is considerable redundancy between the different checks.
    \item Ignoring the $\grad \cdot \vvector$, virial, density maximum, tidal, and infall sink formation criteria in turn. These all had negligible effects except for neglecting the density maximum and virial criteria, which were at the upper envelope of number of sinks formed.
    \item Including a version of the \citet{hubber:2013.sinks} angular momentum return prescription, exerting a net torque $\mathbf{\tau} = \frac{\mathbf{J}_{\rm s}}{t_{\rm acc}}$ upon the surrounding gas to transfer angular momentum from the sink back to the gas (taking $t_{\rm acc}$ to be $500\rm yr$, likely much faster than the actual angular momentum transfer timescale in a protoplanetary disk). This had negligible effects, but may be more pronounced in problems where protostellar disks are well-resolved and angular momentum support is important.
    \item Enforcing the additional criterion of ``collapse along all 3 axes", a stricter version of the $\grad \cdot \vvector$ criterion. We check that all 3 eigenvalues of the symmetric component of $\grad \vvector$ are negative (as opposed to merely their sum $\grad \cdot \vvector$), similar to prescriptions used in \citet{Federrath_2010_sinks} and \citet{gong_2013_athena_sinks}. This had negligible effects.
\end{itemize}

\section{Resolution dependence of IMF statistics with various mass cuts}
\label{sec:masscut}
\begin{figure}
    \centering
    \includegraphics[width=\columnwidth]{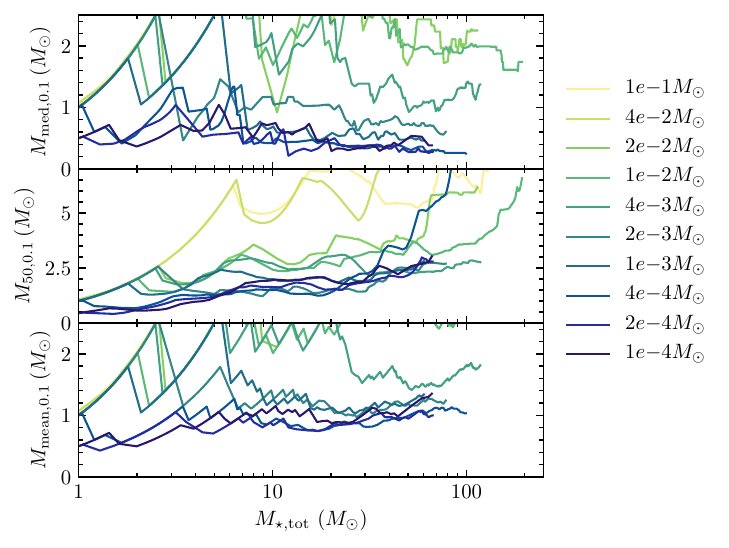}\vspace{-0.8cm}
    \caption{Effect of numerical resolution upon various IMF statistics in a GMC simulation with cooling, MHD, and jets, as in Fig. \ref{fig:jets_convergence}, but computed after cutting masses $<0.1\msun$.}
    \label{fig:imfcut1}
\end{figure}

\begin{figure}
    \centering
    \includegraphics[width=\columnwidth]{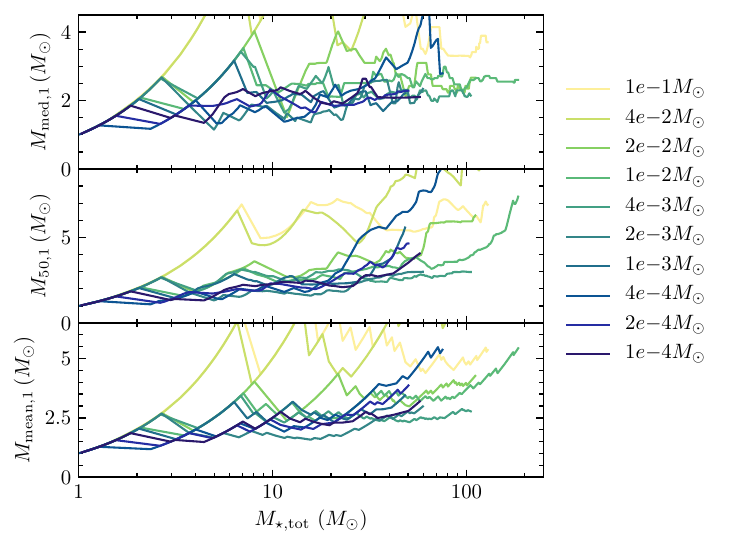}\vspace{-0.8cm}
    \caption{Effect of numerical resolution upon various IMF statistics in a GMC simulation with cooling, MHD, and jets, as in Fig. \ref{fig:jets_convergence}, but computed after cutting masses $<1\msun$.}
    \label{fig:imfcut2}
\end{figure}
In \S \ref{sec:jets} we perform a resolution study of a $2000\msun$ GMC with mass resolution ranging from $0.1-10^{-4}\msun$ with cooling, MHD, and protostellar jet physics enabled, and found that the predicted IMF statistics stablilized at sufficient resolution (Figure \ref{fig:jets_convergence}). In Figures \ref{fig:imfcut1} and \ref{fig:imfcut2} we remake the relevant panels from Fig. \ref{fig:jets_convergence} while cutting stars $<0.1\msun$ and $<1\msun$ respectively, to determine the resolution requirements for statistics computed on different mass ranges of the IMF. Cutting at $<0.1\msun$ (Fig \ref{fig:imfcut1}), a mass resolution of $\approx 2 \times 10^{-3}\msun$ appears marginally sufficient to predict the mean stellar mass, and $10^{-3}\msun$ is marginally sufficient to predict the median. Cutting at $1\msun$ (Fig. \ref{fig:imfcut2}), $0.01\msun$ is sufficient for all three statistics. This suggests that the effect of numerical resolution is simply to impose a lower completeness limit on the predicted IMF, without seriously affecting larger masses (to a point). Rigorous comparisons with the observed IMF should ideally take both observational and numerical incompleteness functions into consideration.

\section{Dust opacity fits}
\label{appendix:dustopacity}
For the opacities used in our RHD treatment of the IR band (\S\ref{sec:radiation}), we fit to results from \citet{semenov_2003} for the `porous 5-layered sphere' composition as a function of both dust temperature $T_{\rm dust}$, which determines the dust composition, and radiation temperature $T_{\rm rad}$, which affects the opacity seen by the radiation. We assume a dust sublimation temperature of $1500\rm K$, above which we assume dust to be absent and the opacity to be zero. Otherwise, if $T_{\rm dust} < 1500 \rm K$,  we use the fit
\begin{equation}
    \kappa_{\rm dust,IR} = f_{\rm d} \exp\left(0.57 \max\left(x-7,0\right)\right) \exp \left(c_1 + c_2 x + c_3 x^2 + c_4 x^3 +  c_4 x^4\right),
\end{equation}
where $x=4 \log_{\rm 10} \left(T_{\rm rad}/\rm K\right) - 8$, $f_{\rm d}$  is the local dust-to-gas ratio and the coefficients $\mathbf{c}$ vary with the dust temperature range as
\begin{equation}
    \mathbf{c} = \begin{cases} 
  \left(0.728, 0.751, - 0.0722, - 0.0116 ,0.00249 \right) &  T_{\rm dust} < 160 \rm K \\
  \left(0.166,0.701, -0.0423, - 0.0113 , 0.00213 \right) &  160\rm K \leq T_{\rm dust} < 275 \rm K \\
  \left(0.0358, 0.684, -0.0379, - 0.0113 , 0.00213 \right) &  275\rm K \leq T_{\rm dust} < 425 \rm K \\
    \left(-0.766, 0.571, -0.0123, - 0.0104 , 0.00198\right) &  425\rm K \leq T_{\rm dust} < 680 \rm K \\
    \left(-2.24, 0.812, 0.0801, 0.00862 , -0.00272\right) &  680\rm K \leq T_{\rm dust} < 1500 \rm K \\

   \end{cases}
\end{equation}

\section{Bonus}
\begin{figure*}
    \centering
    \includegraphics[width=\textwidth]{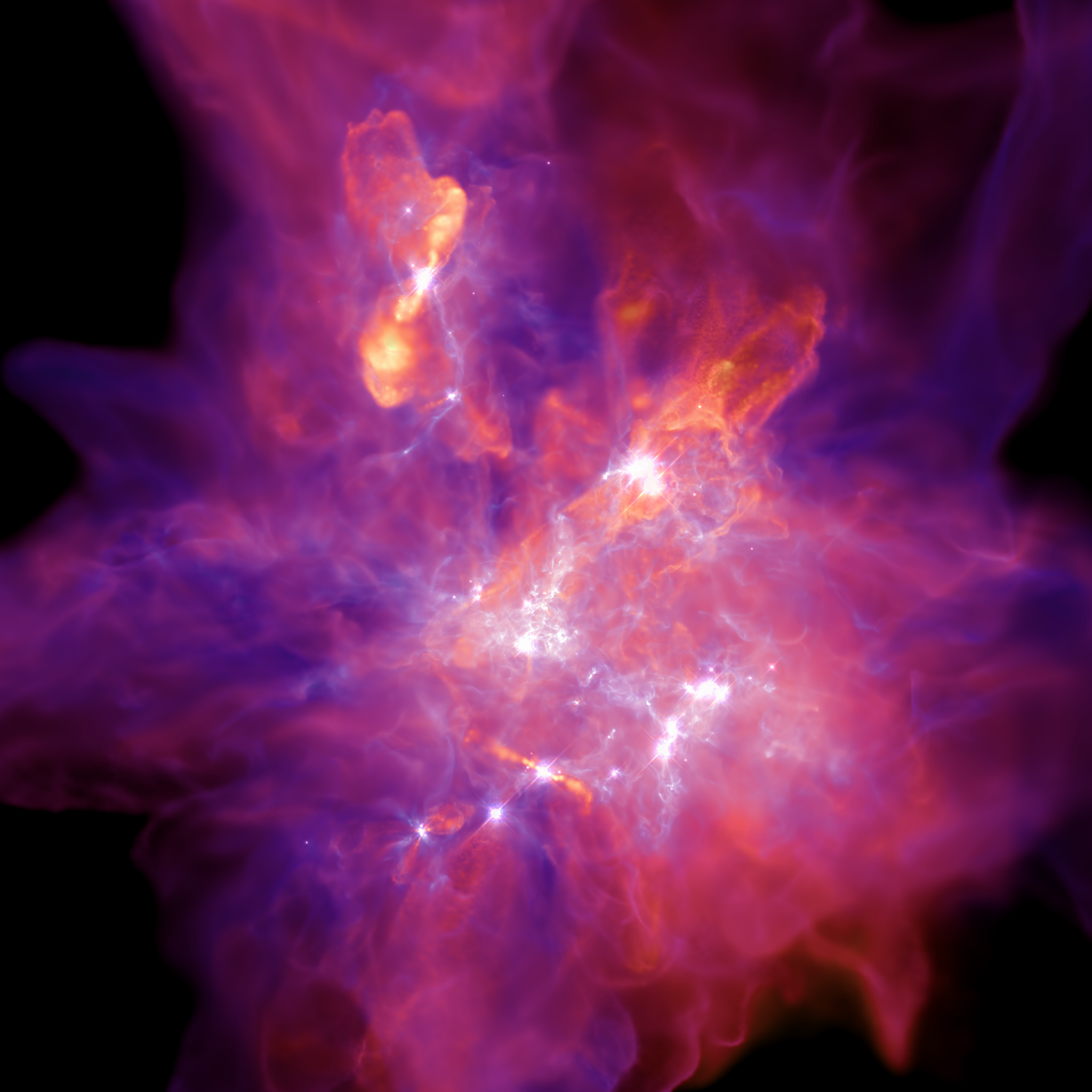}
    \caption{The ``Anvil of Creation", the first STARFORGE simulation to combine jet, wind, radiation, and supernova feedback in concert. Mock stellar point-spread functions thanks to {\small Fresco} \citep{steven_rieder_2019_3553805}.}
    \label{fig:anvil}
\end{figure*}
Wow, you made it all the way down here? Congratulations, please enjoy the non-scientific image in Figure \ref{fig:anvil}. 

\bsp	
\label{lastpage}
\end{document}